\pdfoutput=1
\documentclass[11pt,a4paper]{article} 
\usepackage{jheppub}
\usepackage{preprintcover}

\usepackage[section]{placeins}
\usepackage{float} 
\usepackage{atlasphysics}           
\usepackage{url}
\usepackage{subfigure}
\usepackage{epstopdf}
\usepackage{rotating}
\usepackage{multirow}
\usepackage{xspace}
\usepackage{hyperref}
\hypersetup{
  colorlinks=true,
  linkcolor=red,
  citecolor=red,
  urlcolor=blue
}

\def\pythia{{\sc Pythia}\xspace}
\def\herwig{{\sc Herwig}\xspace}
\def\herwigpp{{\sc Herwig++}\xspace}

\def\sherpa{{\sc Sherpa}\xspace}

\def\alpgen{{\sc Alpgen}\xspace}

\def\jimmy{{\sc Jimmy}\xspace}

\newcommand{\FastJet}       {{\sc FastJet}\xspace}

\renewcommand{\KT} {\ensuremath{k_{t}}}
\newcommand{\antikt}        {anti-\KT\xspace}
\newcommand{\ca}            {Cambridge-Aachen\xspace}
\newcommand{\nsub}          {$N$-subjettiness\xspace}

\newcommand{\dij}           {\ensuremath{\sqrt{d_{ij}}}\xspace}
\newcommand{\DOneTwo}       {\ensuremath{\sqrt{d_{12}}}\xspace}
\newcommand{\DTwoThr}       {\ensuremath{\sqrt{d_{23}}}\xspace}
\newcommand{\tauN}          {\ensuremath{\tau_{N}}\xspace}
\newcommand{\tauThrTwo}     {\ensuremath{\tau_{32}}\xspace}
\newcommand{\tauTwoOne}     {\ensuremath{\tau_{21}}\xspace}

\newcommand{\Rx}            {\ensuremath{\rho^X}\xspace}
\newcommand{\rxdata}        {\ensuremath{r^X_{\rm data}}\xspace}
\newcommand{\rxmc}        {\ensuremath{r^X_{\rm MC}}\xspace}

\newcommand{\Npv}           {\ensuremath{N_{\rm PV}}\xspace}

\newcommand{\seventev}      {\ensuremath{  7~\TeV}\xspace}

\newcommand{\sqsseven}      {\ensuremath{\sqrt{s} = \seventev}\xspace}
\newcommand{\pp}            {\ensuremath{pp}\xspace}

\newcommand{\commentout}[1]{}

\PreprintCoverPaperTitle{Jet mass and substructure of inclusive jets
  in \sqsseven \\ \pp collisions with the ATLAS experiment}
\PreprintIdNumber{CERN-PH-EP-2012-031}
\PreprintCoverAbstract{Recent studies have highlighted the potential of jet
  substructure techniques to identify the hadronic decays of boosted
  heavy particles. These studies all rely upon the assumption that the
  internal substructure of jets generated by QCD radiation is well
  understood. In this article, this assumption is tested on an
  inclusive sample of jets recorded with the ATLAS detector in 2010,
  which corresponds to 35 \ipb\ of \pp\ collisions delivered by the
  LHC at \sqsseven. In a subsample of events with single \pp\
  collisions, measurements corrected for detector efficiency and
  resolution are presented with full systematic uncertainties. Jet
  invariant mass, \KT\ splitting scales and \nsub variables are
  presented for \antikt\ $R=1.0$ jets and \ca\ $R=1.2$ jets. Jet
  invariant-mass spectra for \ca\ $R=1.2$ jets after a splitting and
  filtering procedure are also presented. Leading-order parton-shower
  Monte Carlo predictions for these variables are found to be broadly
  in agreement with data. The dependence of mean jet mass on
  additional \pp\ interactions is also explored.}
\PreprintJournalName{JHEP}

\title{Jet mass and substructure of inclusive jets
  in \sqsseven \\ \pp collisions with the ATLAS experiment}

\author{The ATLAS Collaboration}

\abstract{Recent studies have highlighted the potential of jet
  substructure techniques to identify the hadronic decays of boosted
  heavy particles. These studies all rely upon the assumption that the
  internal substructure of jets generated by QCD radiation is well
  understood. In this article, this assumption is tested on an
  inclusive sample of jets recorded with the ATLAS detector in 2010,
  which corresponds to 35 \ipb\ of \pp\ collisions delivered by the
  LHC at \sqsseven. In a subsample of events with single \pp\
  collisions, measurements corrected for detector efficiency and
  resolution are presented with full systematic uncertainties. Jet
  invariant mass, \KT\ splitting scales and \nsub variables are
  presented for \antikt\ $R=1.0$ jets and \ca\ $R=1.2$ jets. Jet
  invariant-mass spectra for \ca\ $R=1.2$ jets after a splitting and
  filtering procedure are also presented. Leading-order parton-shower
  Monte Carlo predictions for these variables are found to be broadly
  in agreement with data. The dependence of mean jet mass on
  additional \pp\ interactions is also explored.  }

\begin{document}
\maketitle
\flushbottom


\section{Introduction}
\label{sec:introduction}
The ATLAS experiment observes proton-proton (\pp) collisions provided
by the Large Hadron Collider (LHC). The outcome of these
collisions is frequently the production of large numbers of
hadrons. In order to understand these collisions, studies usually
group hadrons into jets defined by one of a number of standard
algorithms~\cite{Catani:1993hr,Ellis:1993tq,Dokshitzer:1997in,Wobisch:1998wt,Cacciari:2008gp,Salam:2007xv,Blazey:2000qt}.
The variables most often used in analyses are the jet direction
and momentum transverse to the beam (\pt). However the jets remain composite
objects and their masses and internal substructure contain
additional information.

One strong motivation for studies of the internal substructure of jets
is that at the LHC particles such as $W$ and $Z$ bosons and top quarks
are produced abundantly with significant Lorentz boosts. The same may
also be true for new particles produced at the LHC. When such
particles decay hadronically, the products tend to be collimated in a
small area of the detector. For sufficiently large boosts, the
resulting hadrons can be clustered into a single jet. Substructure
studies offer a technique to extract these single jets of interest
from the overall jet background. Such techniques have been found
promising for boosted $W$ decay identification, Higgs searches and
boosted top identification amongst others~\cite{boost2010}. However,
many of these promising approaches have never been tested with
collision data and rely on the assumption that the internal structure
of jets is well modelled by parton-shower Monte Carlo approaches. It
is therefore important to measure some of the relevant variables in a
sample of jets to verify the expected features.

In this paper, measurements are made with an inclusive sample of
high-transverse momentum jets produced in proton-proton collisions
with a centre-of-mass energy ($\sqrt{s}$) of 7 TeV. This is a natural
continuation of the studies in previous
experiments~\cite{Aaltonen:2011pg, ZEUS:subjet2009, CDF:subjet2005,
  ZEUS:subjet2004, ZEUS:subjet2003}. It also complements previous
ATLAS studies~\cite{PhysRevD.83.052003} probing the shape of jets
reconstructed with the \antikt algorithm~\cite{Cacciari:2008gp} with
smaller radius parameters $R=$ 0.4 and 0.6.

This study focuses on two specific jet algorithms that are likely to
be of interest for future searches: \antikt jets with an $R$-parameter
of 1.0 and \ca~\cite{Dokshitzer:1997in,Wobisch:1998wt} jets with
$R=1.2$. Jets are required to be at high-transverse momentum
($\pt>200\GeV$) and central in rapidity\footnote{ATLAS uses
  a right-handed coordinate system with its origin at the nominal
  interaction point (IP) in the centre of the detector and the
  $z$-axis along the beam pipe. The $x$-axis points from the IP to the
  centre of the LHC ring, and the $y$-axis points upward. Cylindrical
  coordinates $(r,\phi)$ are used in the transverse plane, $\phi$
  being the azimuthal angle around the beam pipe. The pseudorapidity
  is defined in terms of the polar angle $\theta$ as
  $\eta=-\ln\tan(\theta/2)$. Rapidity is $y = 1/2 \times ln((E +
  p_z)/(E - p_z))$.} ($|y|<2$). The normalised cross-section as a function of
jet mass, taken from the jet four-momentum, is measured for both these
algorithms. In addition to the mass, two sets of substructure
variables, \KT\ splitting scales~\cite{Butterworth:2002tt} and \nsub\
ratios~\cite{Thaler:2010tr}, are measured. For the \ca\ jets, the mass
distribution after a substructure splitting and filtering
procedure~\cite{Butterworth:2008iy} is also presented.

\section{Definitions}
\label{sec:substructure}

\subsection{Jet algorithms}
Jets are constructed using two infrared and collinear-safe
recombination jet algorithms, \antikt\ and \ca, as implemented in the
\FastJet package~\cite{fastjet}. Both act by iteratively merging the
nearest objects in the event but with different definitions of
``distance''. The \antikt\ jet algorithm builds jets with a very
regular shape, while the \ca\ algorithm builds less regular jets. This
analysis uses $R$-parameters of 1.0 and 1.2, for \antikt\ and \ca\
jets respectively, in line with recent studies of heavy boosted
objects. All discussion of jet algorithms in this paper assumes that
the constituents from which the jet is composed are recombined using
four-vector addition. Jets in the real or simulated detector are
constructed from clusters of energy deposited in the calorimeter (see
Section~\ref{sec:selection}). Jets in the ``true'' final state, or
``hadron level'', are defined as jets made from all particles with a
proper lifetime longer than 10 ps, including neutrinos.

\subsection{The splitting and filtering procedure}
The ``splitting and filtering'' procedure aims to identify relatively
hard, symmetric splittings in a jet that contribute significantly to
the jet's invariant mass. This procedure is taken from recent Higgs
search studies~\cite{Butterworth:2008iy,ATLASHV}. The parameters are
tuned to maximise sensitivity to a Standard Model Higgs boson decaying
to \bbbar, but this procedure is suitable generally for identifying
two-body decay processes. The effect of the procedure is to search for
jets where clustering the constituents with \ca\ combines two
relatively low mass objects to make a much more massive object. This
indicates the presence of a heavy particle decay. The procedure then
attempts to retain only the constituents believed to be related to the
decay of this particle. Because the procedure itself uses the \ca\
algorithm, it is most natural to apply it to jets originally found
with this algorithm.

Each stage in the clustering combines two objects $j_1$ and $j_2$ to
make another object $j$. Use definitions $v =
\frac{\mathrm{min}(p^2_{Tj1}, p^2_{Tj2})}{m^2_j} \delta R^2_{j1,j2}$
and $\delta R_{j1,j2} = \sqrt{\delta y^2_{j1,j2} + \delta
  \phi^2_{j1,j2}}$, where $\delta y$ and $\delta \phi$ are the
differences in rapidities and azimuthal angles respectively. The
procedure takes a jet to be the object $j$ and applies the following:

\vspace{0.5cm}

\begin{enumerate}

\item Undo the last clustering step of $j$ to get $j_1$ and
  $j_2$. These are ordered such that their mass has the property
  $m_{j1}>m_{j2}$. If $j$ cannot be unclustered (i.e. it is a single
  particle) or $\delta R_{j1,j2} < 0.3$ then it is not a suitable
  candidate, so discard this jet.

\item If the splitting has $m_{j1} / m_j <\mu$ (large change in jet
  mass) and $v > v_{\mathrm{cut}}$ (fairly symmetric) then continue,
  otherwise redefine $j$ as $j_1$ and go back to step 1. Both $\mu$
  and $v$ are parameters of the algorithm.

  \item Recluster the constituents of the jet with the Cambridge-Aachen
        algorithm with an $R$-parameter of $R_{filt} = \mathrm{min}(0.3,
        \delta R_{j1,j2} / 2)$ finding $n$ new subjets $s_1$, $s_2 \ldots
        s_n$ ordered in descending \pt.

  \item Redefine the jet as the sum of subjet four-momenta $\sum_{i=1}^{\mathrm{min}(n,3)} s_i$.

\end{enumerate}

\vspace{0.5cm}

The algorithm parameters $\mu$ and $v_{\mathrm{cut}}$ are taken as
$0.67$ and $0.09$ respectively~\cite{ATLASHV}.

\vspace{0.5cm}

The $\mu$ cut attempts to identify a hard structure in the
distribution of energy in the jet, which would imply the decay of a
heavy particle. The cut on $v$ further helps by suppressing very
asymmetric decays of the type favoured by splittings of quarks and
gluons. A notable modification of the original
procedure~\cite{Butterworth:2008iy} in this paper is the addition of
the $\delta R_{j1,j2}$ cut in step 1. This cut is applied because with
current techniques the correction for detector resolution at angular
scales below 0.3 is not well controlled. Steps 3 and 4
filter out some of the particles in the candidate jet, the aim being
to retain particles relevant to the hard process while reducing the
contribution from effects like underlying event and pile-up. The
4-vector after step 4 can be treated like a new jet. This new jet has
a \pt\ and mass less than or equal to those of the original jet.

\subsection{\KT\ splitting scales, \dij}
The \KT\ splitting scales are defined by reclustering the constituents
of the jet with the \KT\ recombination algorithm~\cite{Catani:1993hr, Ellis:1993tq}. The
\KT-distance of the final clustering step can be used to define a splitting
scale variable \DOneTwo:

$$\sqrt{d_{12}} = \mathrm{min}(p_{Tj1},p_{Tj2}) \times \delta R_{j1,j2},$$

\vspace{0.5cm}

\noindent where $1$ and $2$ are the two jets before the final
clustering step~\cite{Butterworth:2002tt}. The ordering of clustering
in the \KT\ algorithm means that in the presence of a two-body heavy
particle decay the final clustering step will usually be to combine
the two decay products. The parameter \DOneTwo can therefore be used
to distinguish heavy particle decays, which tend to be more
symmetric, from the largely asymmetric splittings of quarks and
gluons. The expected value for a heavy particle decay is approximately
$m/2$, whereas inclusive jets will tend to have values $\sim \pt/10$,
although with a tail extending to high values. The variable
$\sqrt{d_{23}}$ is defined analogously but for the two objects
combined in the penultimate clustering step.

\subsection{\nsub}
The \nsub\ variables \tauN~\cite{Thaler:2010tr} are designed to be
smooth, continuous observables related to the subjet
multiplicity. Intuitively, the variables can be thought of as
answering the question: ``How much does this jet look like $N$
different subjets?'' The variable \tauN is calculated by clustering
the constituents of the jet with the \KT\ algorithm and requiring $N$
subjets to be found. These $N$ subjets define axes within the jet around
which the jet constituents may be concentrated.  The variables \tauN
are then defined as the following sum over all constituents $k$ of the
jet:
\begin{eqnarray}
  \tauN &=& \frac{1}{d_{0}} \sum_k p_{T,k} \times \mathrm{min} (\delta R_{1,k}, \delta R_{2,k}, \ldots, \delta R_{N,k}) \label{eq:tauN} \\
  d_{0} &=& \sum_k p_{T,k} R,
\label{eq:tauN:d0}
\end{eqnarray}
where $\delta R_{i,k}$ is the distance from the subjet $i$ to the
constituent $k$ and $R$ is the $R$-parameter of the original jet algorithm.

Using this definition, \tauN describes how well the substructure of
the jet is described by $N$ subjets by assessing the degree to which
constituents are localized near the axes defined by the \KT\
subjets. For two- and three-body decays, respectively, the ratios
$\tau_{2}/\tau_{1}$ and $\tau_{3}/\tau_{2}$ have been shown to provide
excellent discrimination for hadronic decays of $W$-bosons and boosted
top quarks~\cite{Thaler:2011gf}. These ratios will be referred to as
$\tau_{21}$ and $\tau_{32}$ respectively. These variables mostly fall
within the range 0 to 1. As an example, $\tau_{21} \simeq 1$
corresponds to a jet which is narrow and without substructure;
$\tau_{21} \simeq 0$ implies a jet which is much better described by
two subjets than one. Similarly low values of $\tau_{32}$ imply a jet
which is much better described by three subjets than two. However, as
can be seen from the definition, adding an additional subjet axis will
tend to reduce the value of \tauN and therefore even narrow jets tend
to have values of $\tau_{21}$ and $\tau_{32}$ slightly less than 1.

\section{The ATLAS Detector}
\label{sec:detector}

The ATLAS detector~\cite{Aad:2008zzm} provides nearly full solid angle
coverage around the collision point with tracking detectors, calorimeters
and muon chambers. Of these subsystems the most relevant to
this study are the inner detector, the barrel and endcap calorimeters,
and the trigger system.

The inner detector is a tracking detector covering the range
$|\eta|<2.5$ and with full coverage in $\phi$. It is composed of a
silicon pixel detector, a silicon microstrip detector and a transition
radiation tracker. The whole system is immersed in a 2 T magnetic
field. The information from the inner detector is used to reconstruct
tracks and vertices.

The barrel and endcap calorimeters cover the regions $|\eta| \lesssim
1.5$ and $1.5 \lesssim |\eta| < 3.2$, respectively. Electromagnetic
measurements are provided by a liquid-argon (LAr) sampling
calorimeter. The granularity of this detector ranges from $\delta \eta
\times \delta \phi = 0.025 \times 0.025$ to $0.1 \times 0.1$. Hadronic
calorimetry in $|\eta| < 1.7$ is provided by a scintillating-tile
detector, while in the endcaps, coverage is provided by a second LAr
system. The granularity of the hadronic calorimetry ranges from $0.1
\times 0.1$ to $0.2 \times 0.2$.

The trigger system~\cite{Aad:2012xs} is composed of three consecutive
levels. Only the Level-1 (L1) trigger is used in this study, with
higher levels not rejecting any events. The L1 trigger is based on
custom-built hardware that processes events with a fixed latency of
2.5 $\mu$s. Events in this analysis are selected based on their L1
calorimeter signature. The L1 calorimeter trigger uses coarse detector
information to identify interesting physics objects above a given
transverse energy ($E_{T}$) threshold. The jet triggers use a sliding
window algorithm taking square $\delta \eta \times \delta \phi = 0.2
\times 0.2$ jet elements as input. The window size is $0.8
\times 0.8$.

\section{Dataset and Reconstruction}
\label{sec:selection}
The data analysed here come from the 2010 \sqsseven\ \pp\ dataset. Data
are used in this study only if the detector conditions were stable,
there was a stable beam present in the LHC, the luminosity was reliably
monitored and the trigger was operational. The selected data set
corresponds to an integrated luminosity of $35.0 \pm 1.1$~\ipb
~\cite{Aad:2011dr,luminosity}.

Events in this analysis are first selected by the L1 calorimeter
trigger system. The efficiency of this trigger was evaluated in data
and found to contain no significant biases for the selection used
here. For the lowest \pt\ bin (200--300~\GeV) a trigger is used which
was only available for part of the dataset. As a result some plots are
presented with the lower integrated luminosity of $2.0 \pm 0.1$~\ipb.

To reject events that are dominated by detector noise or non-collision
backgrounds, events are required to contain a primary vertex
consistent with the LHC beamspot, reconstructed from at least five
tracks with $\pt > 150$~\MeV. Additionally, jets are reconstructed with the \antikt
algorithm using an $R$-parameter of 0.6. Events are discarded if any
such jet with transverse momentum greater than 30 GeV fails to satisfy
a number of quality criteria, including requirements on timing and calorimeter
noise~\cite{Aad:2011he}. This selection
removes approximately 3\% of events in this dataset.

Additional proton-proton collisions (pile-up) can have a significant
impact on quantities like jet mass and
substructure~\cite{boost2010}. The primary results in this paper are
therefore presented only in events where the number of reconstructed
primary vertices ($N_\mathrm{PV}$) composed of at least five tracks is
exactly one. This requirement selects approximately $22\%$ of events
in the 2010 dataset. As vertex finding is highly efficient, this
approach is expected to be very good at rejecting pile-up, and no
additional systematic uncertainties as a result of this requirement
are considered. The effects of pile-up are discussed in more detail in
Section~\ref{sec:pu}.

Calorimeter cells are clustered using a three-dimensional topological
algorithm. These clusters provide a three-dimensional representation
of energy depositions in the calorimeter with a nearest neighbour
noise suppression algorithm~\cite{TopoClusters}. The resulting
clusters are made massless and then classified as either
electromagnetic or hadronic in origin based on their shape, depth and
energy density. Cluster energies are corrected with calibration
constants, which depend on the cluster classification to account for
calorimeter non-compensation~\cite{Aad:2011he}. The clusters are then used as input to a
jet algorithm.

As part of this study, specific calibrations for these jet algorithms
have been devised. Calibrations for the mass, energy and $\eta$ of
jets are derived from Monte Carlo (specifically
\pythia~\cite{pythia}). Hadron-level jets (excluding muons and
neutrinos) are matched to jets reconstructed in the simulated
calorimeter. The matched pairs are used to define functions for these
three variables, dependent on energy and $\eta$, which on average
correct the reconstructed quantities back to the true scale. This
correction is of the order 10-20\% for mass and energy and 0.01 for
$\eta$.

Jets constructed from tracks are used for systematic studies in this
paper. These track-jets are constructed using the same algorithms as
calorimeter jets. The input constituents are inner-detector tracks
originating only from the selected \pp\ collision of interest as
selected by the criteria $\pt > 500 \MeV$, $|\eta| < 2.5$, $|z_{0}| <
5$~mm and $|d_{0}| < 1.5$~mm~\cite{minbiaspaper}. Here $z_0$ and $d_0$
are the longitudinal and transverse impact parameter of the track at
closest approach to the $z$-axis, relative to the primary vertex.

The measurements presented in this paper are for jets that have
$|y|<2$ in four 100~\GeV\ \pt\ bins spanning 200 to 600~\GeV. This
selection is not biased by trigger effects and the jets it selects are
contained entirely within the barrel and end-cap subdetectors.

\section{Monte Carlo Samples}
\label{sec:mc}
Samples of inclusive jet events were produced using several Monte
Carlo (MC) generators including \pythia 6.423~\cite{pythia} and
\herwigpp\ 2.4~\cite{Bahr:2008pv}. These programs implement
leading-order (LO) perturbative QCD (pQCD) matrix elements for $2
\rightarrow 2$ processes. Additionally, \alpgen\ 2.13~\cite{alpgen}
and \sherpa\ 1.2.3~\cite{Gleisberg:2008ta} are used for some
cross-checks. \sherpa\ and \alpgen\ implement $2 \rightarrow n$
processes such as explicit QCD multijet production. Parton-showers are
calculated in a leading-logarithm approximation. Showers are $p_T$
ordered in \pythia and angular ordered in \herwigpp. Fragmentation
into particles is implemented in \pythia following the string
model~\cite{stringmodel} and in \herwigpp\ the
cluster~\cite{clustermodel} model. \alpgen\ is interfaced with
\herwig~\cite{herwig1,herwig2} for parton-shower and fragmentation and
\jimmy~\cite{jimmy} for an underlying event model. Unless otherwise
specified, \pythia samples use the AMBT1
tune~\cite{minbiaspaper}. In some figures the Perugia2010
\pythia tune is used~\cite{Skands:2010ak}, which has been found to
describe jet shapes more accurately at
ATLAS~\cite{PhysRevD.83.052003}. Leading-order parton density
functions are taken from the MRST2007 LO* set~\cite{Martin:2009iq,
Sherstnev:2007nd}, unless stated otherwise. No pile-up was included in
any of these samples.

The MC generated samples are passed through a full
simulation~\cite{SimEPJC} of the ATLAS detector and trigger, based on
GEANT4~\cite{geant4}. The Quark Gluon String Precompound (QGSP) model
is used for the fragmentation of nuclei, and the Bertini cascade
(BERT) model for the description of the interactions of the hadrons in
the medium of the nucleus~\cite{G4Had}.

\section{Detector-level Distributions}

\begin{figure}
  \begin{center}
     \includegraphics[width=0.45\linewidth]{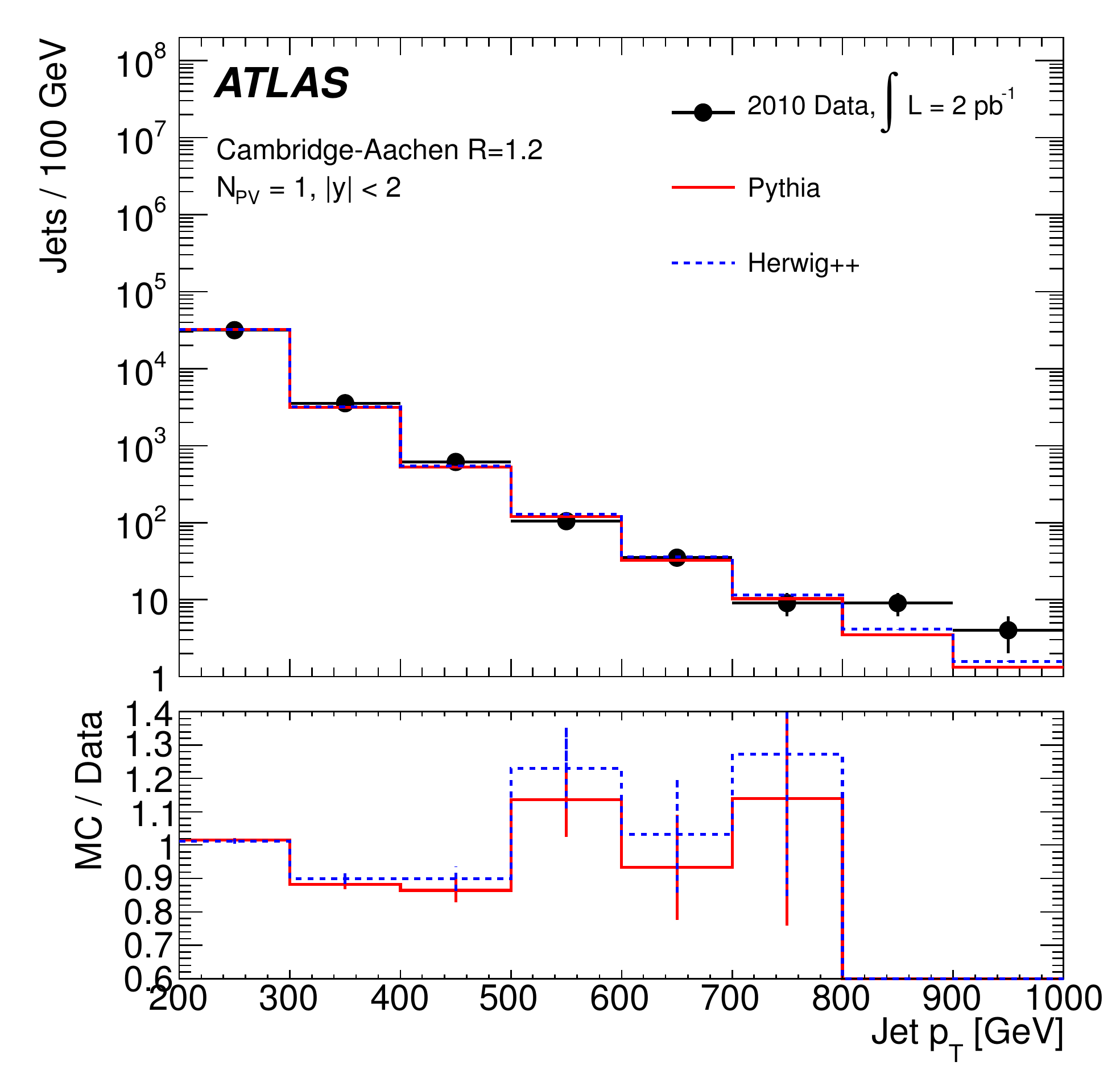}
     \includegraphics[width=0.45\linewidth]{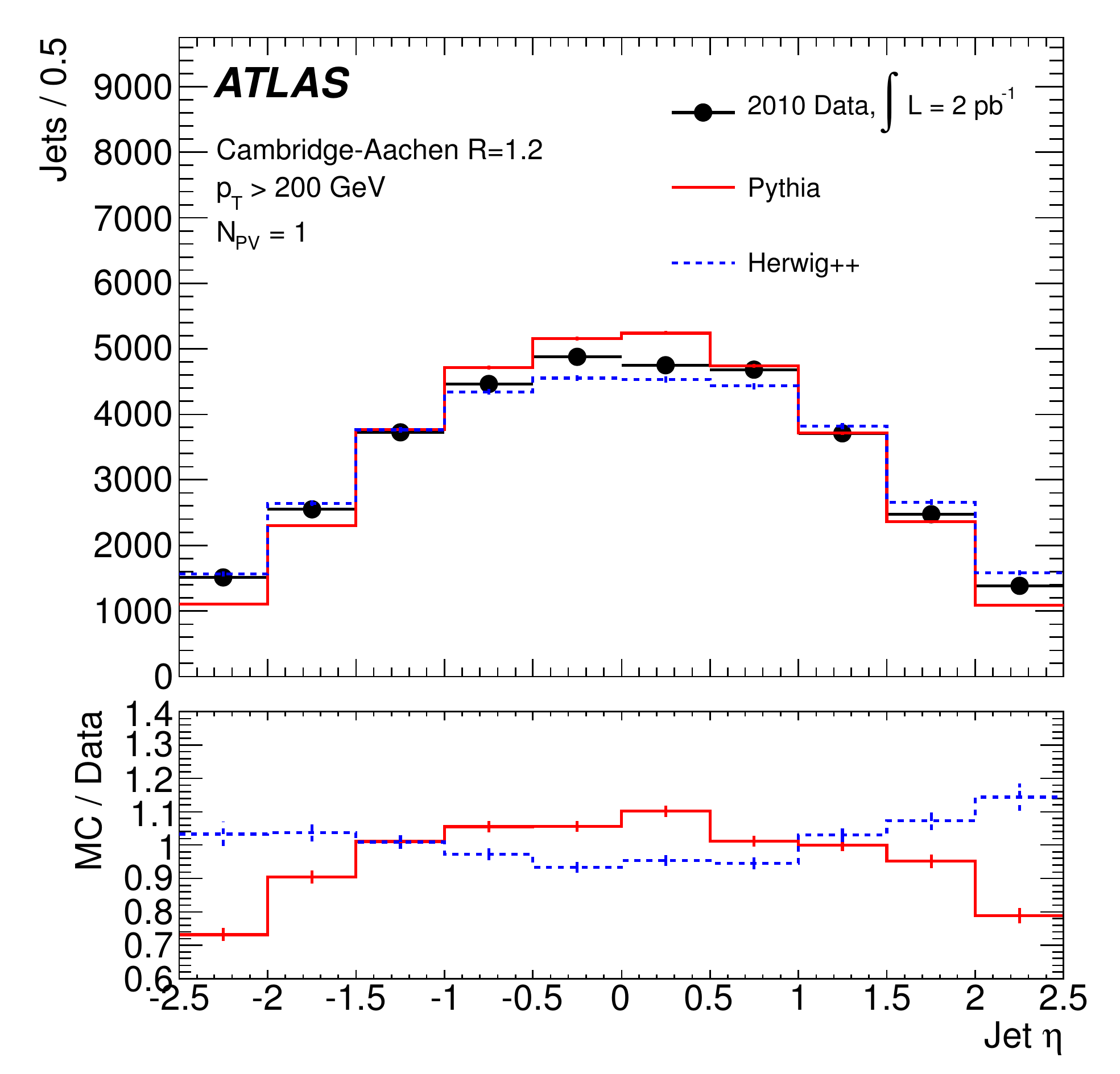}
  
  \caption{\pt\ (left) and $\eta$ distribution (right) of \ca\ $R=1.2$ jets with $\pt
    > 200\GeV$.\label{fig:ca12_pteta}}
  \end{center}
\end{figure}

\begin{figure}
  \begin{center}
     \includegraphics[width=0.45\linewidth]{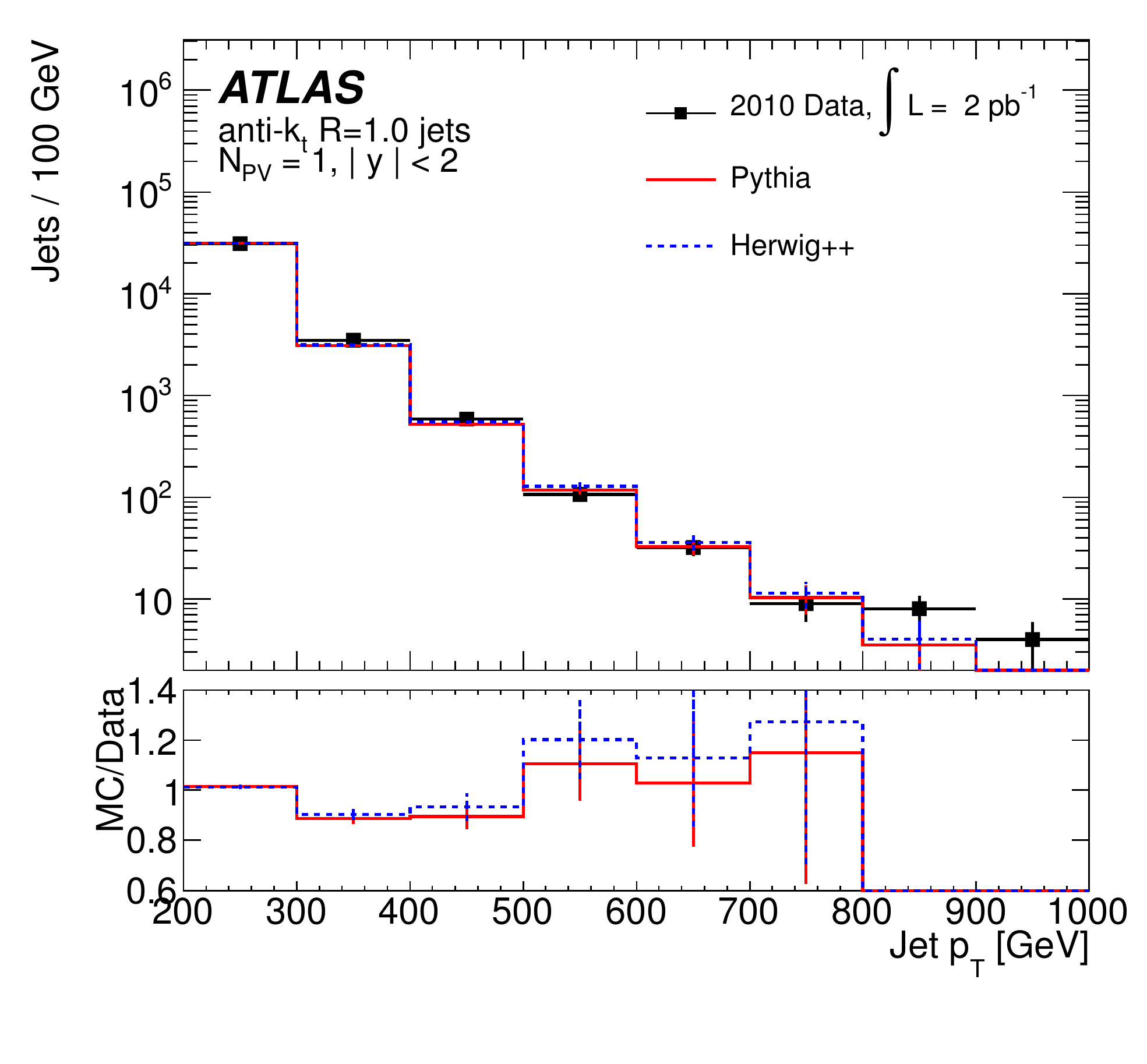}
     \includegraphics[width=0.45\linewidth]{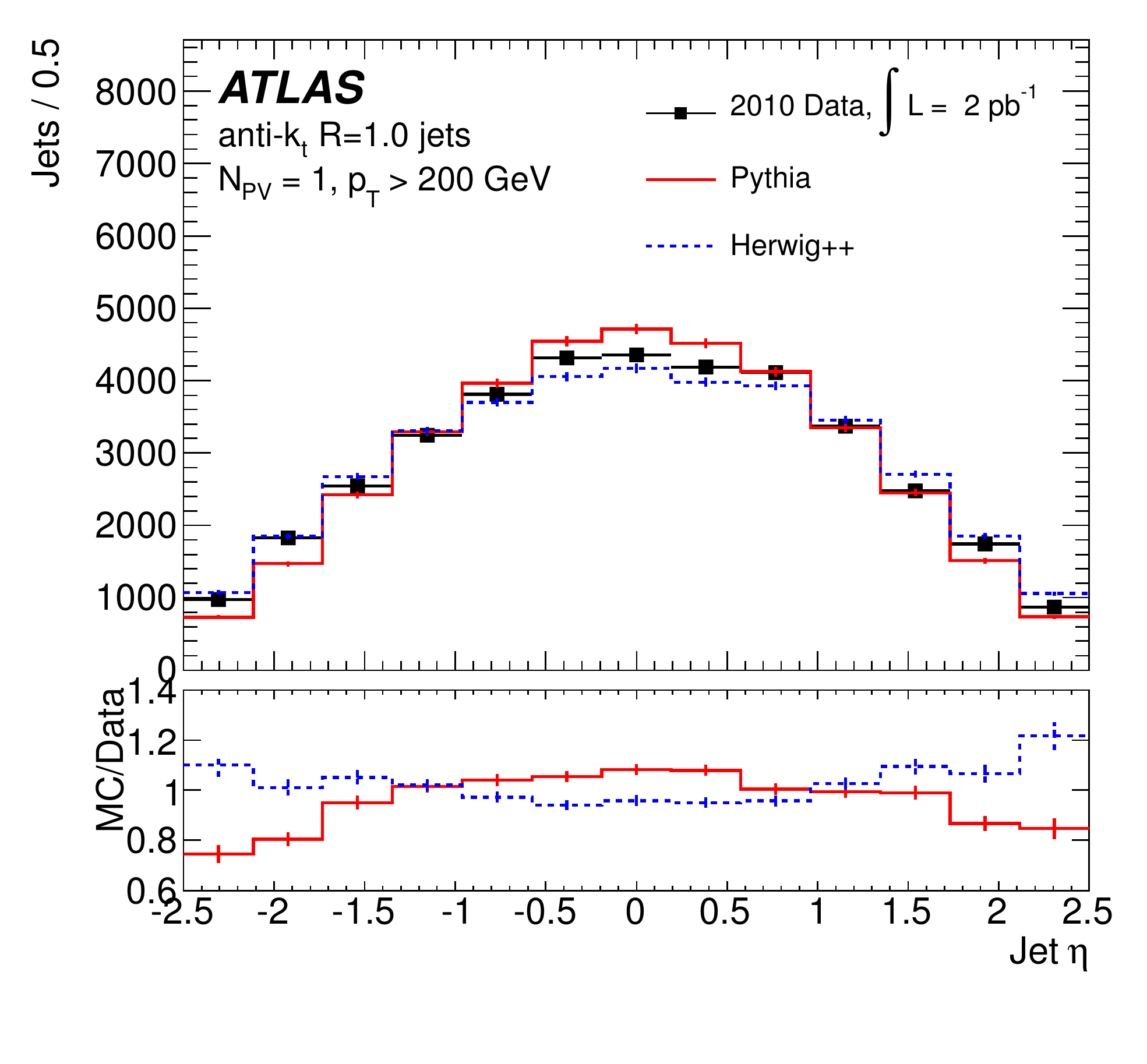}
  
  \caption{\pt\ (left) and $\eta$ distribution (right) of \antikt\ $R=1.0$ jets with
    $\pt > 200\GeV$.\label{fig:akt10_pteta}}
  \end{center}
\end{figure}

\begin{figure}
  \begin{center}
     \includegraphics[width=0.45\linewidth]{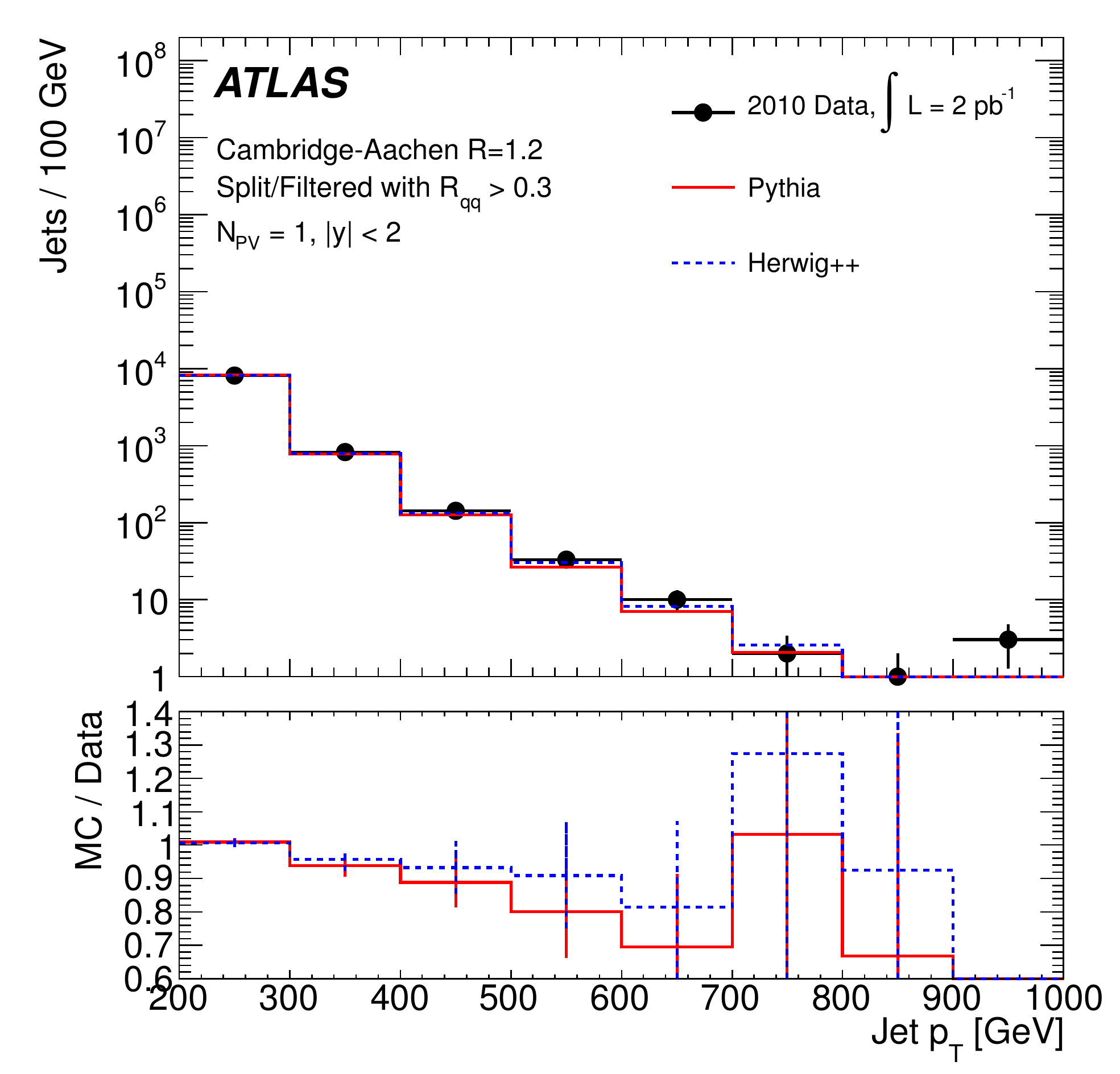}
     \includegraphics[width=0.45\linewidth]{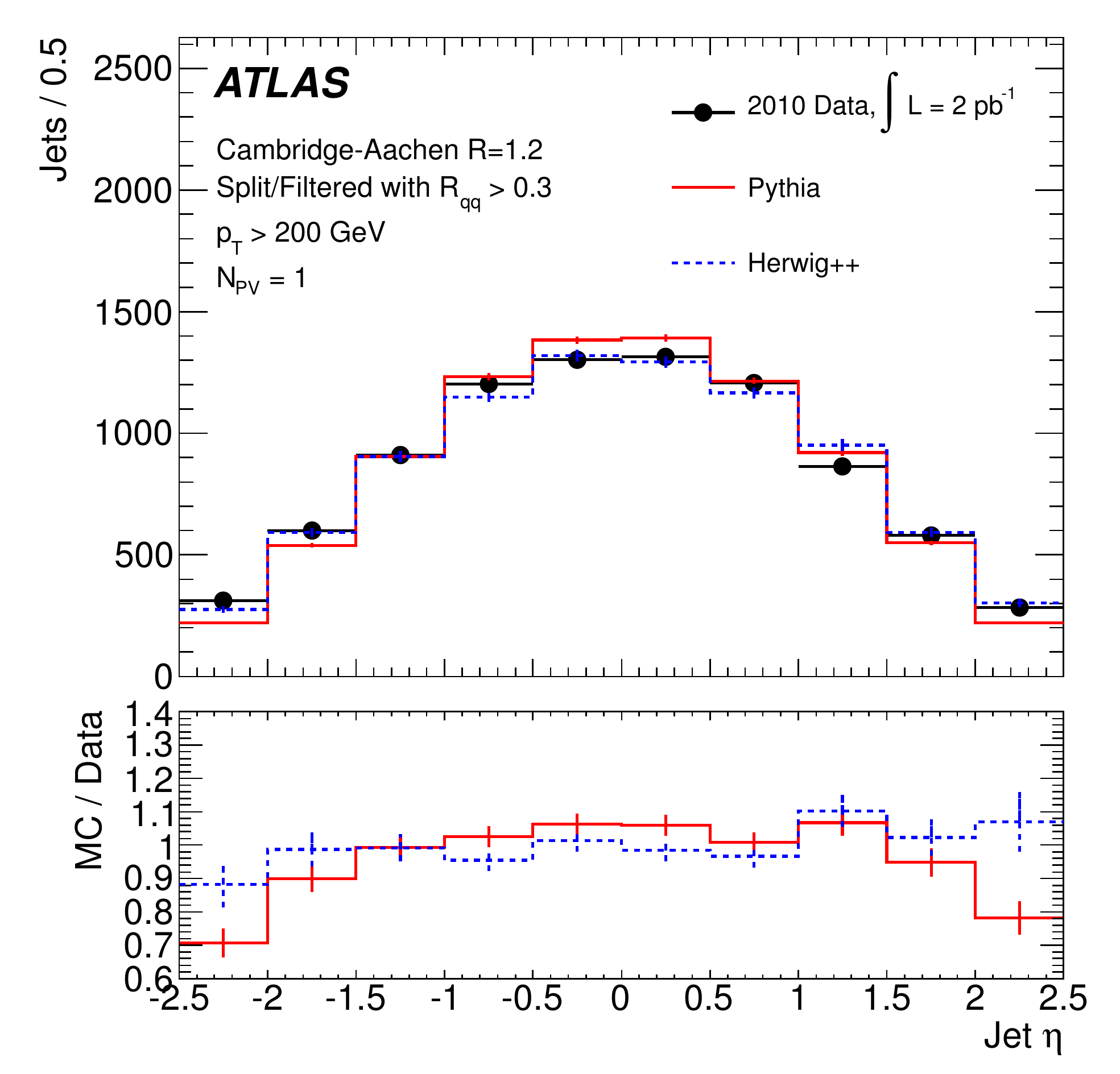}

  \caption{\pt\ (left) and $\eta$ distribution (right) of \ca\ $R=1.2$ jets after splitting and filtering with $\pt > 200\GeV$.  \label{fig:ca12filt_pteta}}
  \end{center}
\end{figure}

\begin{figure}
  \begin{center}

     \includegraphics[width=0.45\linewidth]{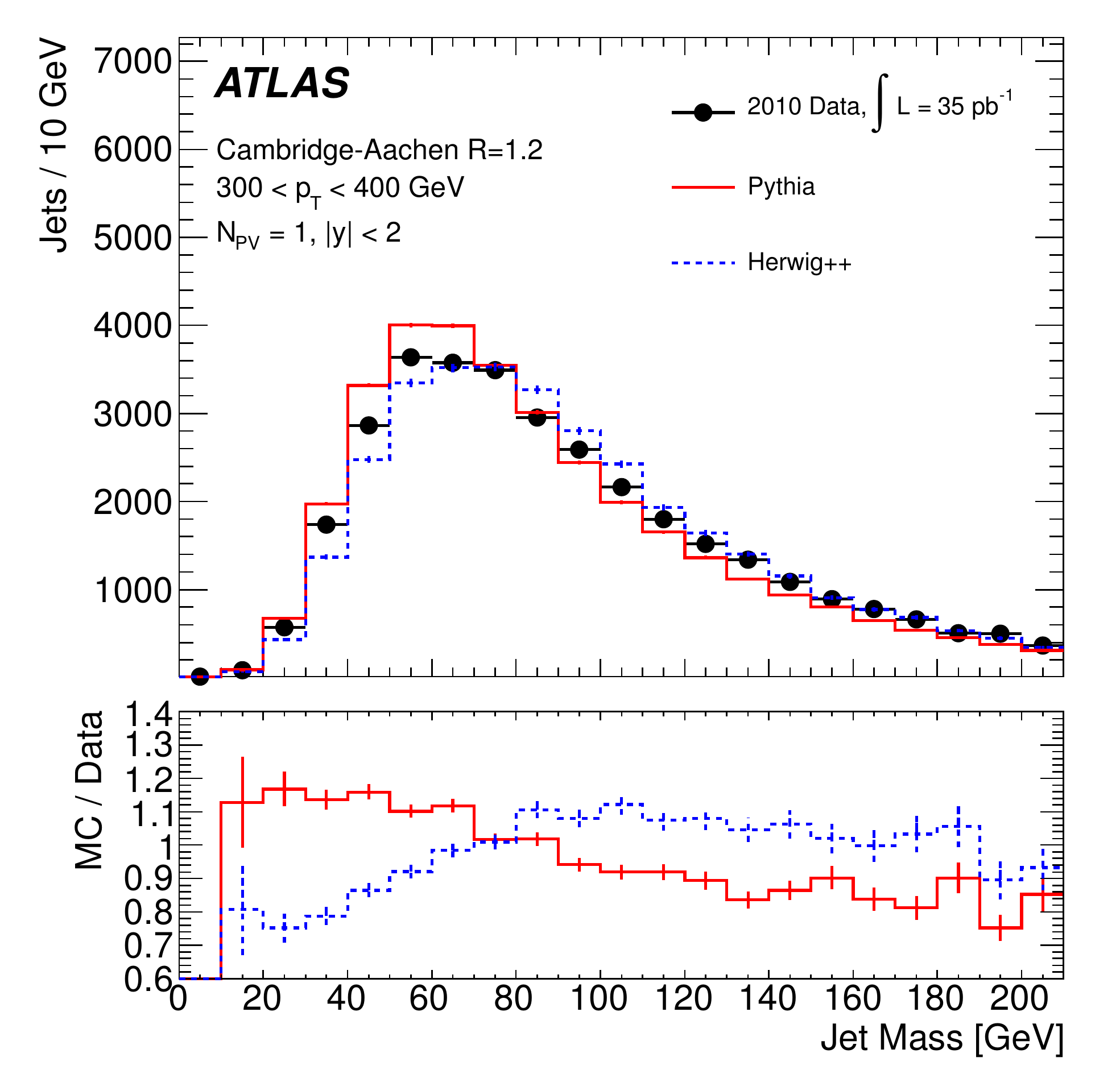}
     \includegraphics[width=0.45\linewidth]{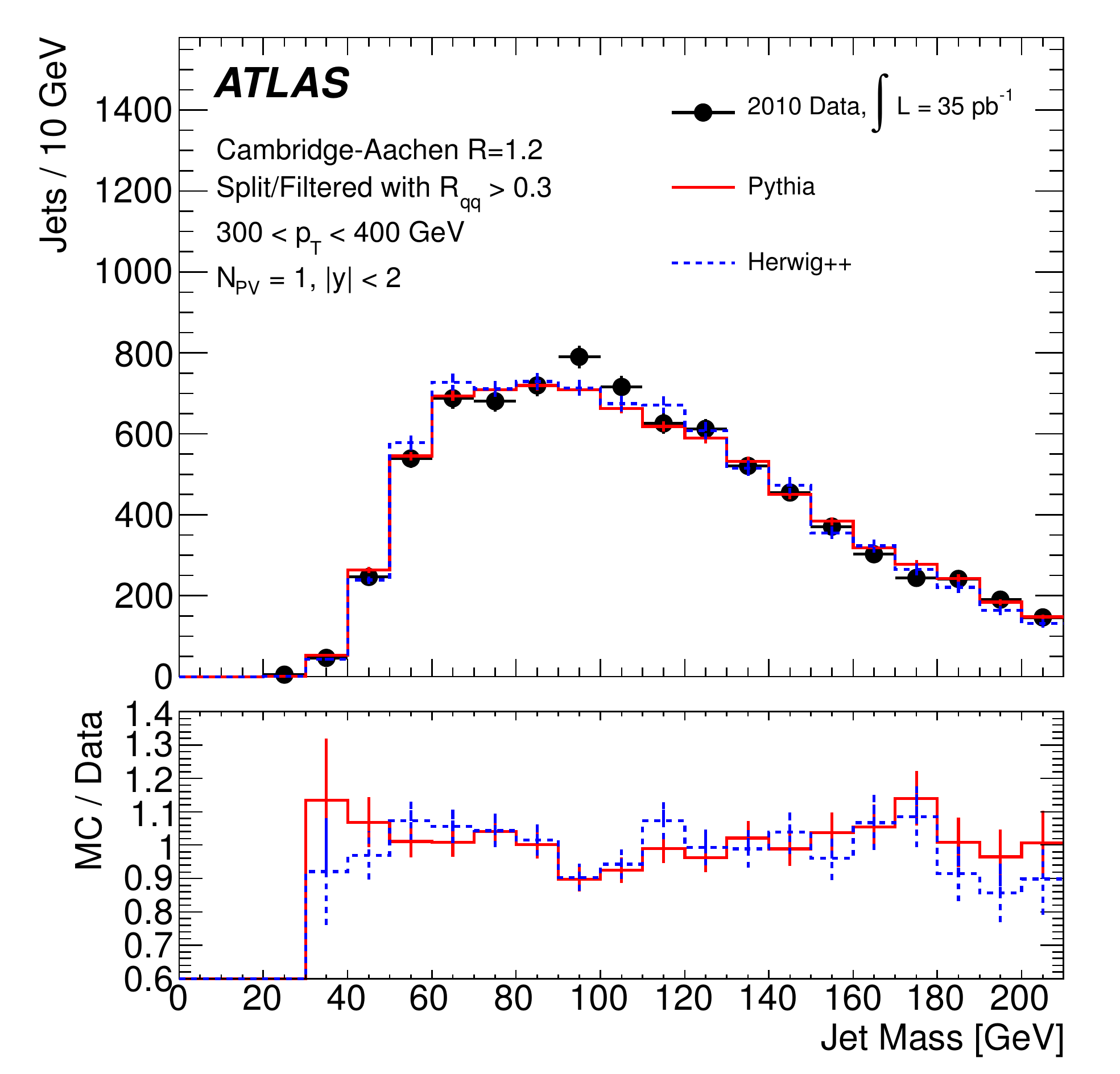}
     \includegraphics[width=0.45\linewidth]{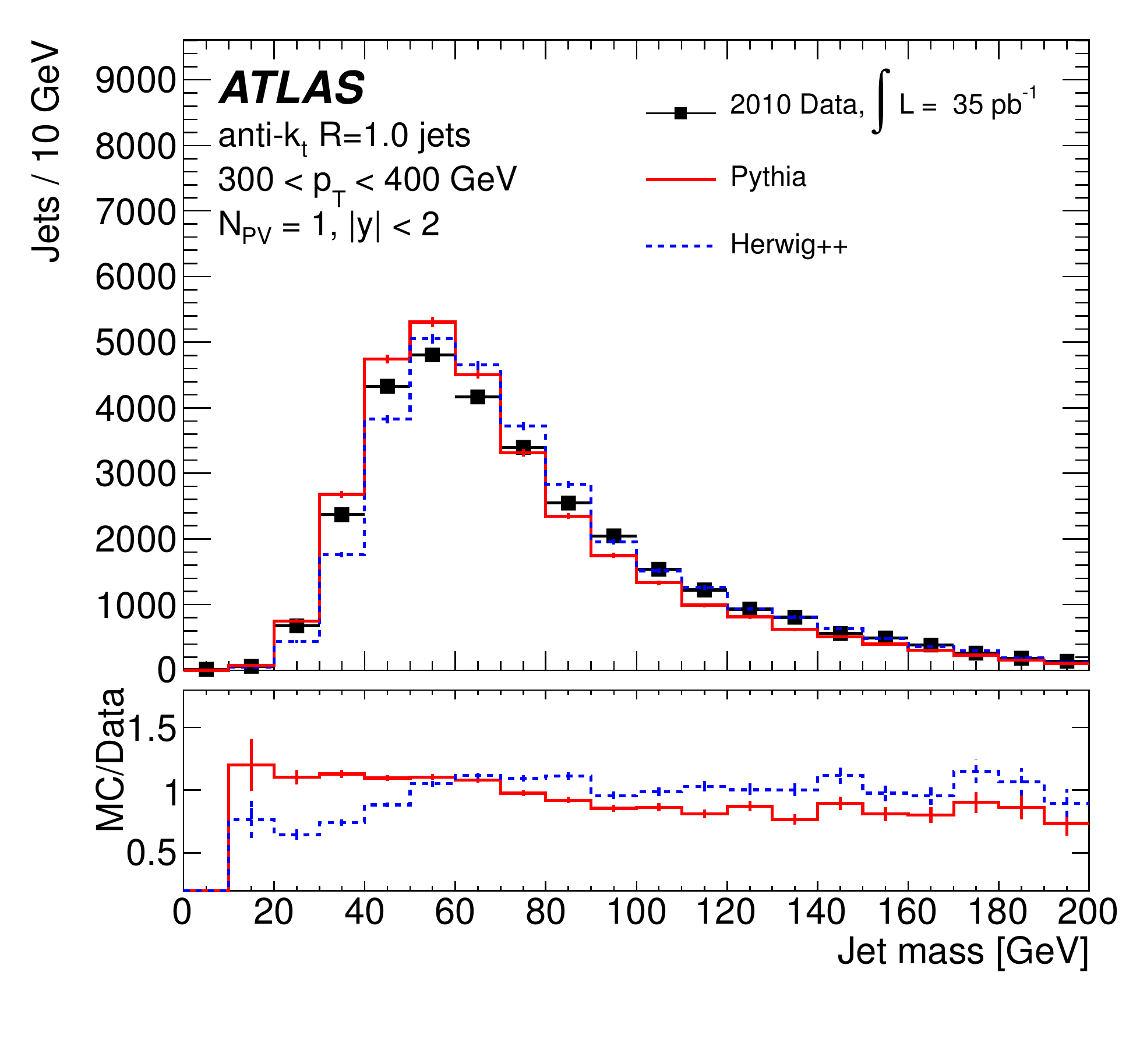}
     \caption{Mass distributions for jets with $|y| < 2.0$ in the
       300--400 \GeV\ \pt\ bin. Jets shown are \ca (top left), \ca\
       after splitting and filtering (top right) and \antikt (bottom).\label{fig:control_mass}}
  \end{center}
\end{figure}

\begin{figure}
  \begin{center}
     \includegraphics[width=0.45\linewidth]{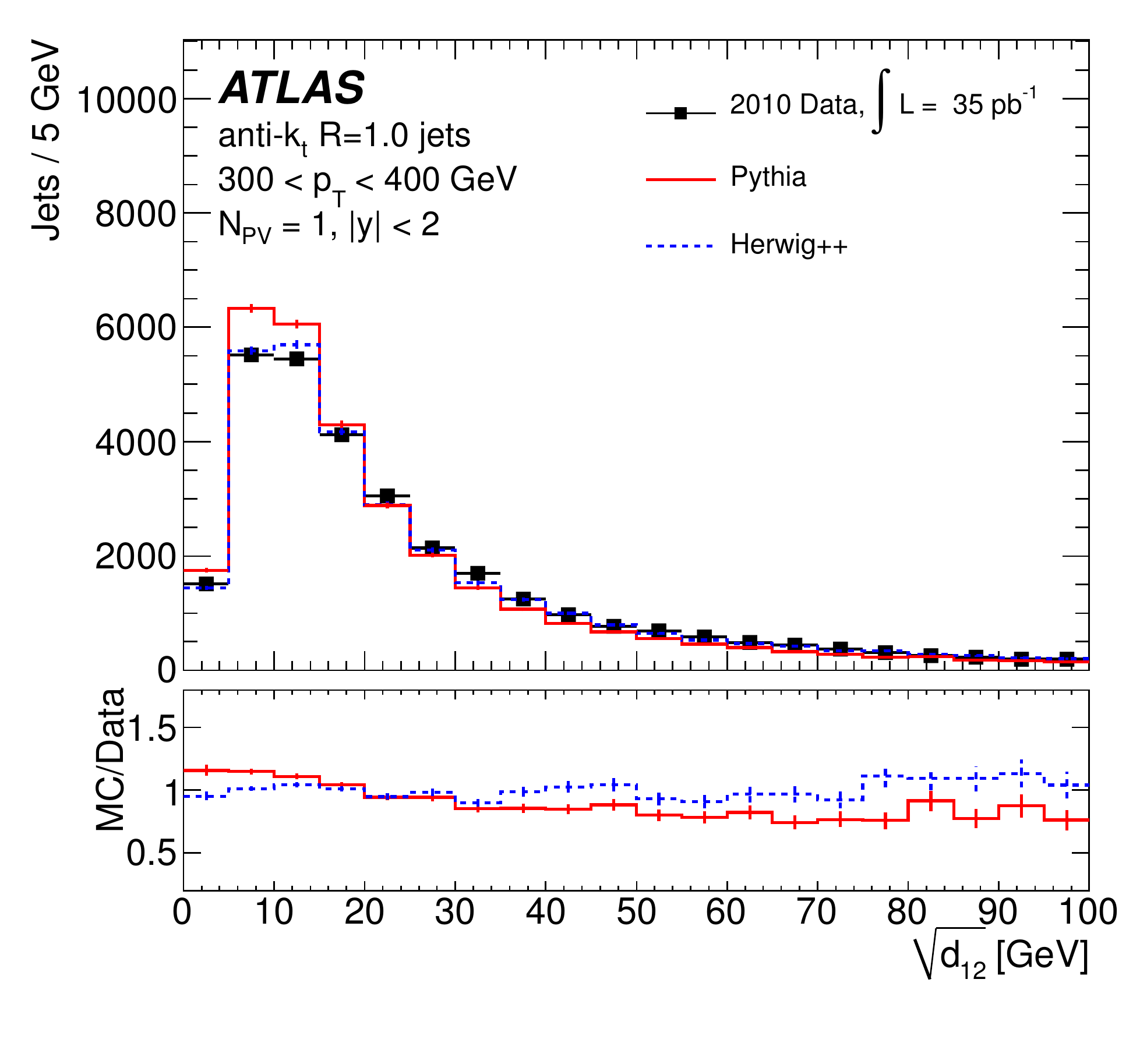}
     \includegraphics[width=0.45\linewidth]{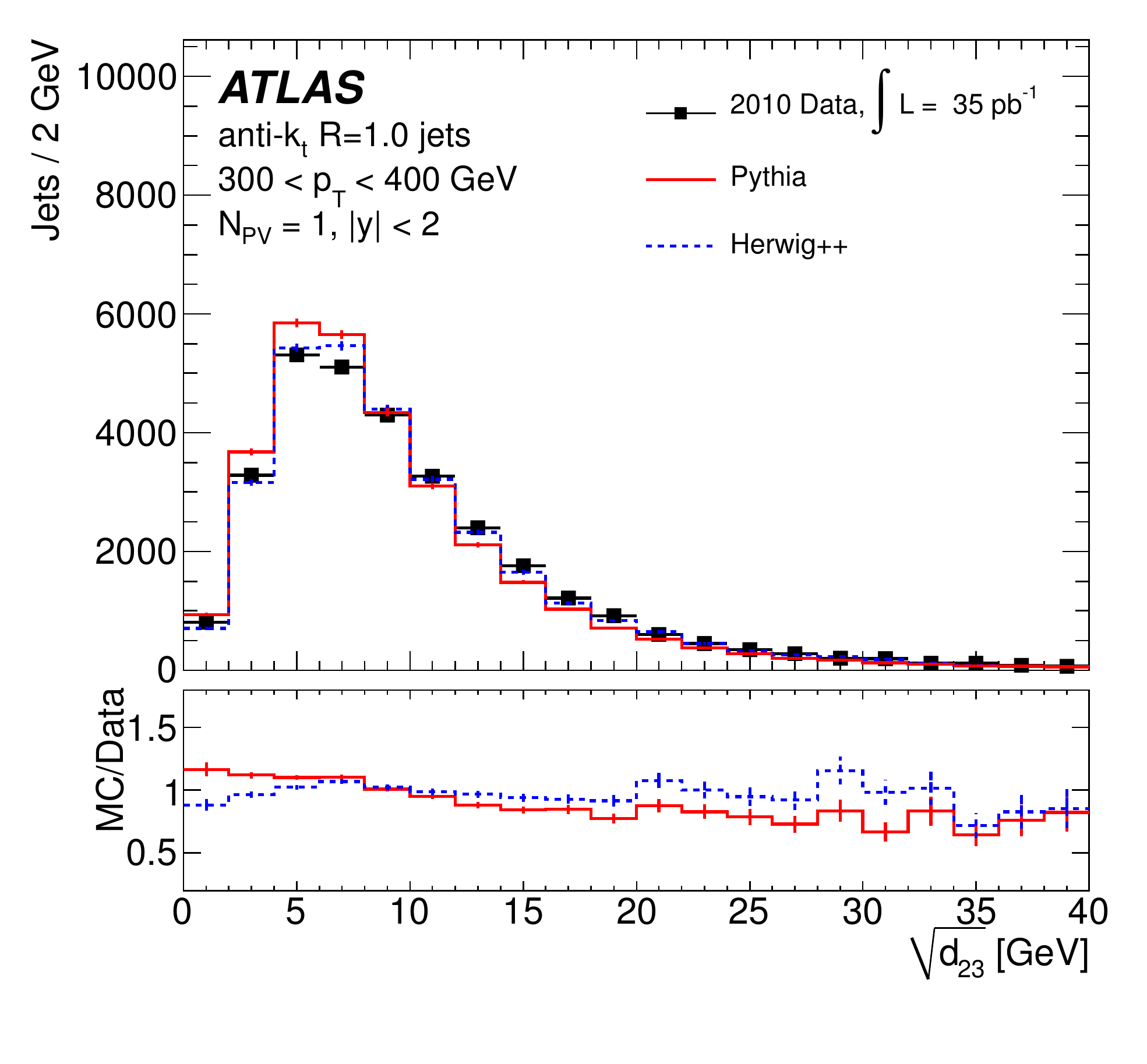}
     \caption{Distributions for \DOneTwo (left) and \DTwoThr (right)
       of \antikt\ $R=1.0$ jets with $|y|
       < 2.0$ in the 300--400 \GeV\ \pt\ bin.\label{fig:control_dij}}
  \end{center}
\end{figure}

\begin{figure}
  \begin{center}
    \includegraphics[width=0.45\linewidth]{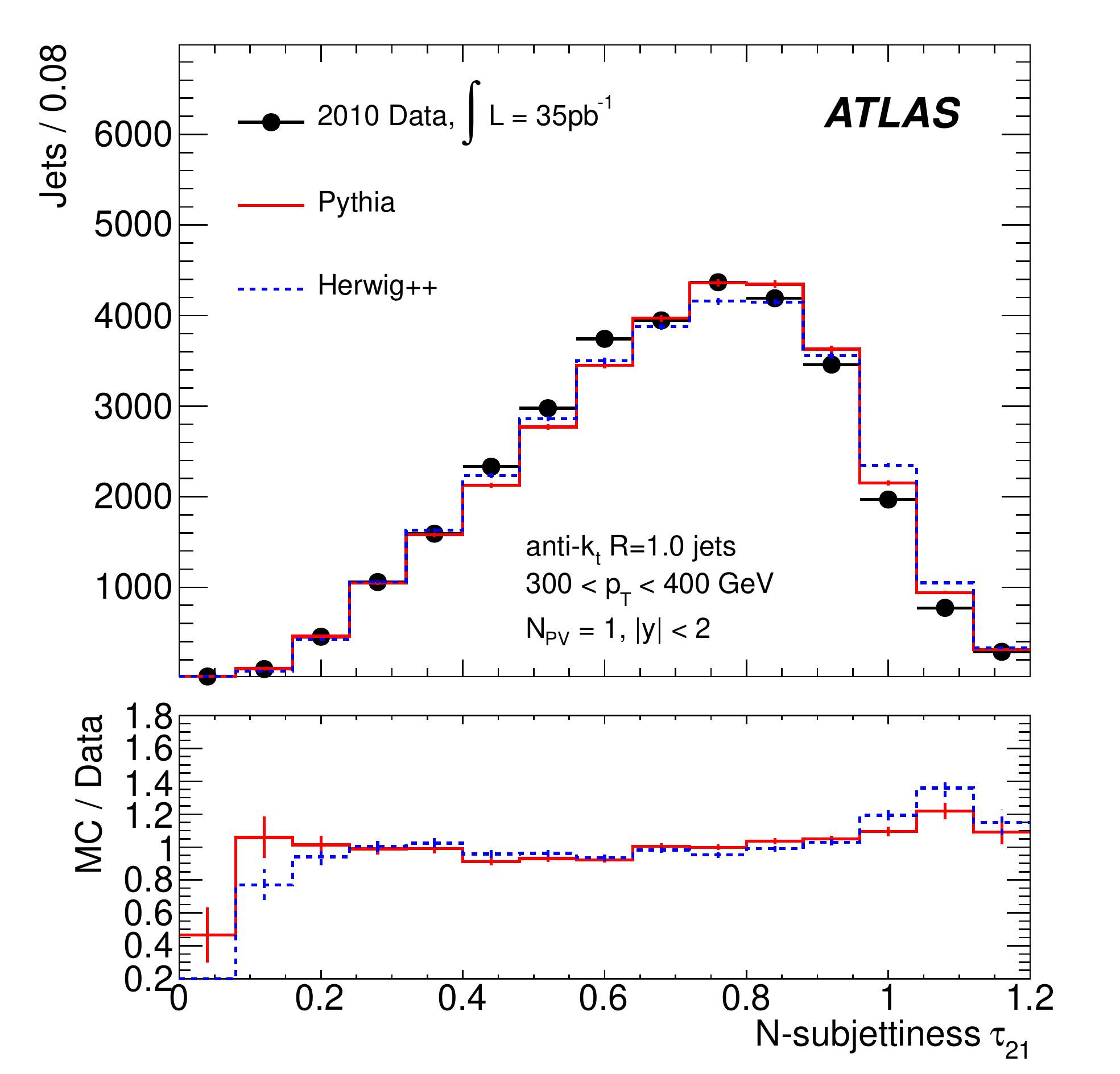}
    \includegraphics[width=0.45\linewidth]{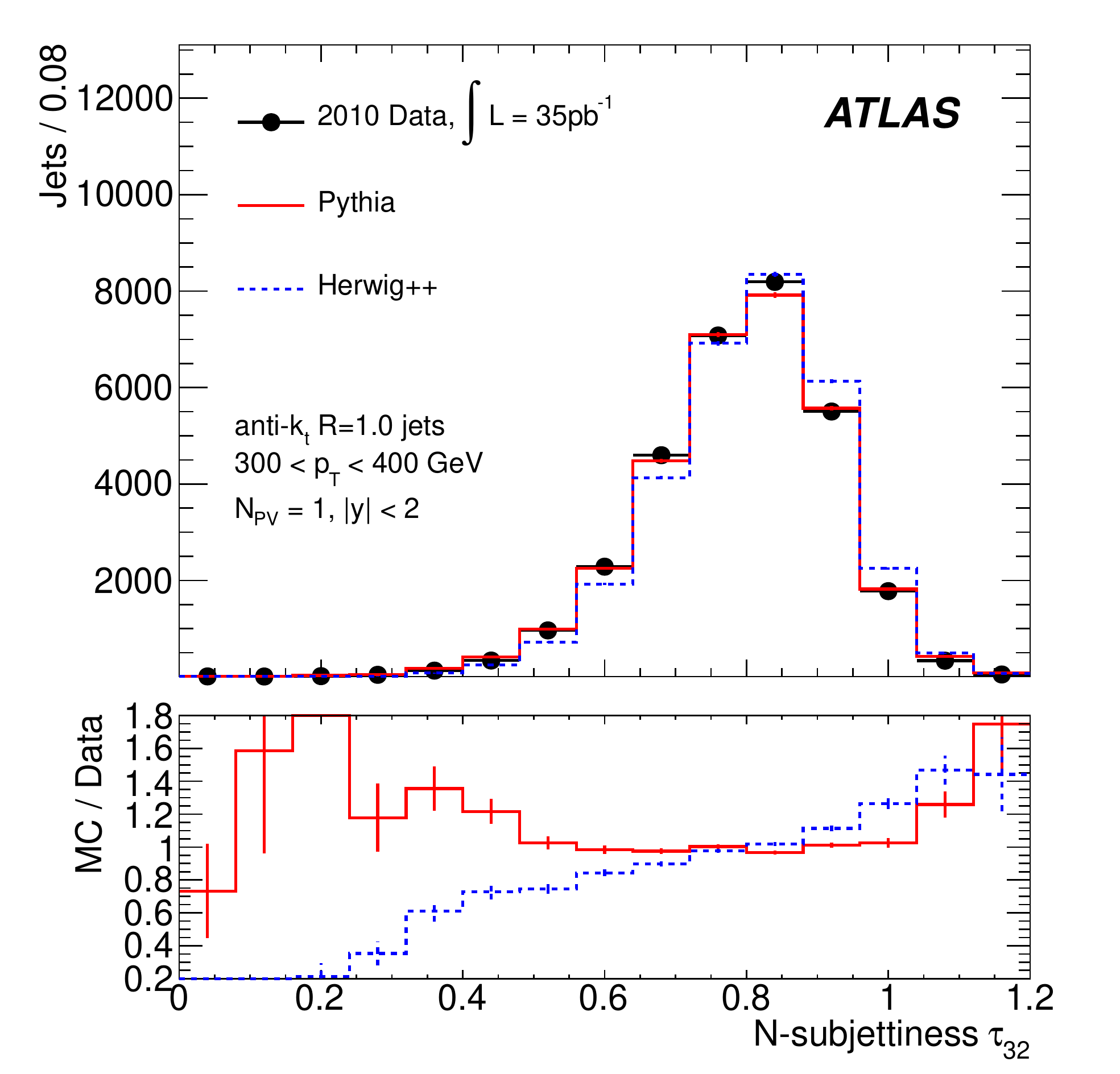}
    \includegraphics[width=0.45\linewidth]{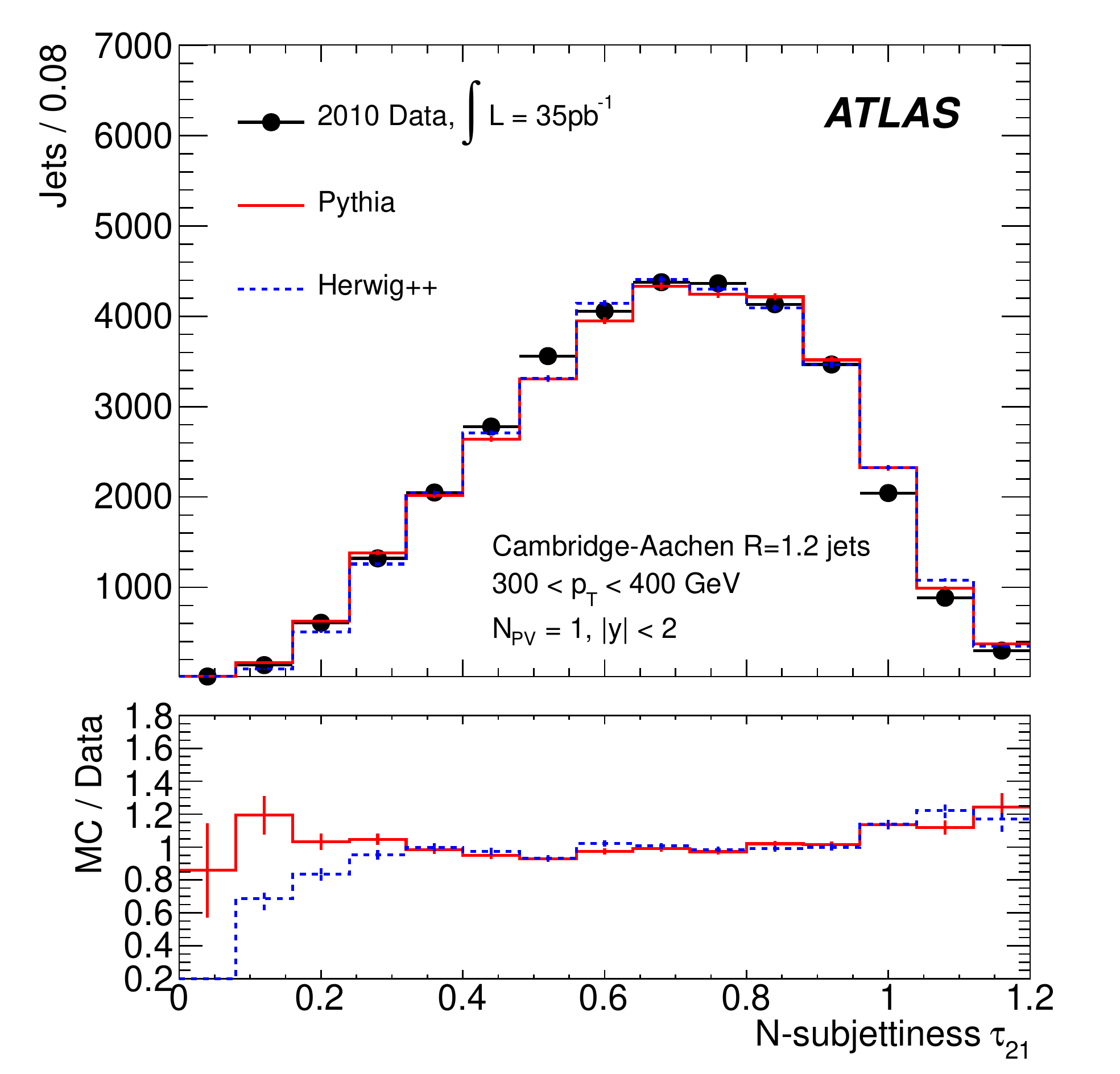}
    \includegraphics[width=0.45\linewidth]{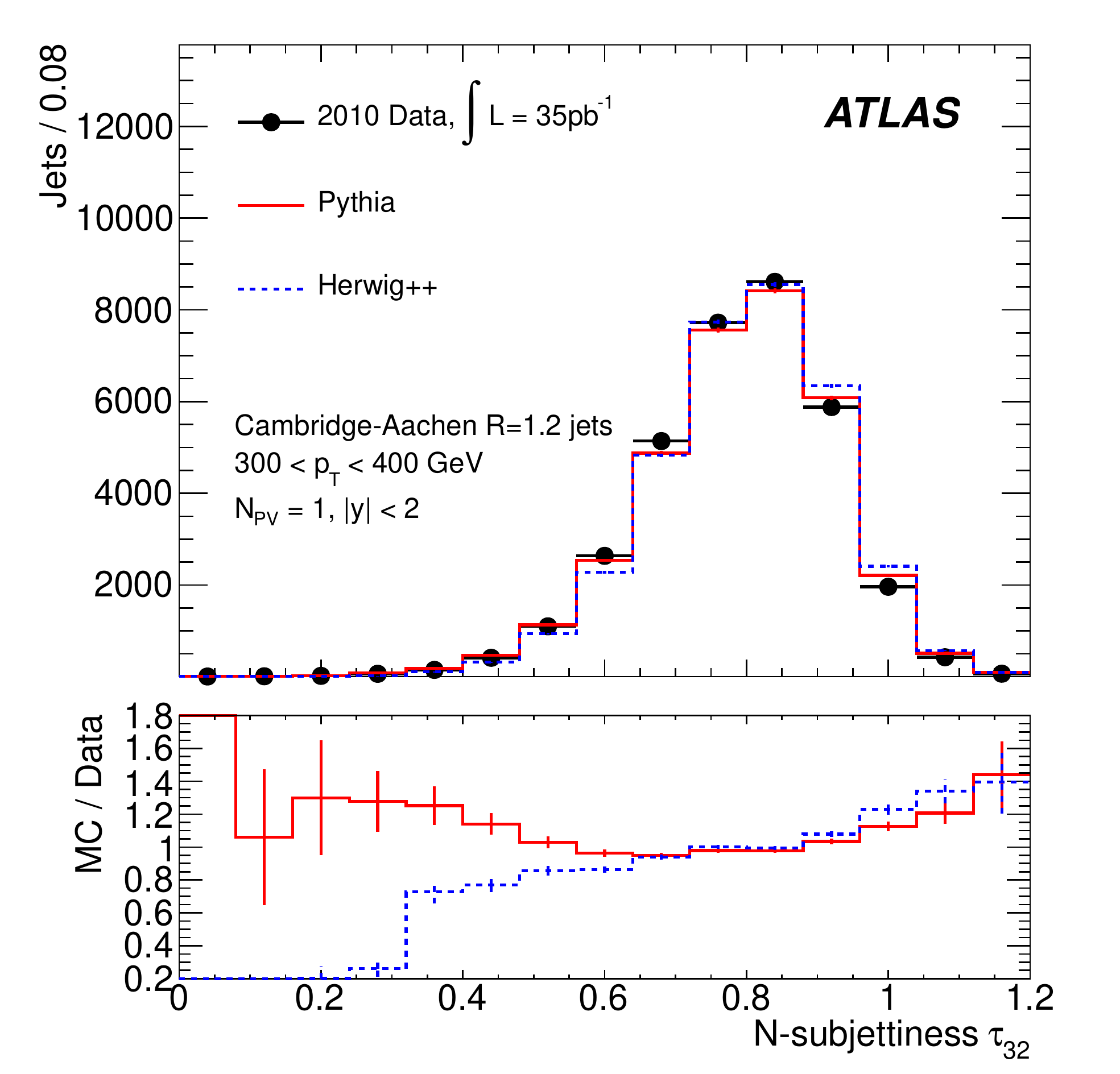}
     \caption{Distributions for \tauTwoOne (left) and \tauThrTwo (right) of jets with $|y|
       < 2.0$ in the 300--400 \GeV \pt\ bin for \antikt\ (top) and \ca\
       jets (bottom).\label{fig:control_tau}}
  \end{center}
\end{figure}

Detector-level distributions for jet \pt, $\eta$, mass, \DOneTwo,
\DTwoThr, \tauTwoOne and \tauThrTwo are shown in
Figures~\ref{fig:ca12_pteta}-\ref{fig:control_tau}. The statistical
uncertainty represented in ratios is that from Monte Carlo and data
added in quadrature. Representative distributions of the substructure
variables are shown for the 300--400~\GeV\ bin only. The Monte Carlo is
normalised to the data separately in each plot. The properties of
these jets are observed to be reasonably well modelled by
leading-order parton-shower Monte Carlo. There are approximately four
times fewer split and filtered jets
(e.g. Figure~\ref{fig:ca12filt_pteta}) because many jets fail the
splitting criteria described above.

\section{Systematic Uncertainties}
\label{sec:calib}
The modelling of the calorimeter response is the biggest systematic
uncertainty for this analysis. The key issue therefore is to validate
the Monte Carlo-based jet calibration described in
Section~\ref{sec:selection}. As the results here use jet algorithms
with larger $R$-parameters, the ATLAS jet energy scale
uncertainty~\cite{Aad:2011he} for \antikt\ $R=0.4$ and 0.6 jets cannot
be applied. The primary systematic uncertainties considered in the
present study are those relating to scales and resolutions, such as
jet \pt\ scale (JES) and jet \pt\ resolution (JER). For each
substructure variable, the scale and resolution of the variable itself
are also considered, for example the jet mass scale (JMS) and jet mass
resolution (JMR).
The scale uncertainties are primarily constrained by in-situ
validation using track-jets. The inner detector and calorimeter have
largely uncorrelated systematic effects, therefore comparison of
variables such as jet mass and energy between the two sub-detectors
allows for some separation of physics and detector effects. This
technique is limited to a precision of around 3-5\% by systematic
uncertainties arising from the inner-detector tracking efficiency and
confidence in Monte Carlo modelling of the relative behaviour of the
charged and neutral components of jets.

Jets composed from tracks are matched to calorimeter-jets if they are
within $\delta R < 0.3$ of each other. The split and filtered
calorimeter-jets are matched to \ca\ $R=1.2$ track-jets. Ratios
are defined between track- and calorimeter-jets for each variable $X$
(\pt, mass, \DOneTwo, \DTwoThr, \tauTwoOne, \tauThrTwo):

\begin{equation}
r^X = \frac{X_{calorimeter-jet}}{X_{track-jet}}
\end{equation}

\begin{figure}
  \begin{center}
  \includegraphics[width=0.45\linewidth]{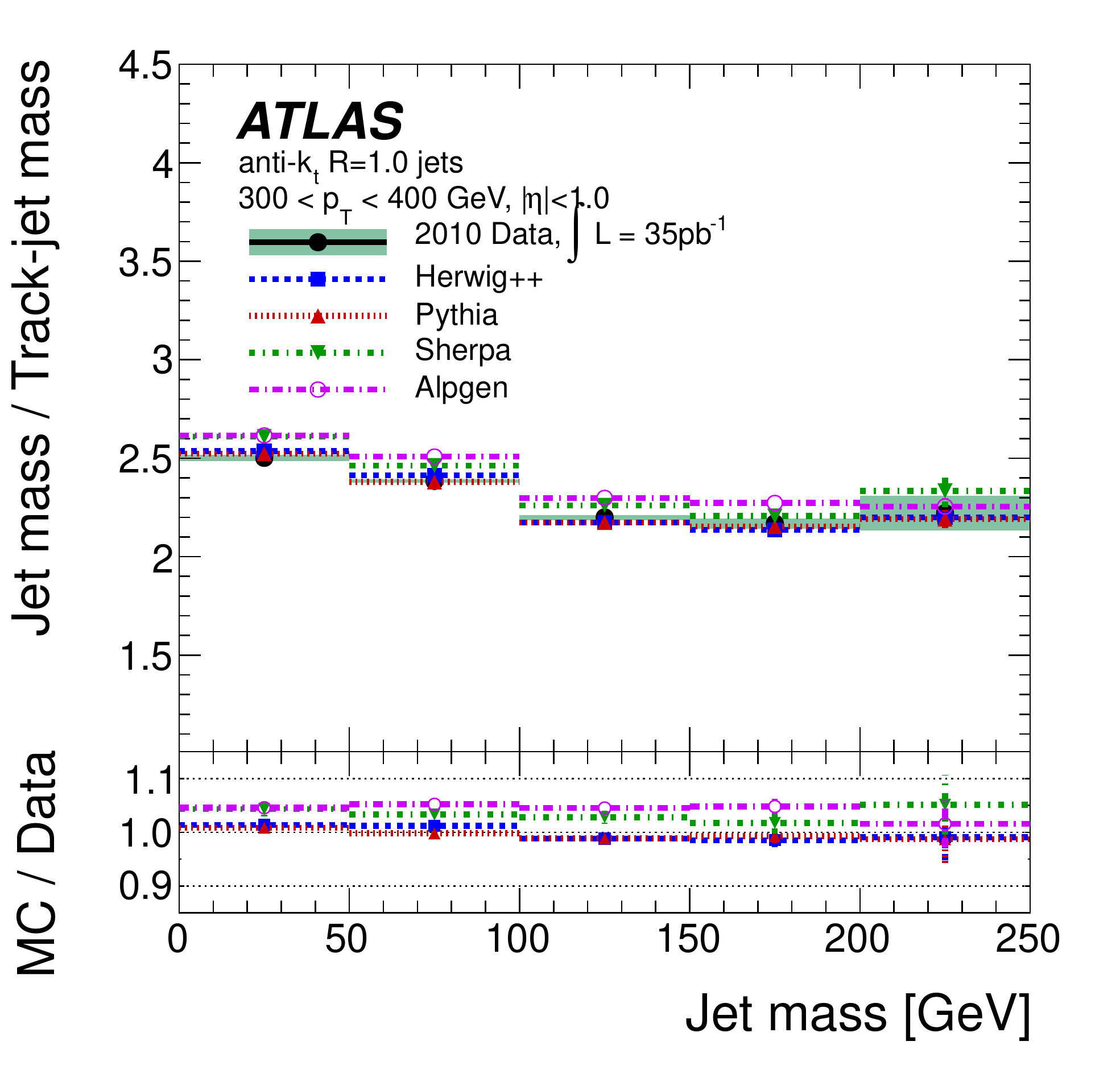}
  \includegraphics[width=0.45\linewidth]{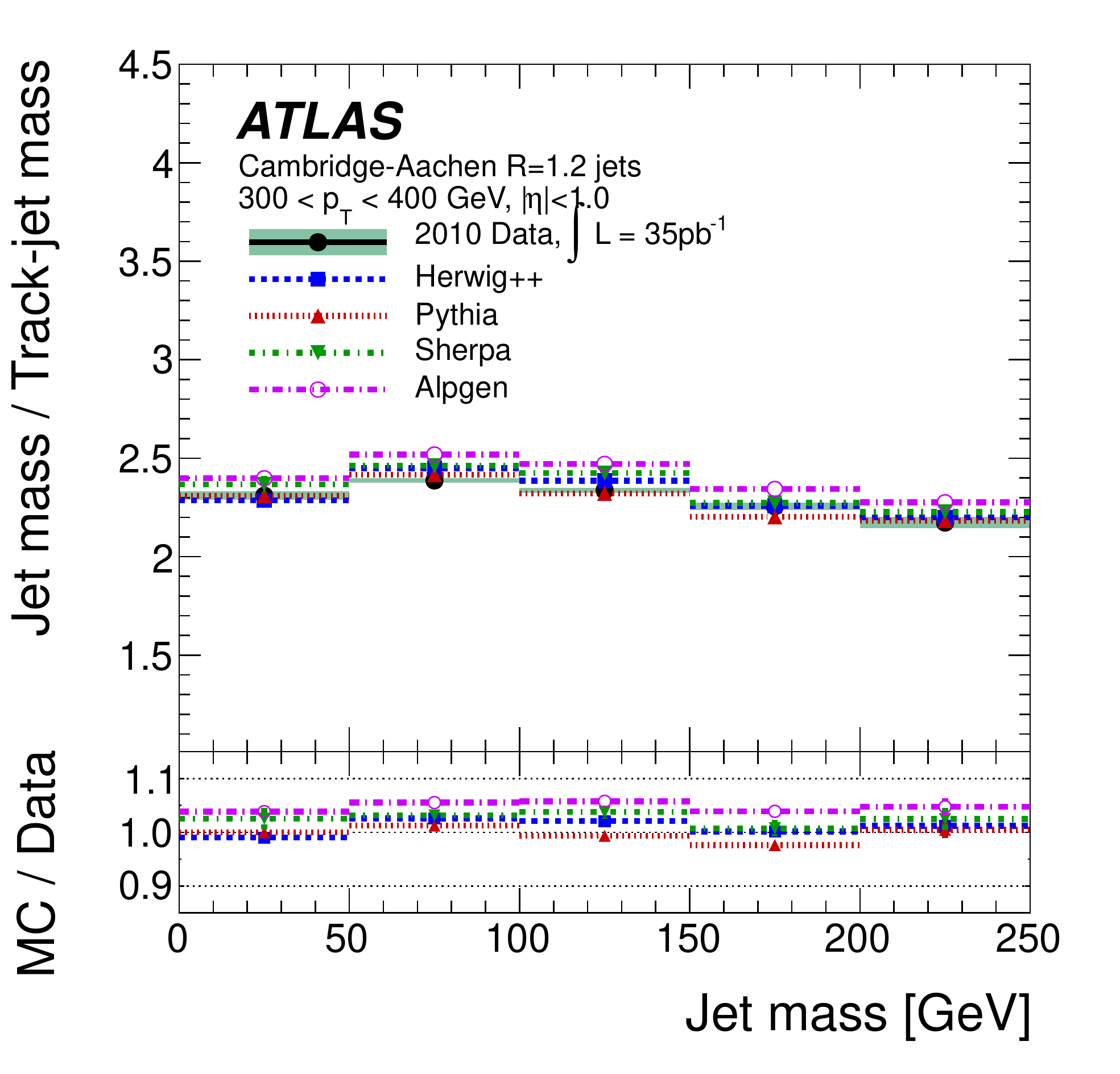}
  \includegraphics[width=0.45\linewidth]{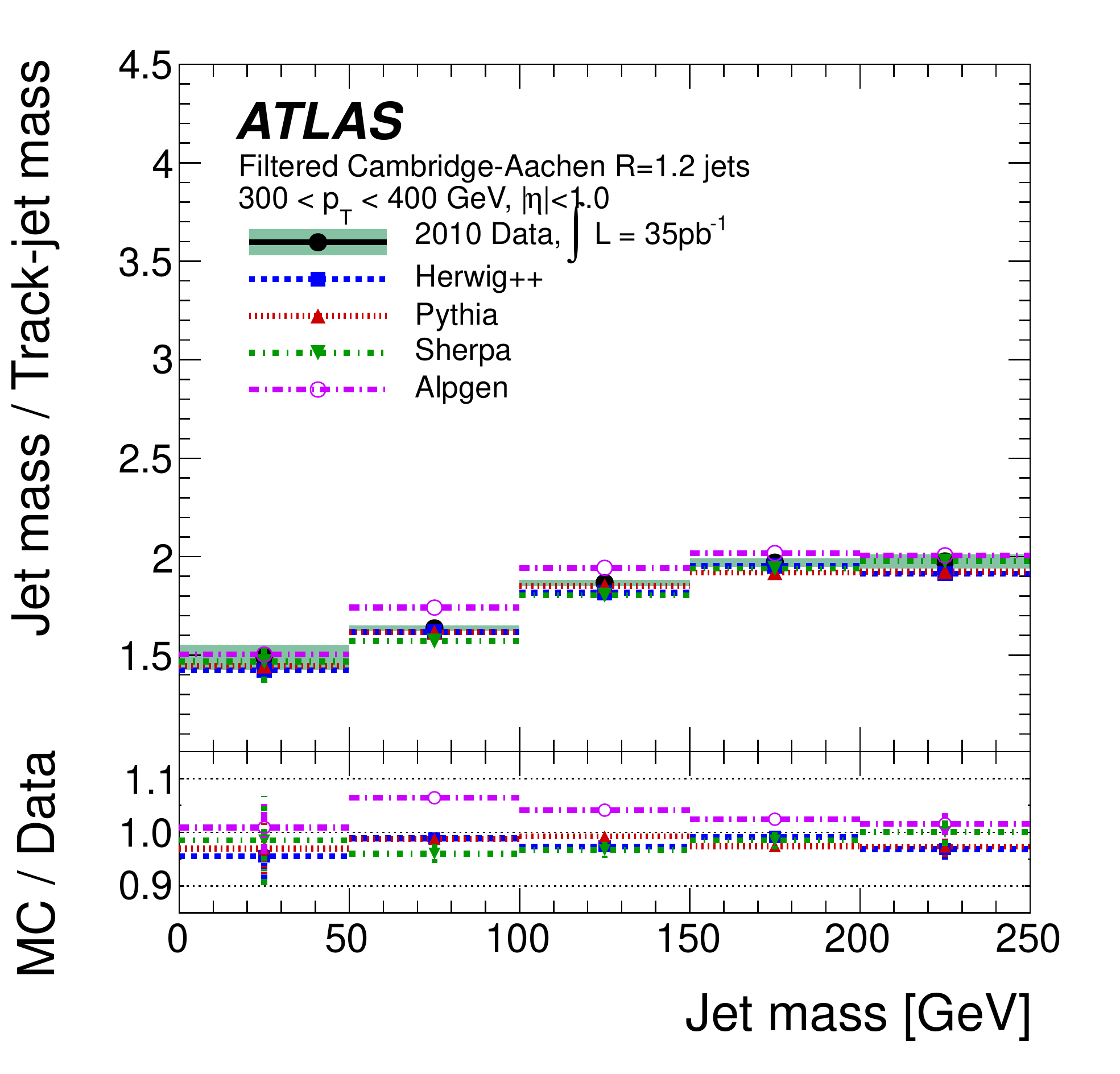}
  \includegraphics[width=0.45\linewidth]{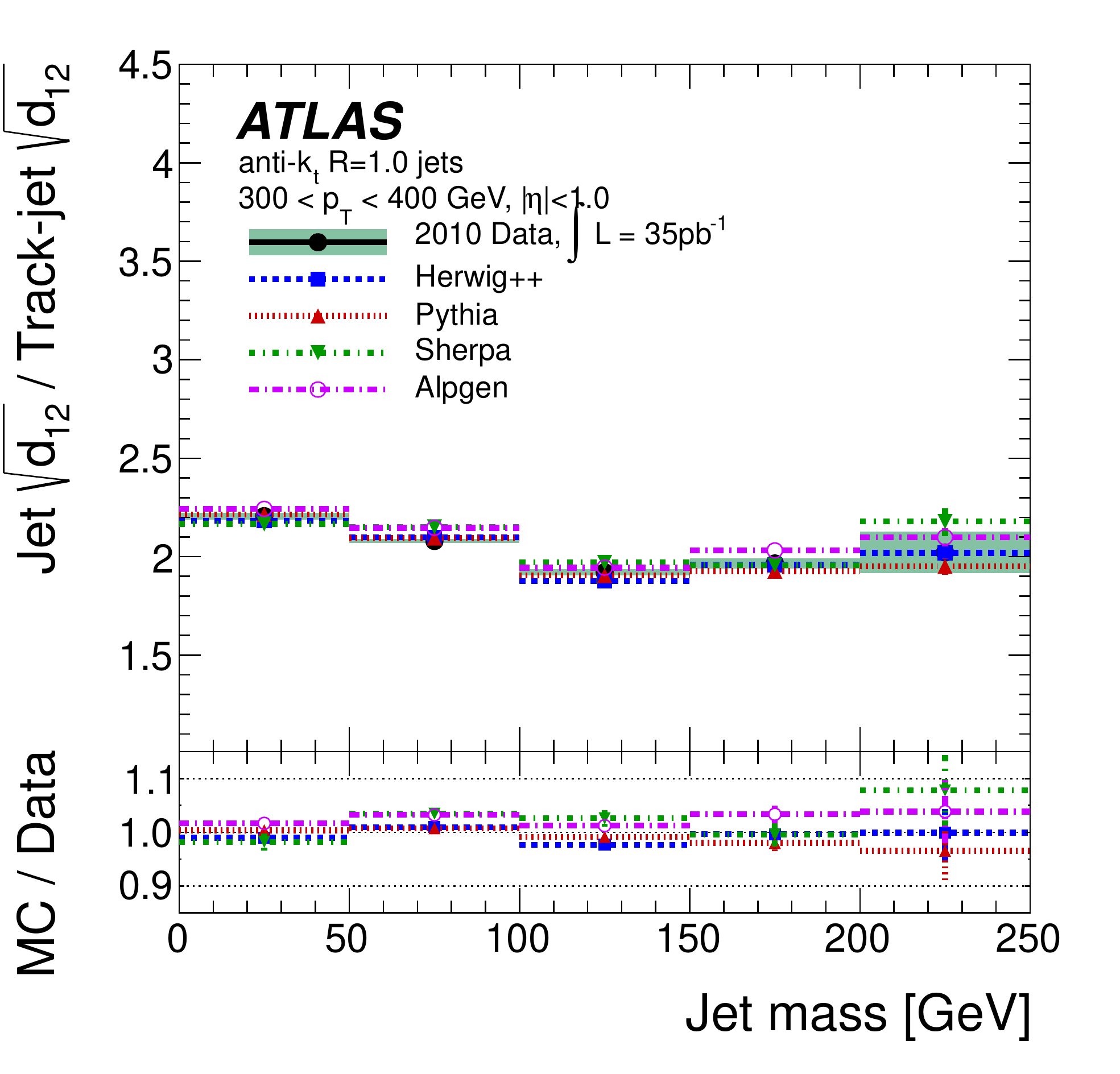}

  \caption{The ratio of a jet property determined by the calorimeter
    to that determined by tracks versus the calorimeter jet mass for
    jets with 300--400~\GeV\ in \pt. Shown are the data and a variety
    of Monte Carlo models. The bottom frame shows the ratio of the
    Monte Carlo models to data. The top left, top right and bottom
    left figures show the ratio for jet mass for three different jet
    algorithms. The bottom-right figure shows the ratios for \DOneTwo\
    in \antikt\ jets.
    \label{fig:systplots}}
  \end{center}
\end{figure}

Example distributions of some of the ratio variables are shown in
Figure~\ref{fig:systplots}. It can be seen that the ratios are in
broad agreement between data and Monte Carlo. To quantify the level of
agreement, double ratios are defined:

\begin{equation}\label{doubleratios}
\Rx = \frac{\rxdata}{\rxmc},
\end{equation}

\noindent
where again, X can be \pt, mass or any of the substructure
variables. The distributions of the variables $X_{calorimeter-jet}$
themselves are not necessarily expected to be correctly modelled by
Monte Carlo. However, if the simulation correctly models the effect of
the detector on these variables, the double ratios \Rx, are expected
to be consistent with unity. Figure~\ref{fig:systplots} also shows
below each plot the corresponding double ratio. In order to account
for possible uncertainties due to different fragmentation and
hadronisation models, these double ratios are also calculated with a
variety of Monte Carlo programs.

Final scale uncertainties are determined by adding in quadrature the
estimated uncertainty on the inner-detector measurement with the
deviation from unity observed in the double ratios. The resulting
scale uncertainties on \pt, mass and substructure variables are in
the range 3-6\%. The highest \pt\ bins contain fewer events and
therefore suffer from statistical fluctuations when calculating the
double ratio deviation. These scale uncertainties tend to dominate the
systematic uncertainties on the final measurements.

As an additional cross-check, Monte Carlo-based tests are used to
determine the dependence of the detector response on a number of
different variables. These include samples produced with modified
detector geometry, different GEANT hadronic physics models and
different Monte Carlo generators. These tests indicate variations of a
similar order of magnitude to those observed in the in-situ
studies. The in-situ track-jet study is limited by inner-detector
acceptance and only extends as far as $|\eta| < 1.0$, which
corresponds to $\simeq 75\%$ of the jets in the measured
distributions. However, the Monte Carlo-based tests also indicate no
strong $\eta$-dependence from any of the different possible types of
mismodelling examined. Based on this, the systematic uncertainty is
applied to the entire sample.

In-situ tests of the JER~\cite{jetResolution7TeV} for \antikt\ jets
with $R=0.4$ and 0.6 indicate that the jet \pt\ resolution predicted
by simulation is in good agreement with that observed in the
data. Here, the resolution uncertainties are taken from the Monte
Carlo tests described above only, primarily because the mass and
substructure variable resolutions are difficult to validate in-situ
with this dataset. From studying the variations in resolution created
by varying the detector geometry, GEANT hadronic physics model and
Monte Carlo generator, resolution uncertainties of around 20\% are
conservatively estimated, except for \tauTwoOne and \tauThrTwo where
they are around 10\%.

\section{Data Correction}
\label{sec:unfolding}
To compare the measurements directly to theoretical predictions the
final distributions in this study are corrected for detector
resolution and acceptance effects. The procedure here is a
matrix-based unfolding technique called Iterative Dynamically
Stabilised (IDS) unfolding~\cite{ids1,ids2}.

In this procedure truth jets and reconstructed jets in Monte Carlo
simulated events are matched using the criterion $\delta R < 0.2$,
which leads to a match for $> 99\%$ of jets. Matched pairs of jets are
used to construct a transfer matrix corresponding to the effect of the
detector. A true jet can be matched with a reconstructed jet that
fails the \pt\ cut and vice-versa. As such, the efficiency for matching a
true jet to a reconstructed jet in the same \pt\ bin is recorded as a
function of the variable of interest. The reverse quantity is also
defined for reconstructed jets. The data are then scaled by the
reconstructed matching efficiency, multiplied by the transfer matrix
and finally divided by the truth matching efficiency. There is also an
iterative optimisation step, where the rows of the matrix are scaled
to match the corrected result. \pythia\ is used to provide the central
value. Each \pt\ bin is unfolded independently. The systematic
uncertainty is assessed by repeating the procedure using \sherpa\
samples.

\section{Results}
\label{sec:results}
\begin{figure}
  \begin{center}
    \includegraphics[width=0.49\linewidth]{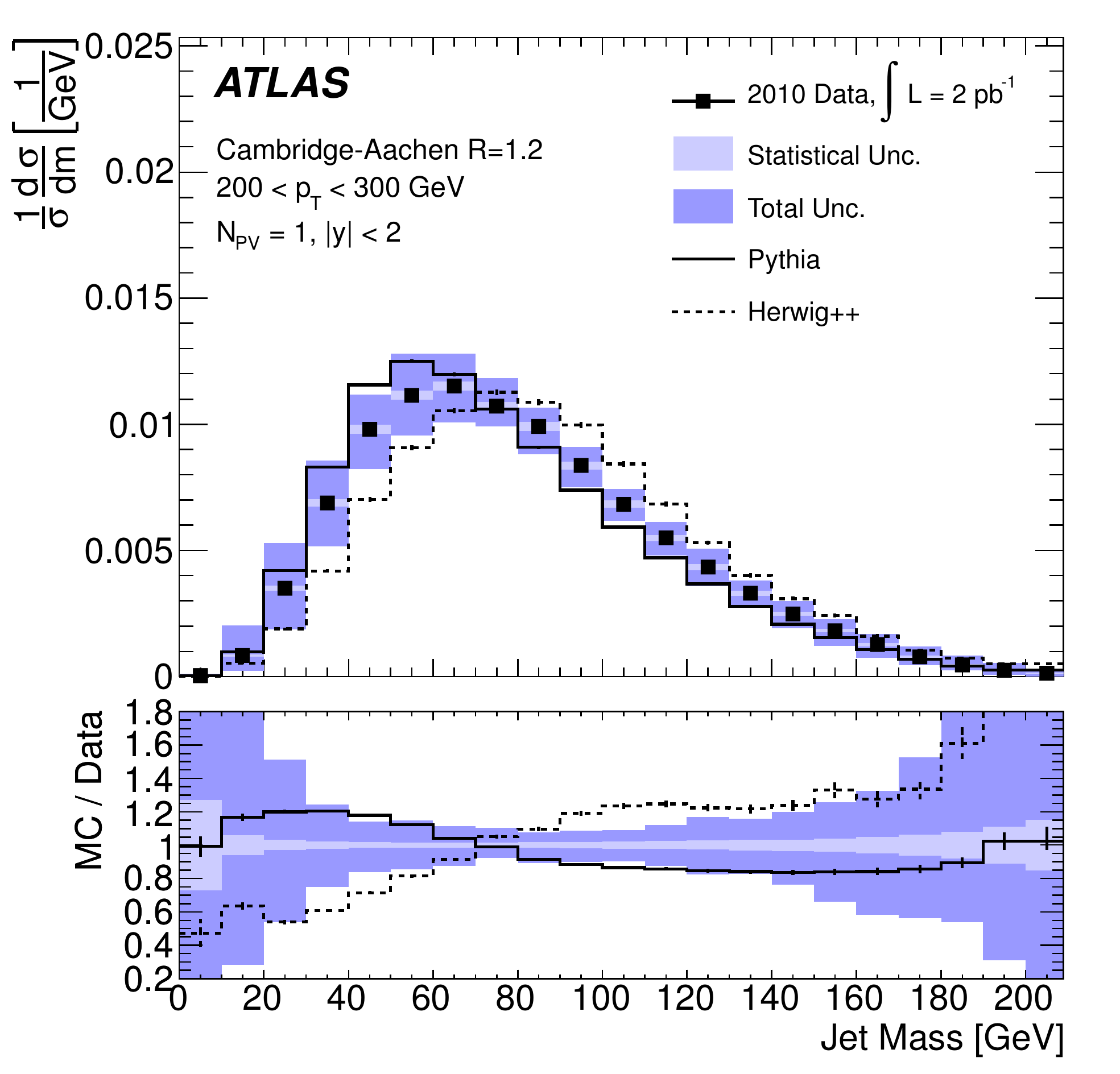}
    \includegraphics[width=0.49\linewidth]{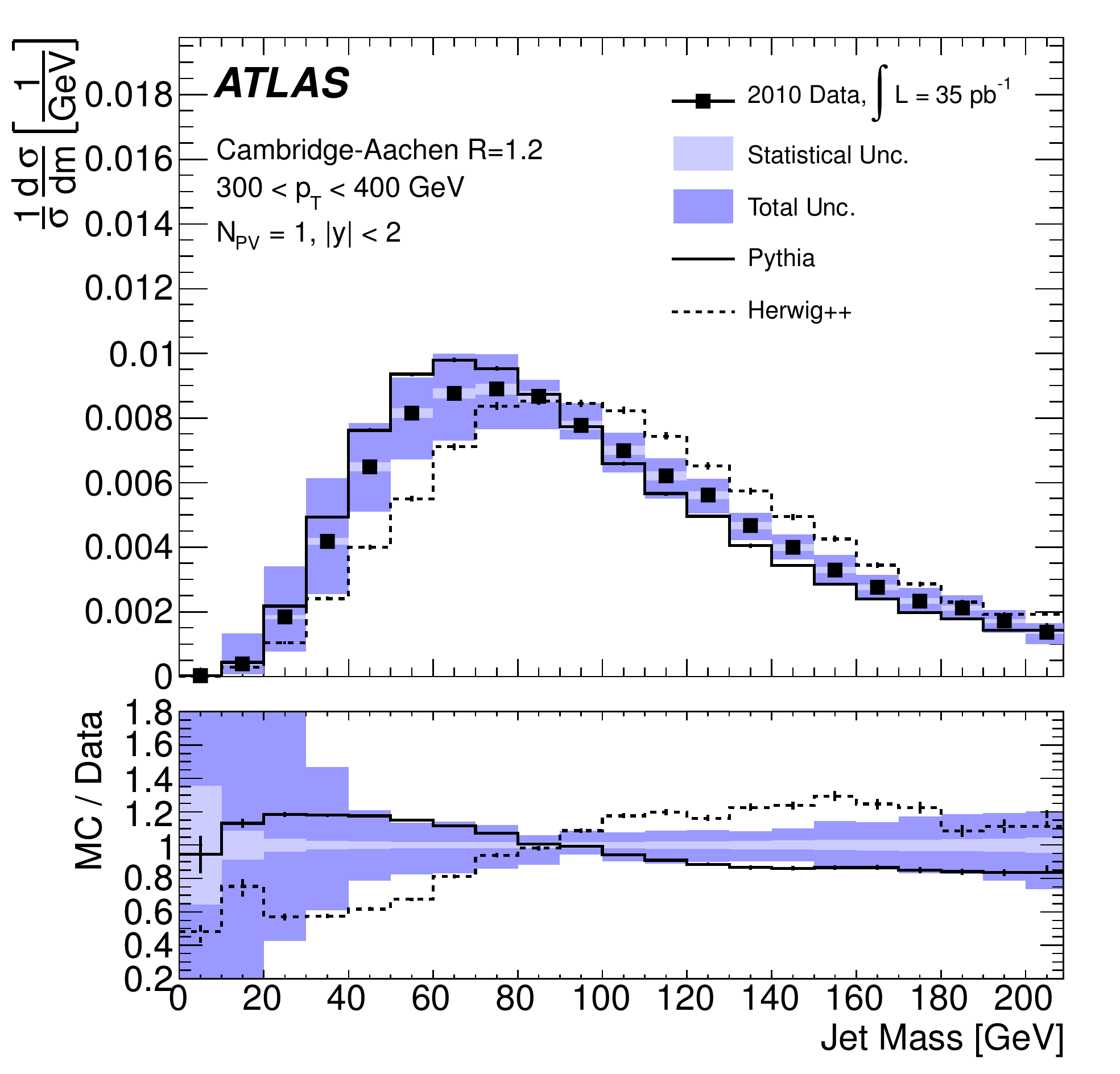}
    \includegraphics[width=0.49\linewidth]{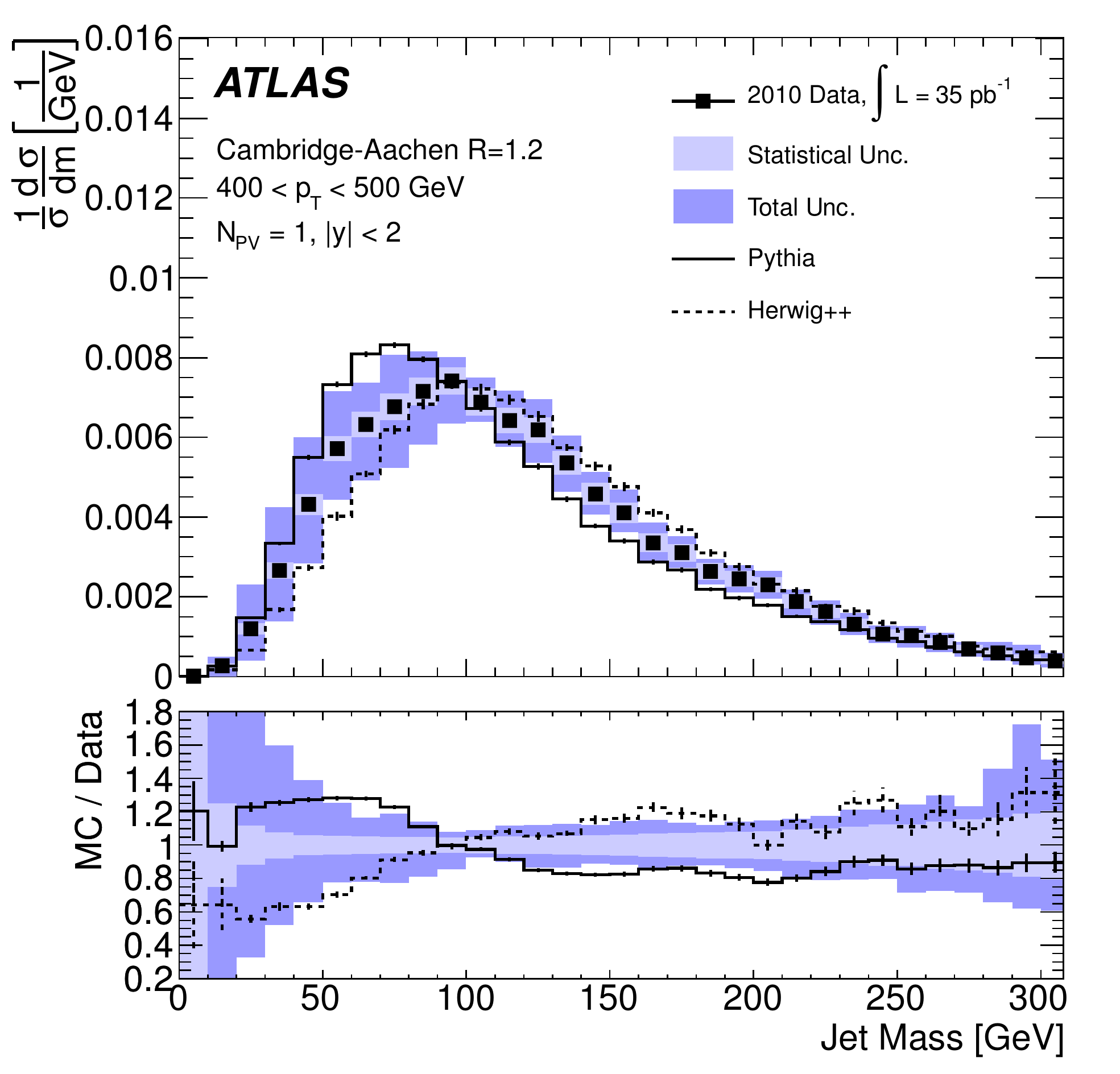}
    \includegraphics[width=0.49\linewidth]{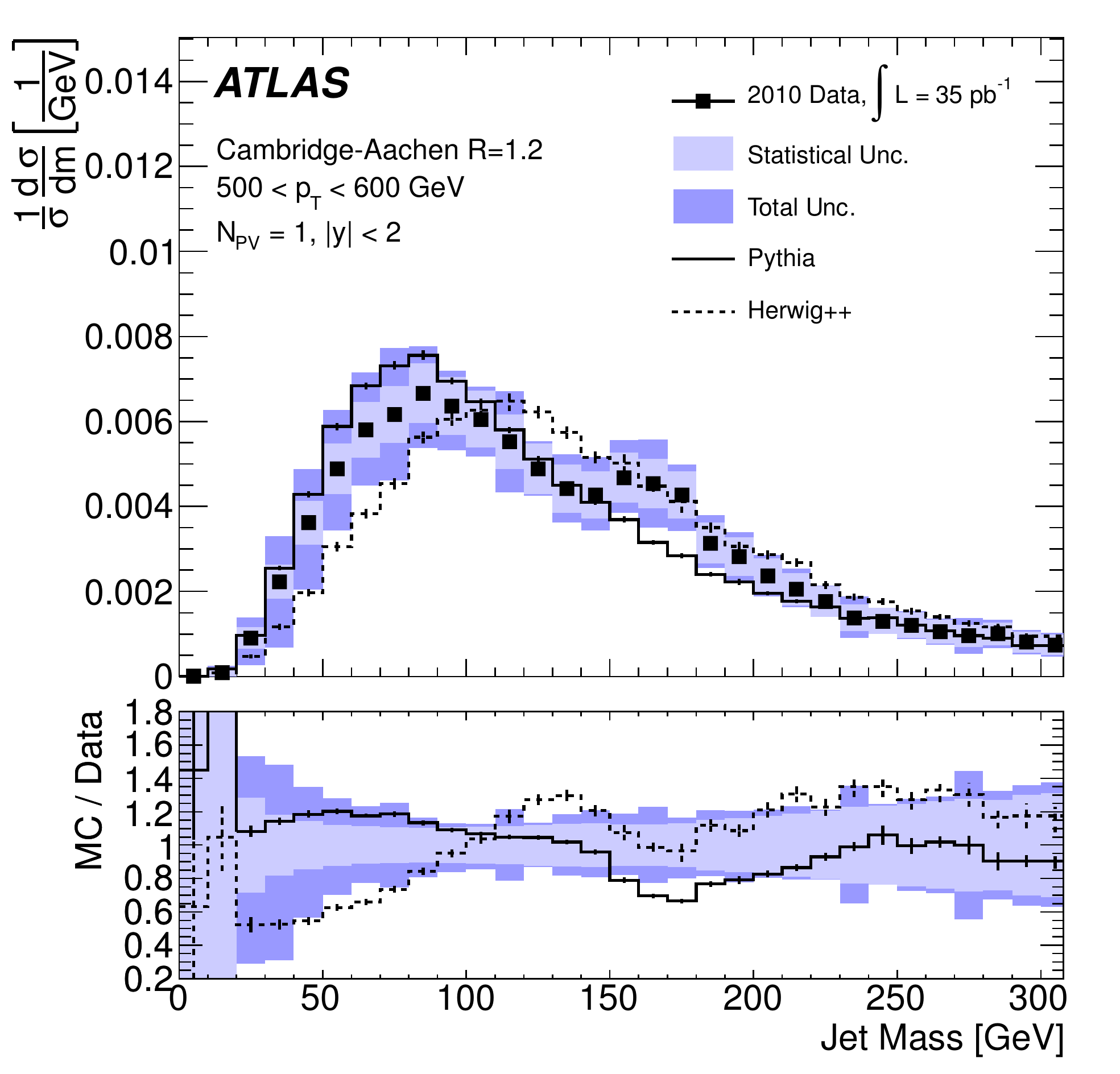}
    \caption{Normalised cross-sections as functions of mass of \ca\ jets
      with $R=1.2$ in four different \pt\ bins.\label{fig:final_ca12mass}}
  \end{center}
\end{figure}

\begin{figure}
  \begin{center}
    \includegraphics[width=0.49\linewidth]{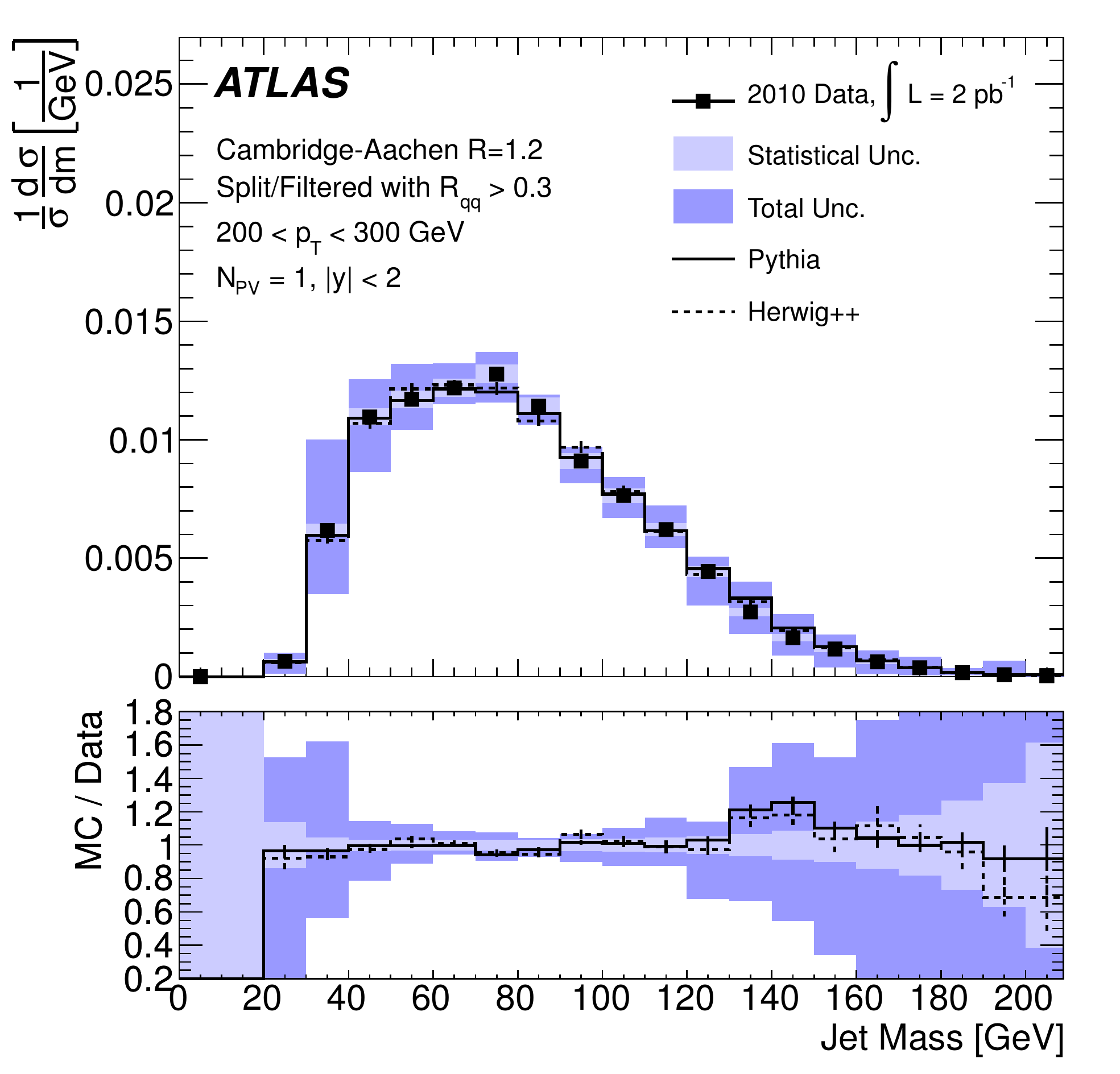}
    \includegraphics[width=0.49\linewidth]{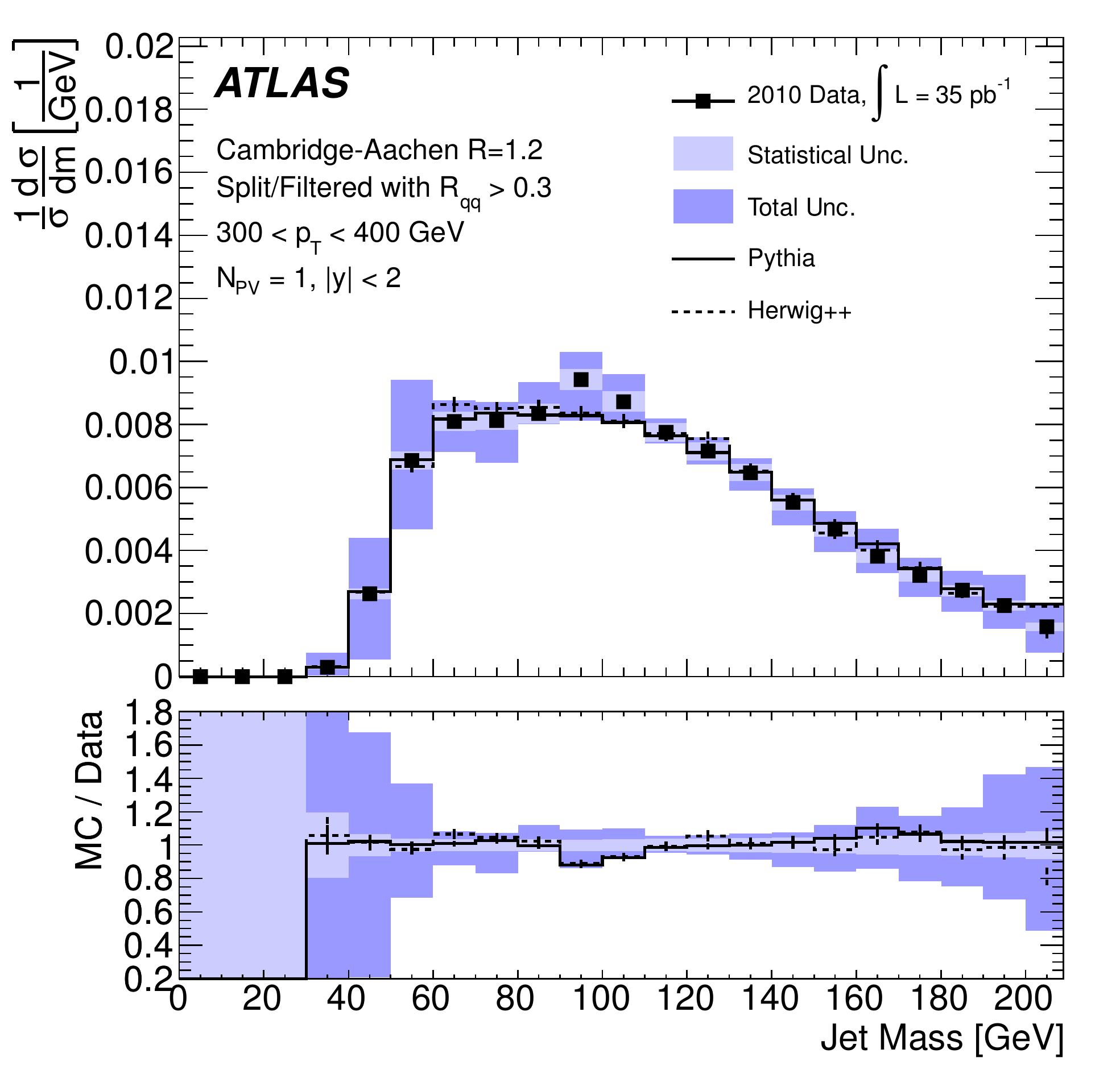}
    \includegraphics[width=0.49\linewidth]{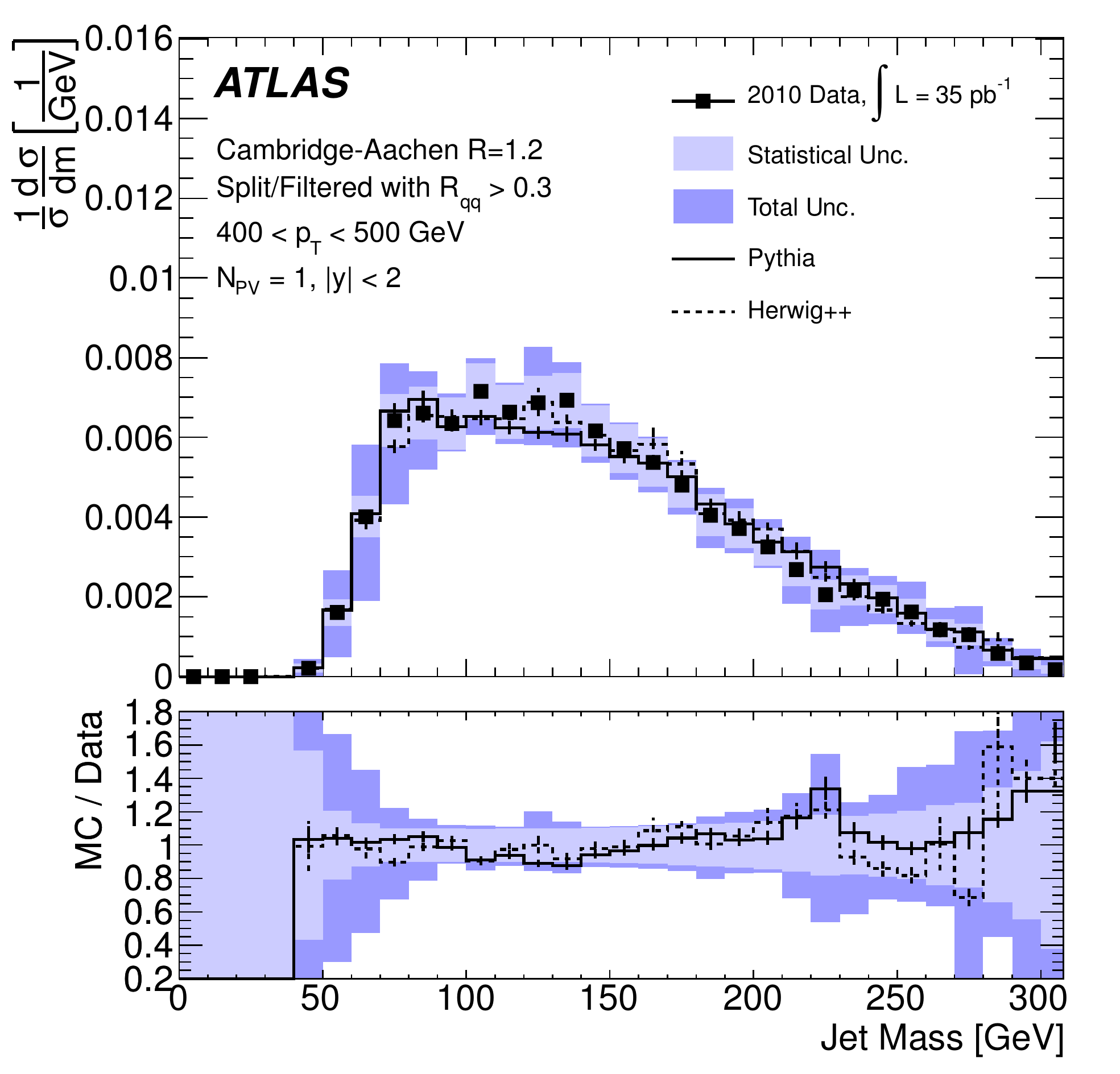}
    \includegraphics[width=0.49\linewidth]{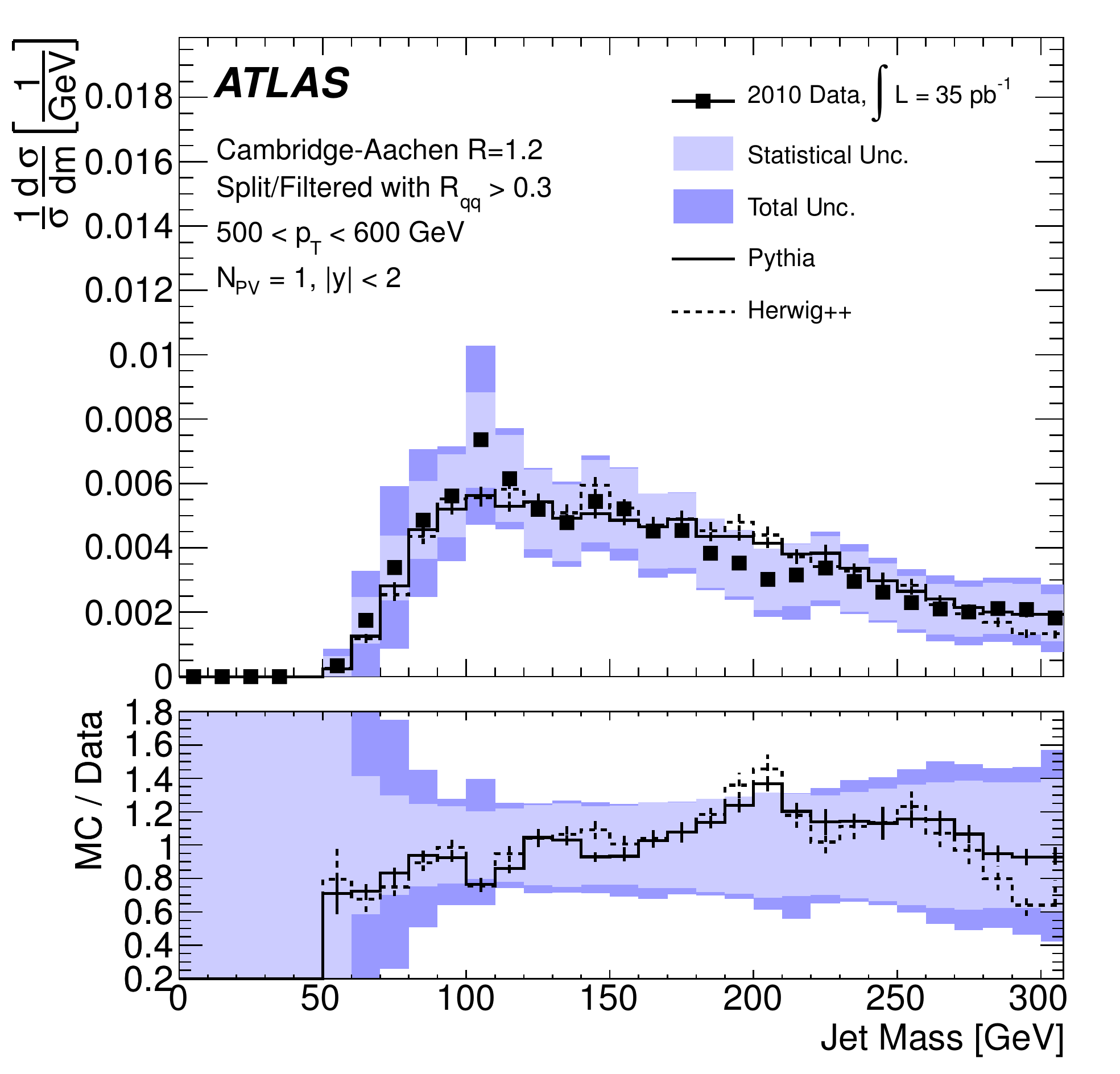}
    \caption{Normalised cross-sections as functions of mass of \ca\ jets with
      $R=1.2$ after splitting and filtering in four different \pt\ bins.\label{fig:final_ca12filtmass}}
  \end{center}
\end{figure}

\begin{figure}
  \begin{center}
    \includegraphics[width=0.49\linewidth]{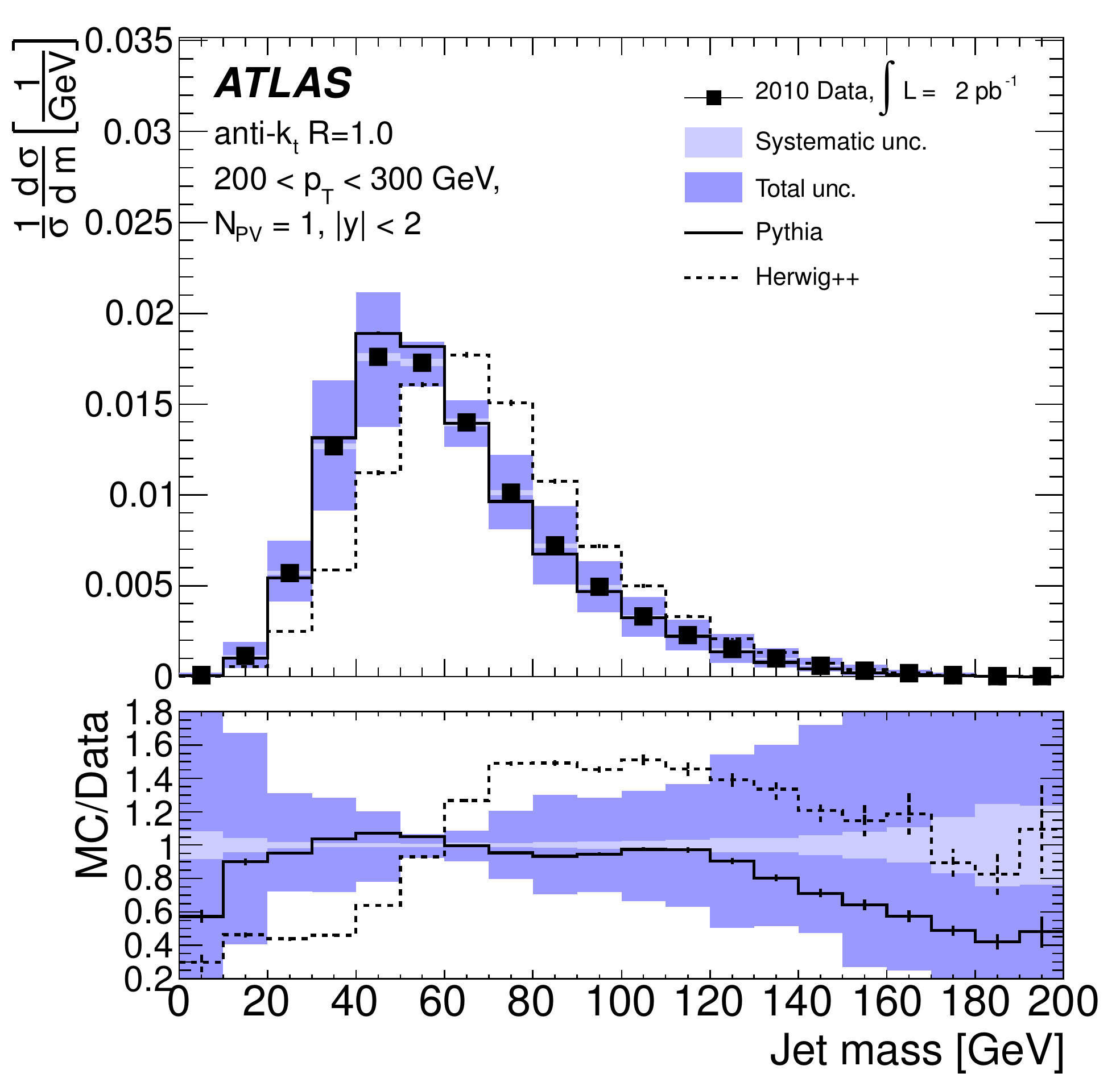}
    \includegraphics[width=0.49\linewidth]{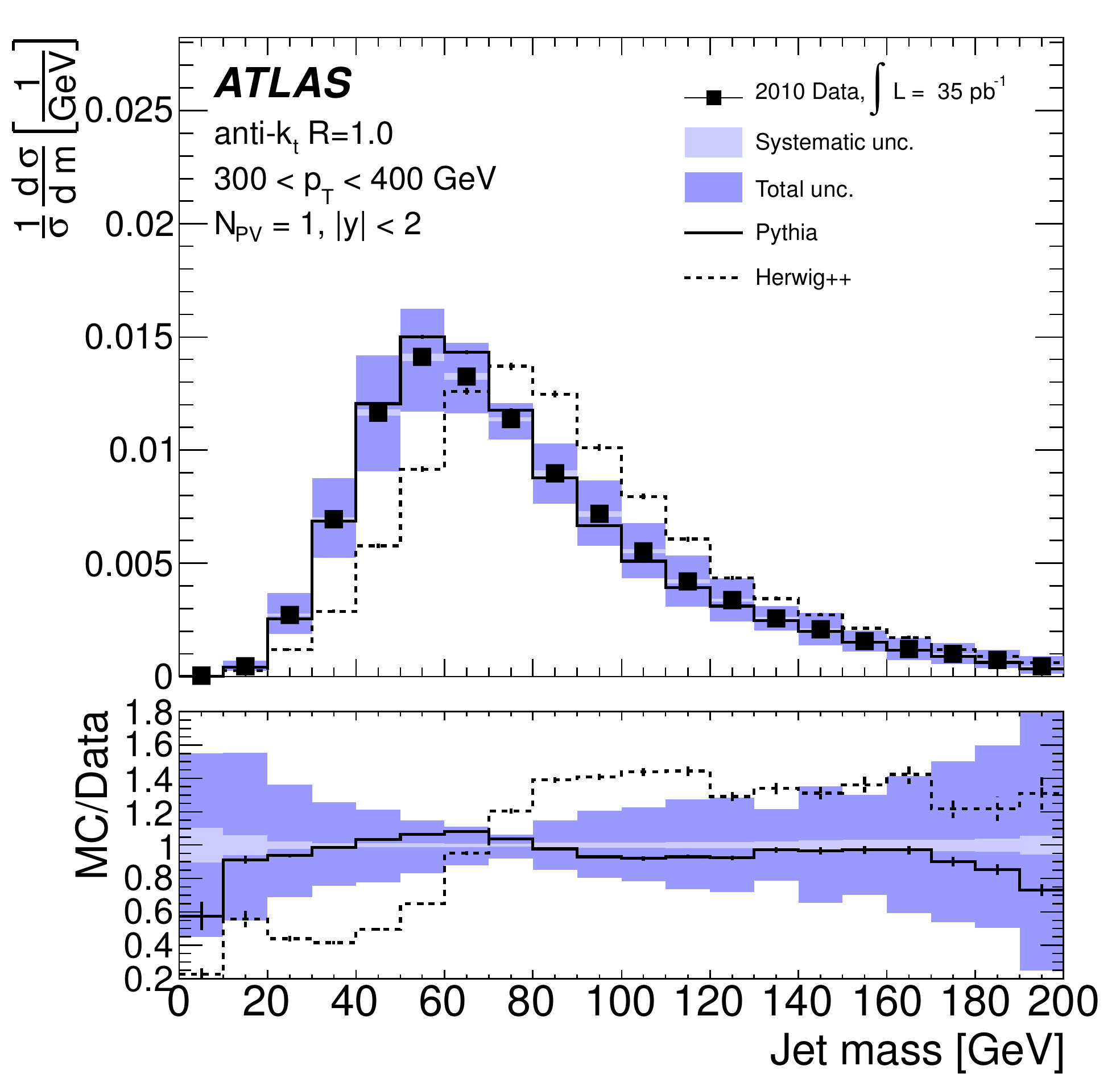}
    \includegraphics[width=0.49\linewidth]{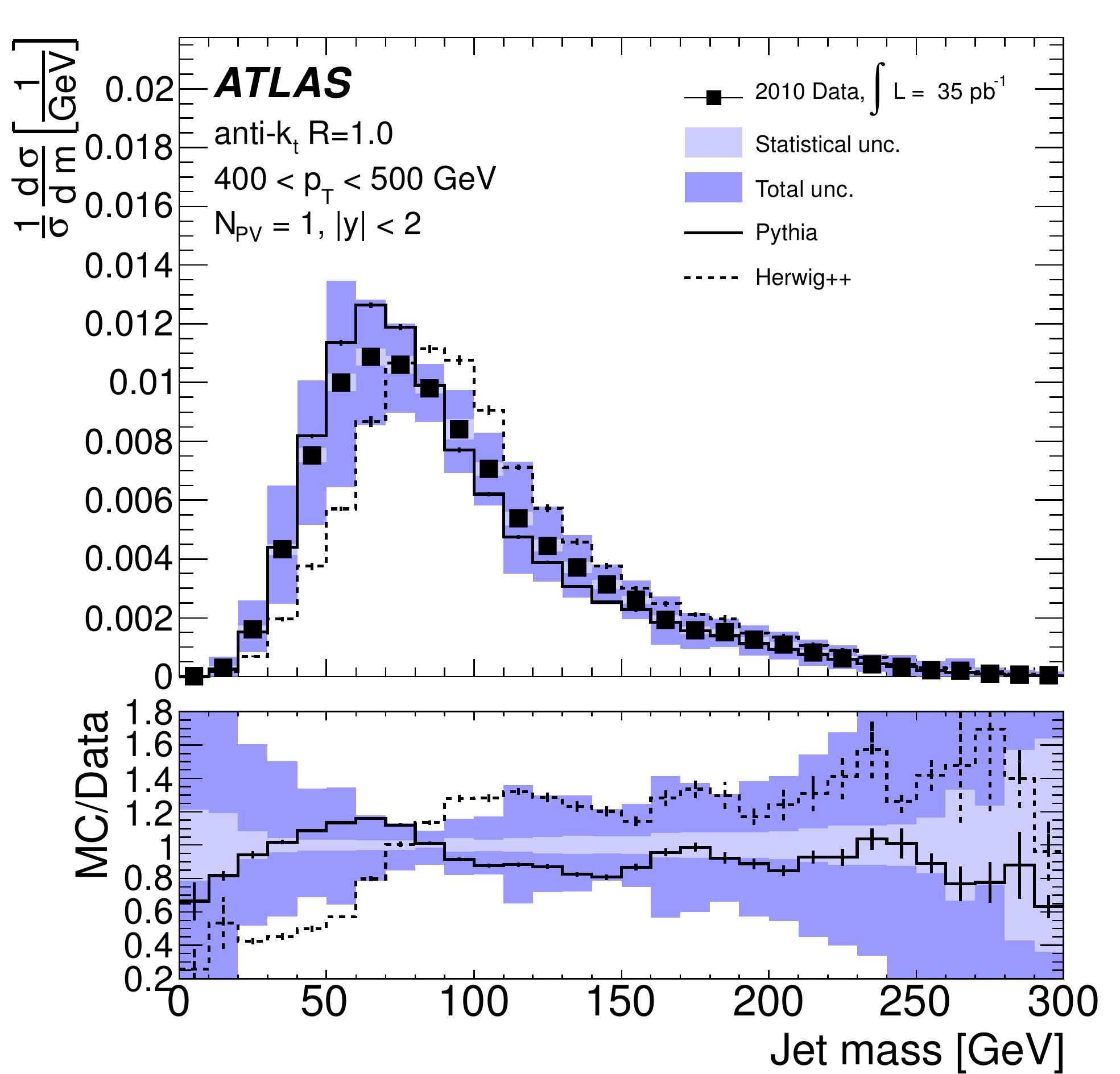}
    \includegraphics[width=0.49\linewidth]{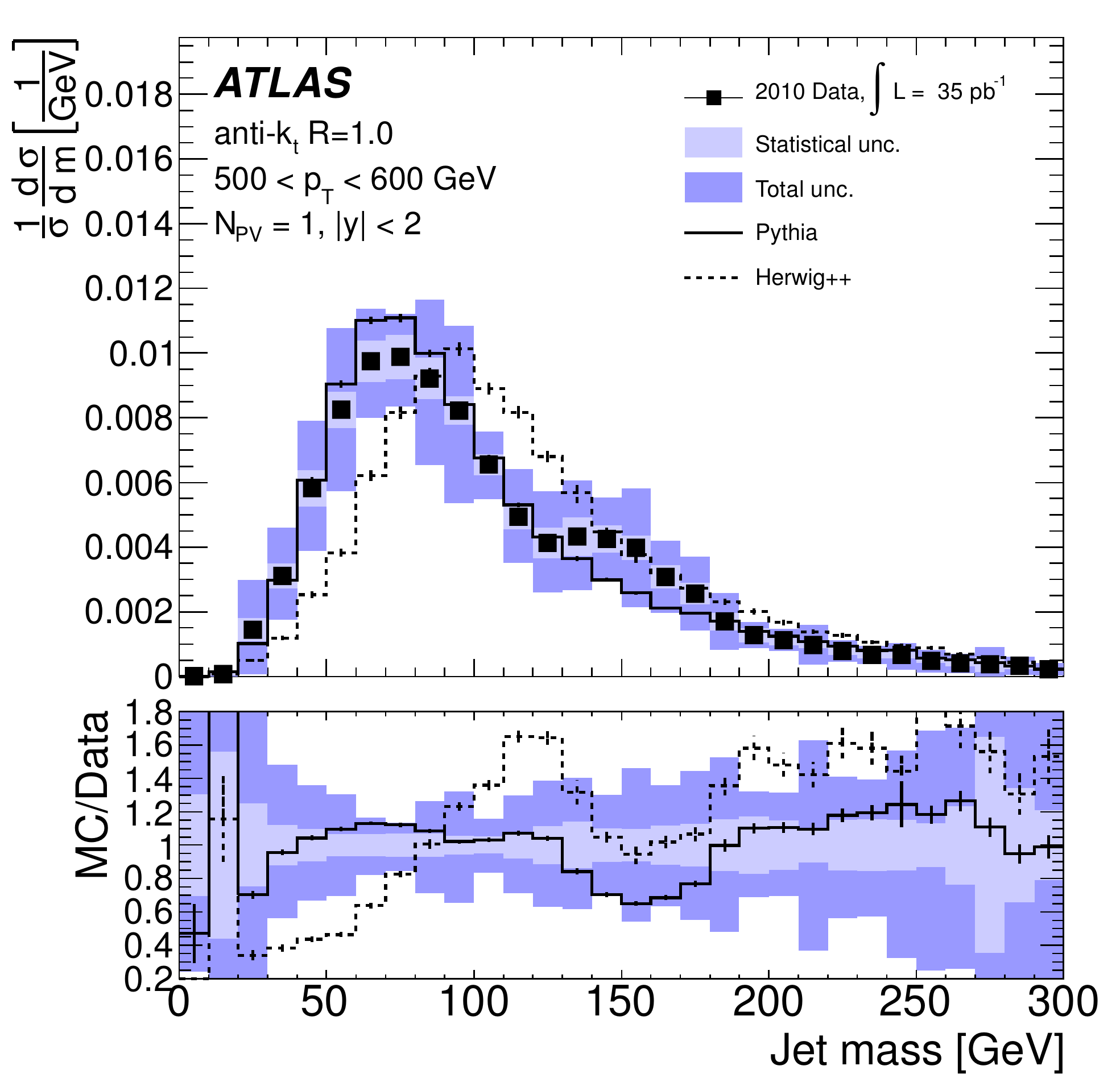}
    \caption{Normalised cross-sections as functions of mass of \antikt\ jets with
      $R=1.0$ in four different \pt\ bins.\label{fig:final_akt10mass}}
  \end{center}
\end{figure}

\begin{figure}
  \begin{center}
    \includegraphics[width=0.49\linewidth]{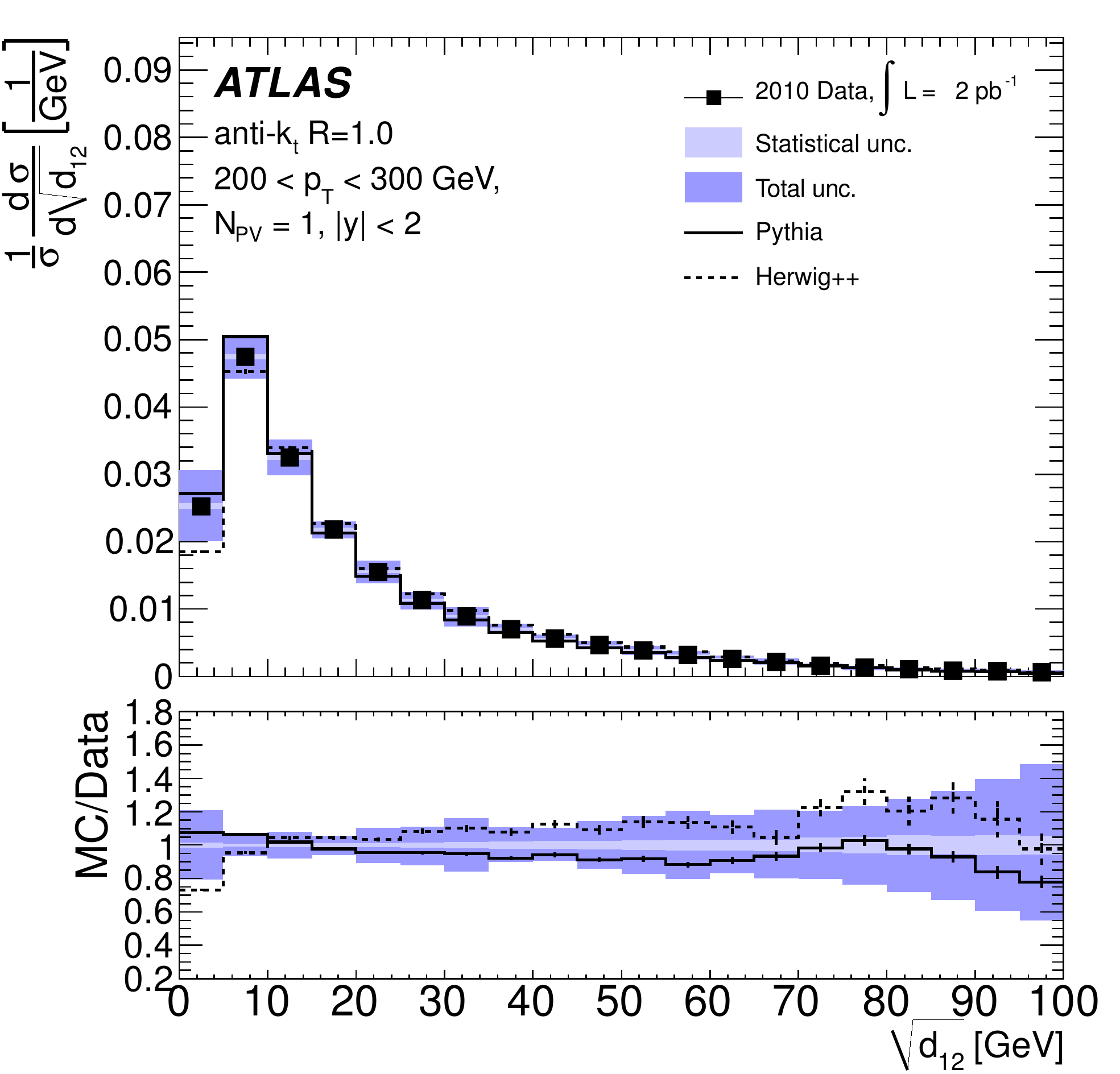}
    \includegraphics[width=0.49\linewidth]{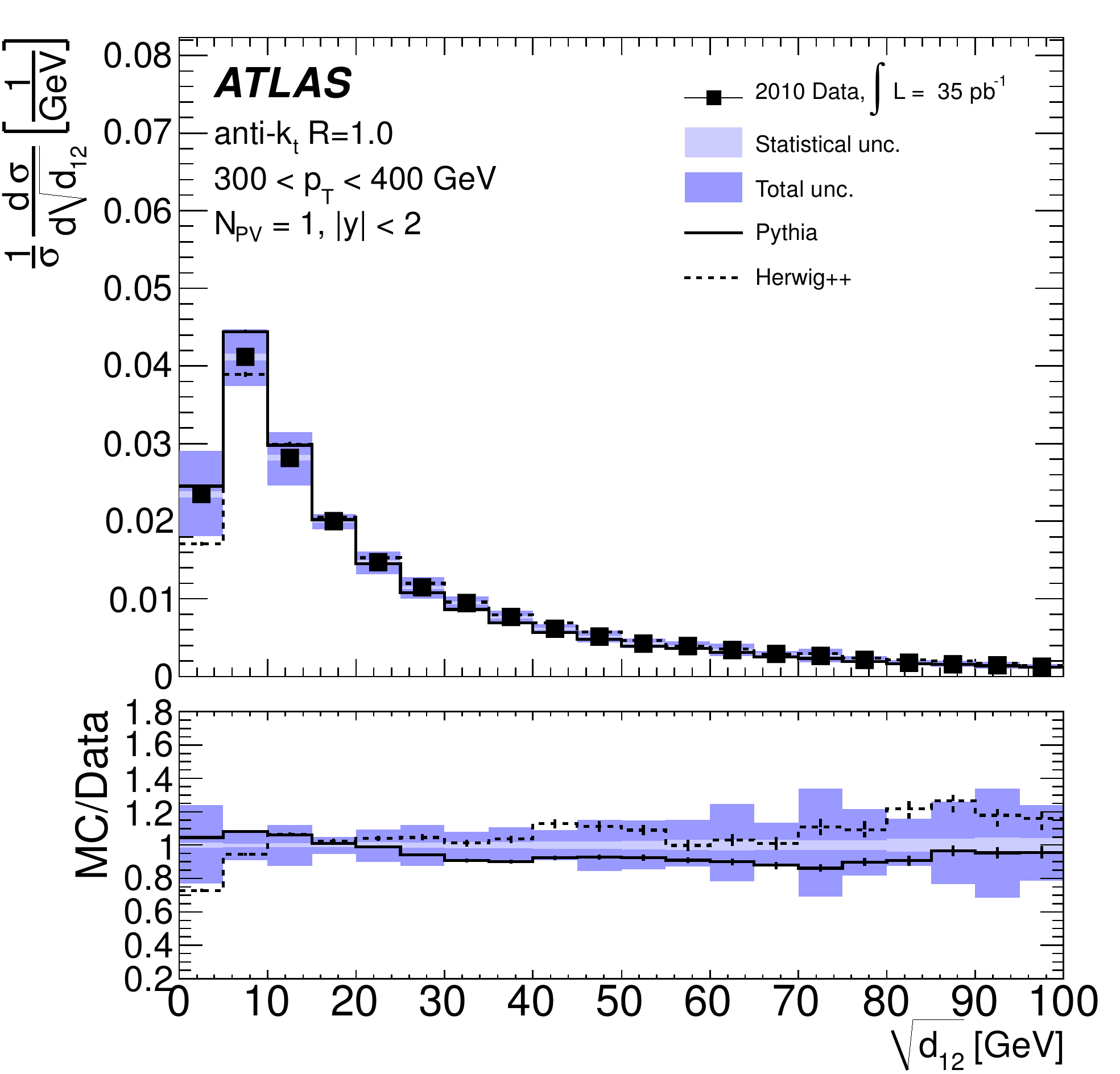}
    \includegraphics[width=0.49\linewidth]{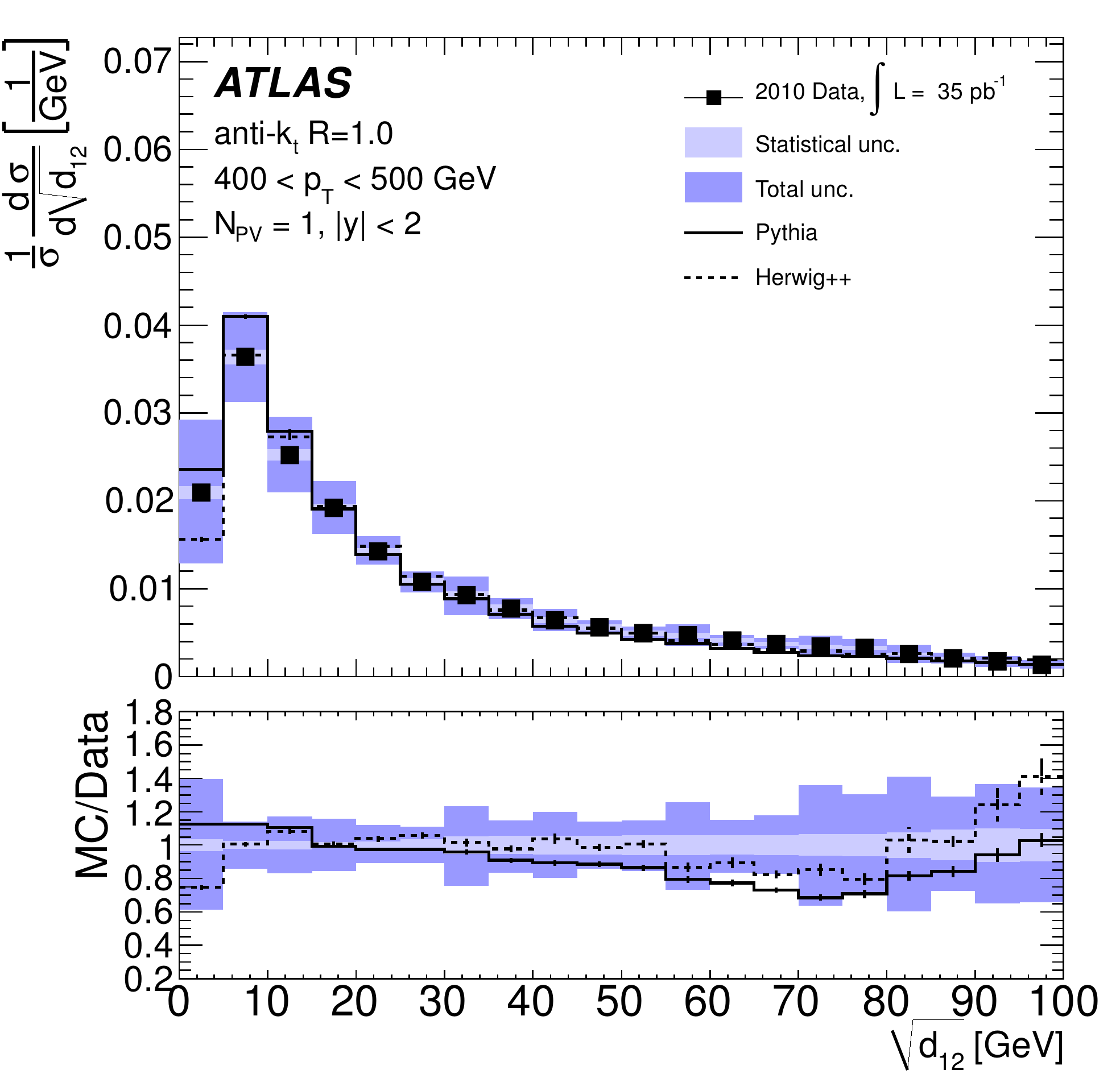}
    \includegraphics[width=0.49\linewidth]{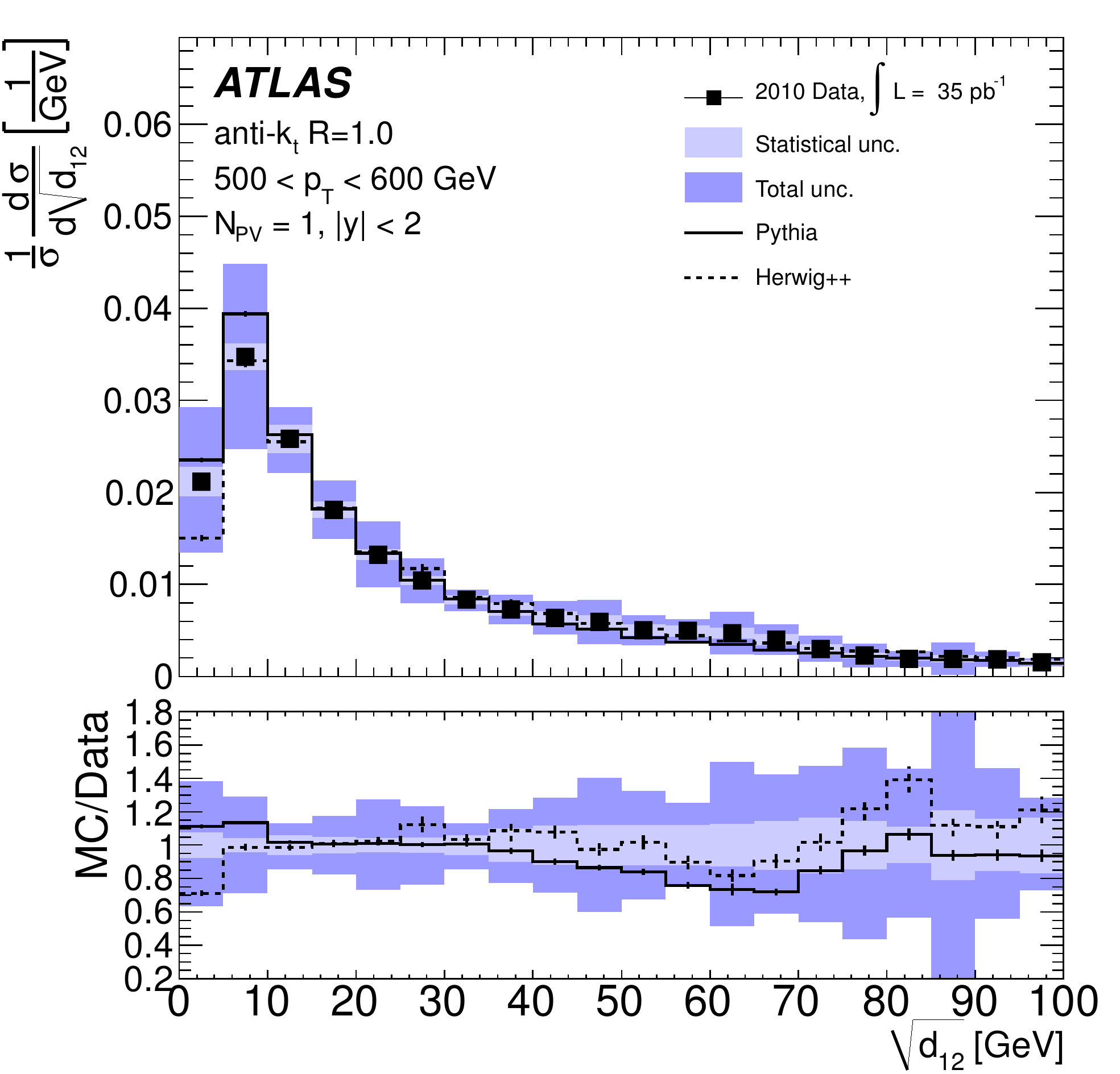}
    \caption{Normalised cross-sections as functions of \DOneTwo of \antikt\ jets with
      $R=1.0$ in four different \pt\ bins.\label{fig:final_akt10d12}}
  \end{center}
\end{figure}

\begin{figure}
  \begin{center}
    \includegraphics[width=0.49\linewidth]{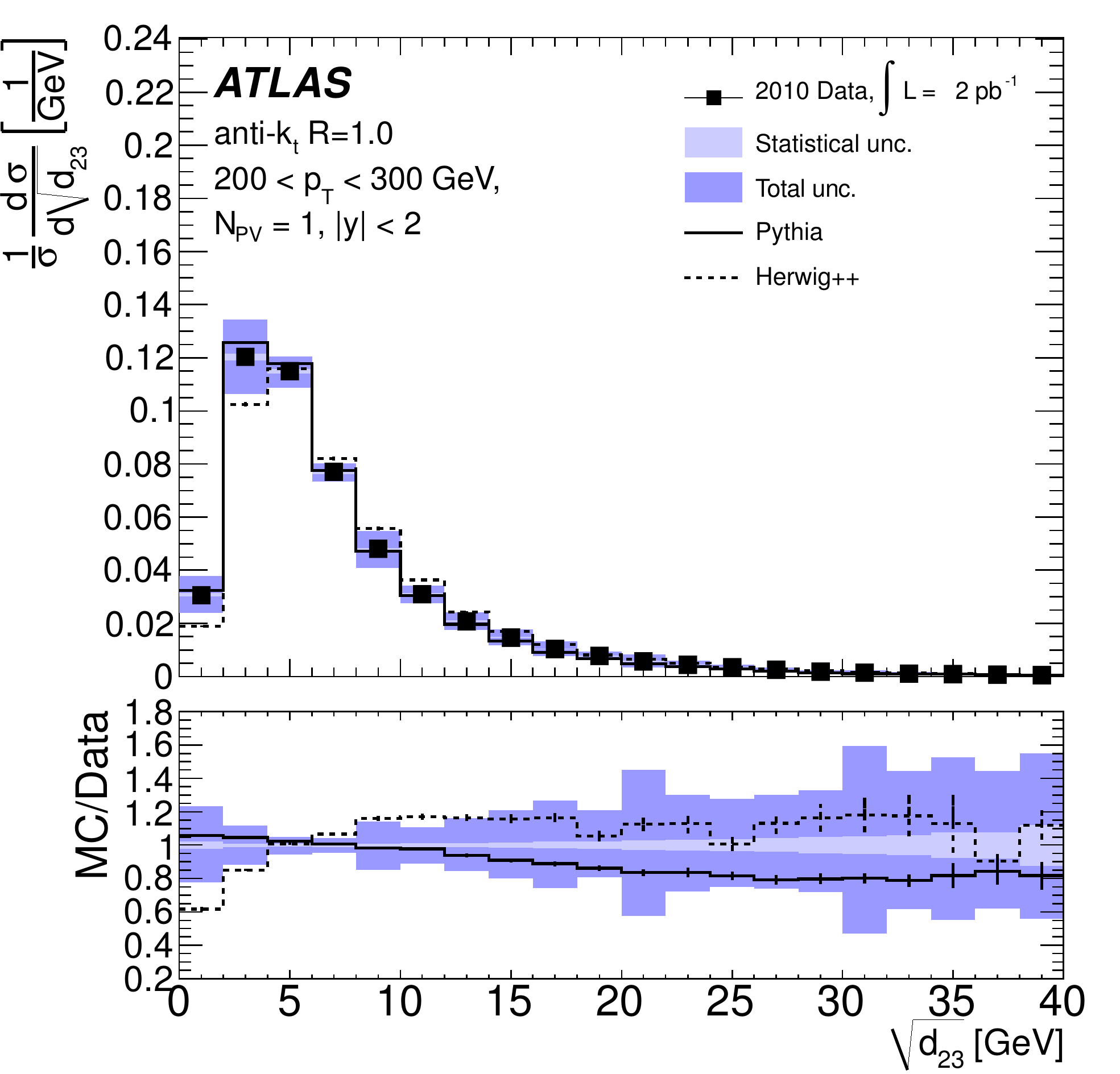}
    \includegraphics[width=0.49\linewidth]{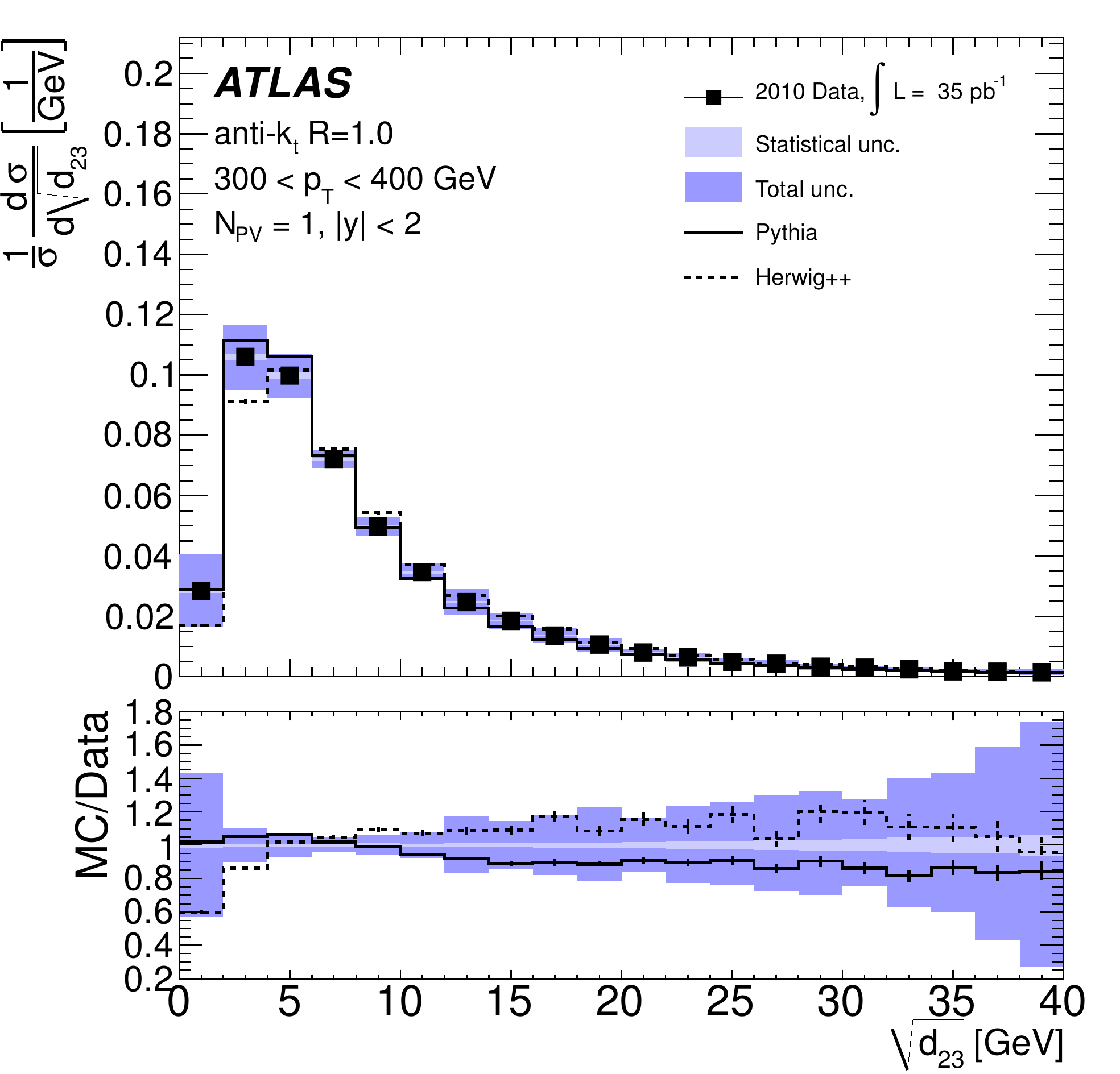}
    \includegraphics[width=0.49\linewidth]{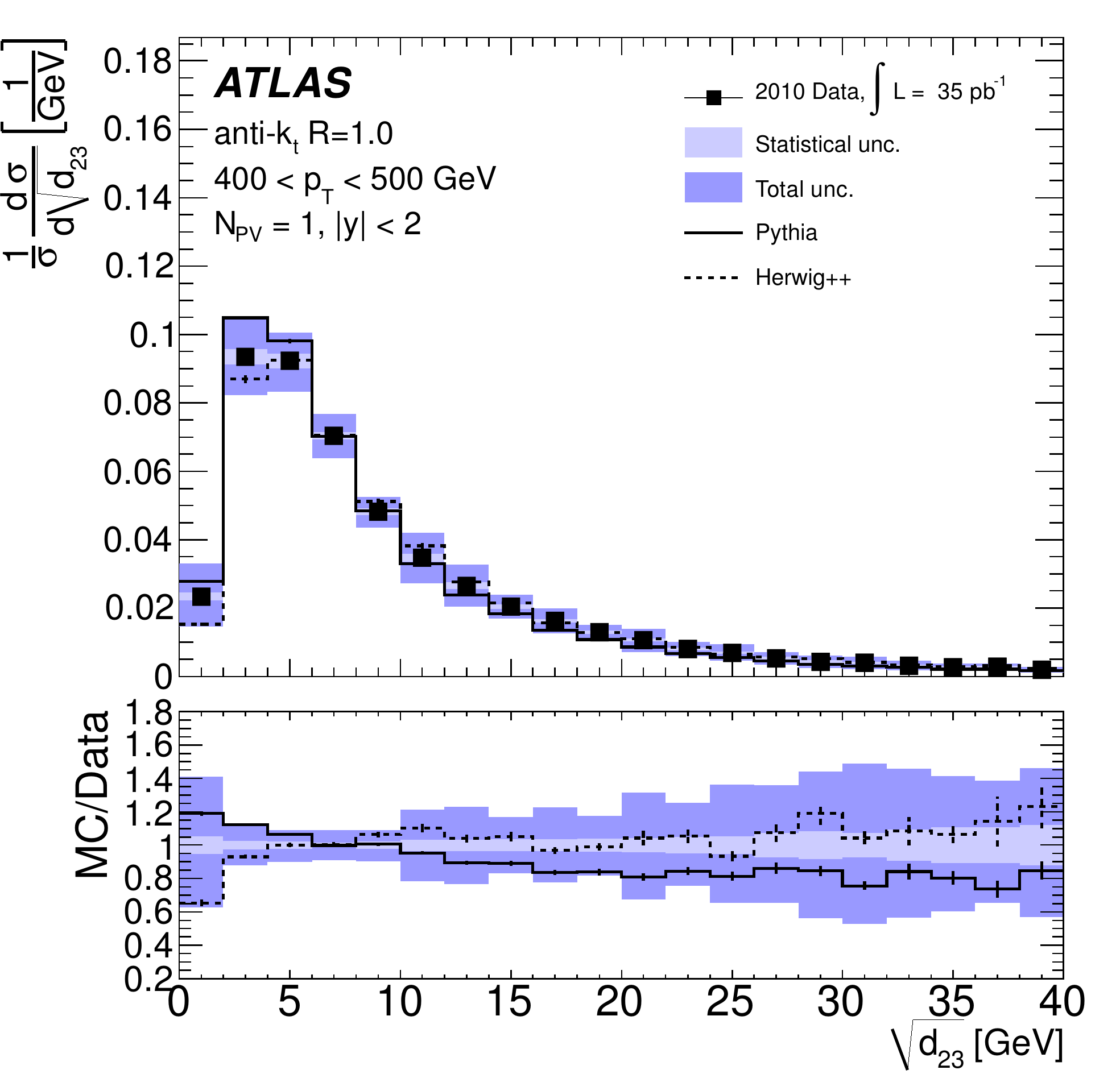}
    \includegraphics[width=0.49\linewidth]{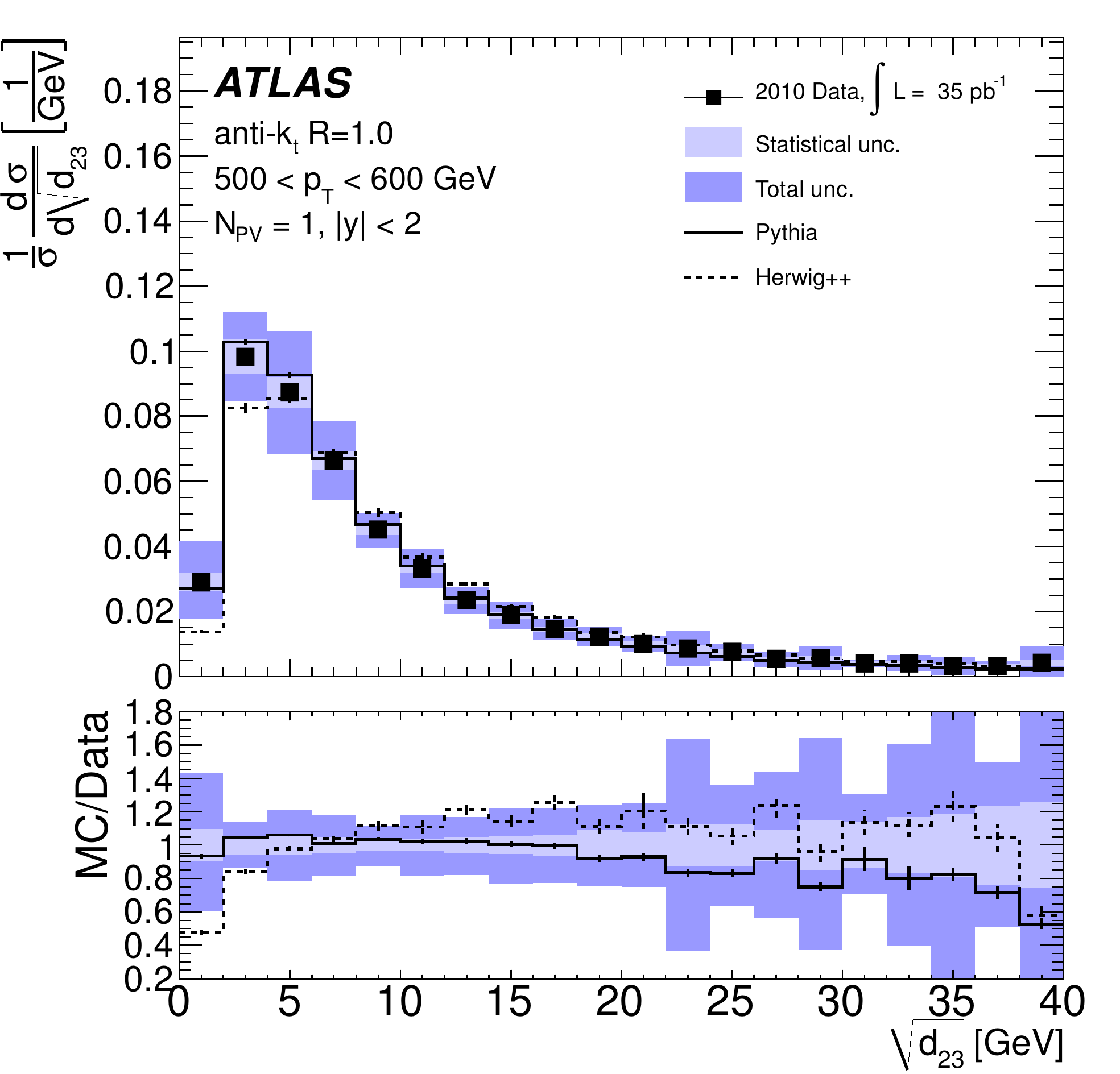}
    \caption{Normalised cross-sections as functions of \DTwoThr of \antikt\ jets with
      $R=1.0$ in four different \pt\ bins.\label{fig:final_akt10d23}}
  \end{center}
\end{figure}

\begin{figure}
  \begin{center}
    \includegraphics[width=0.49\linewidth]{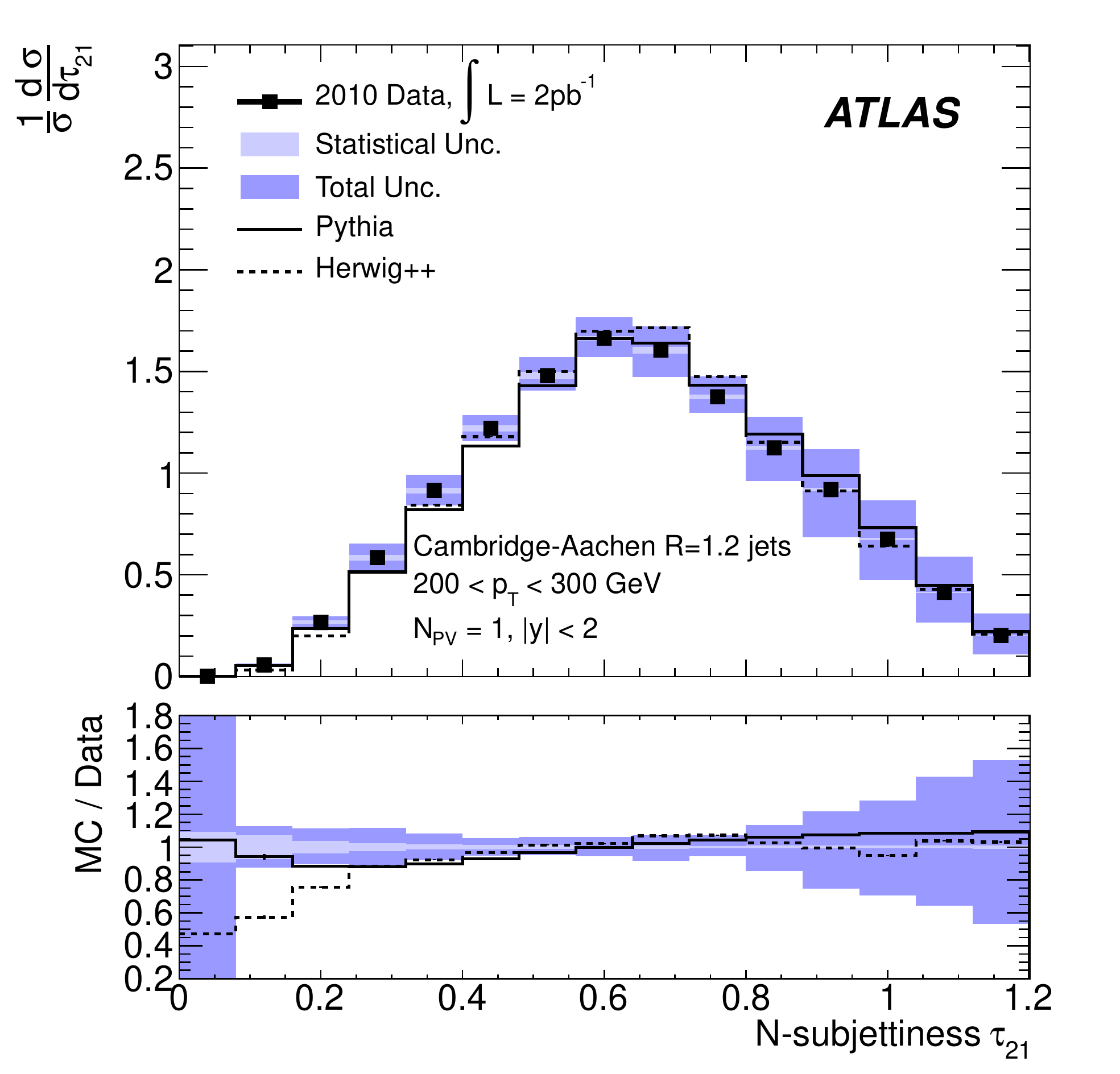}
    \includegraphics[width=0.49\linewidth]{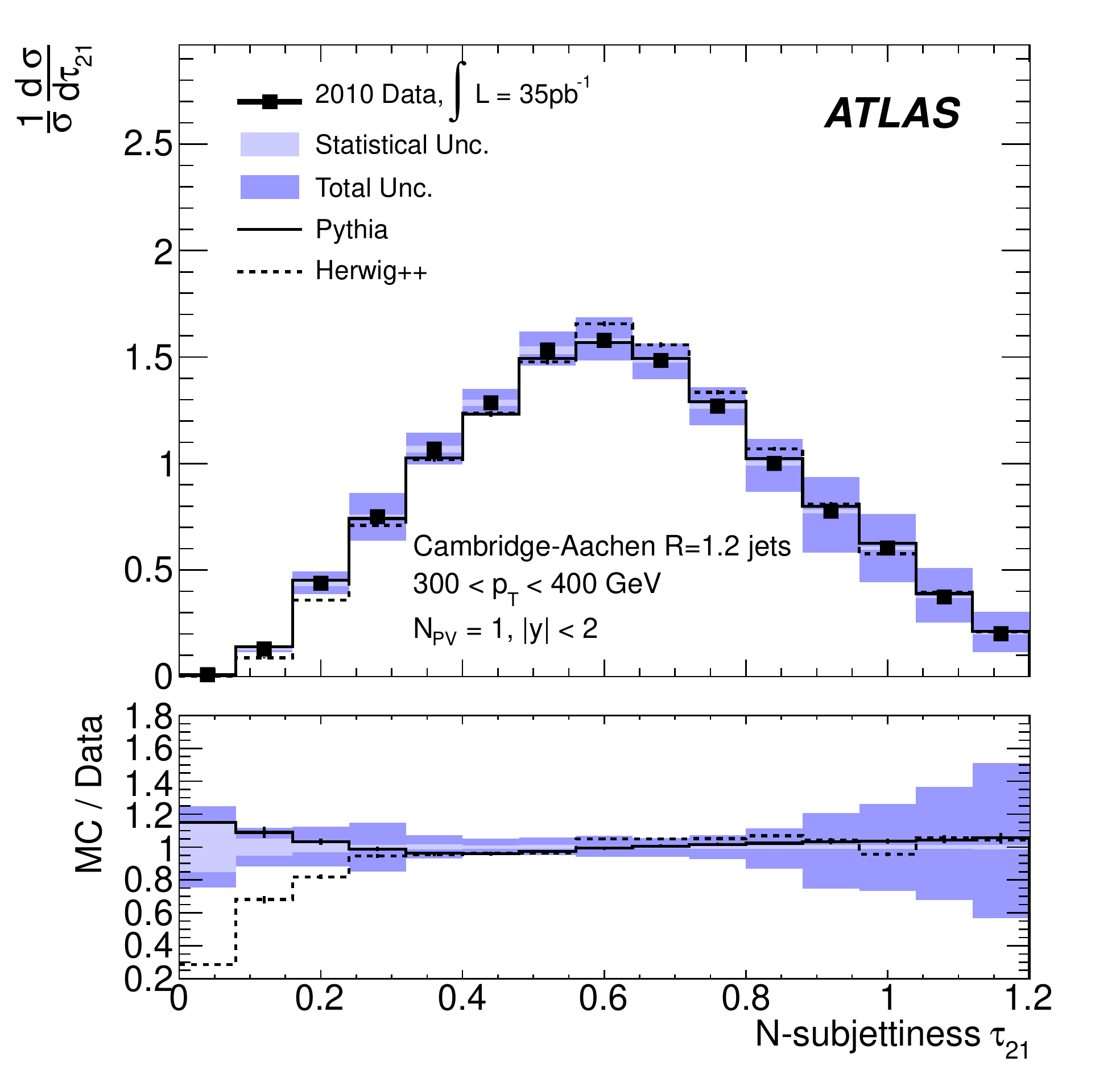}
    \includegraphics[width=0.49\linewidth]{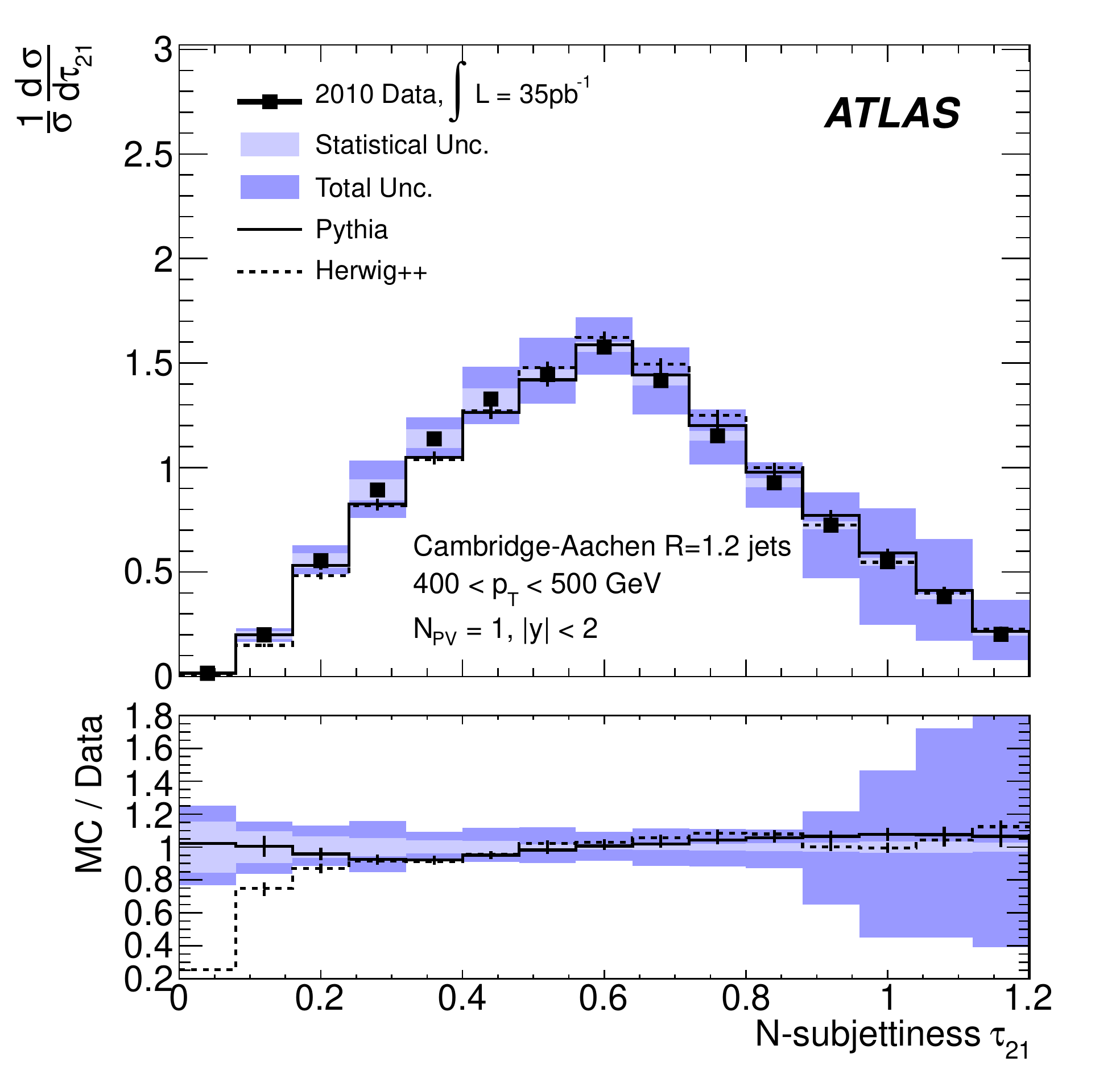}
    \includegraphics[width=0.49\linewidth]{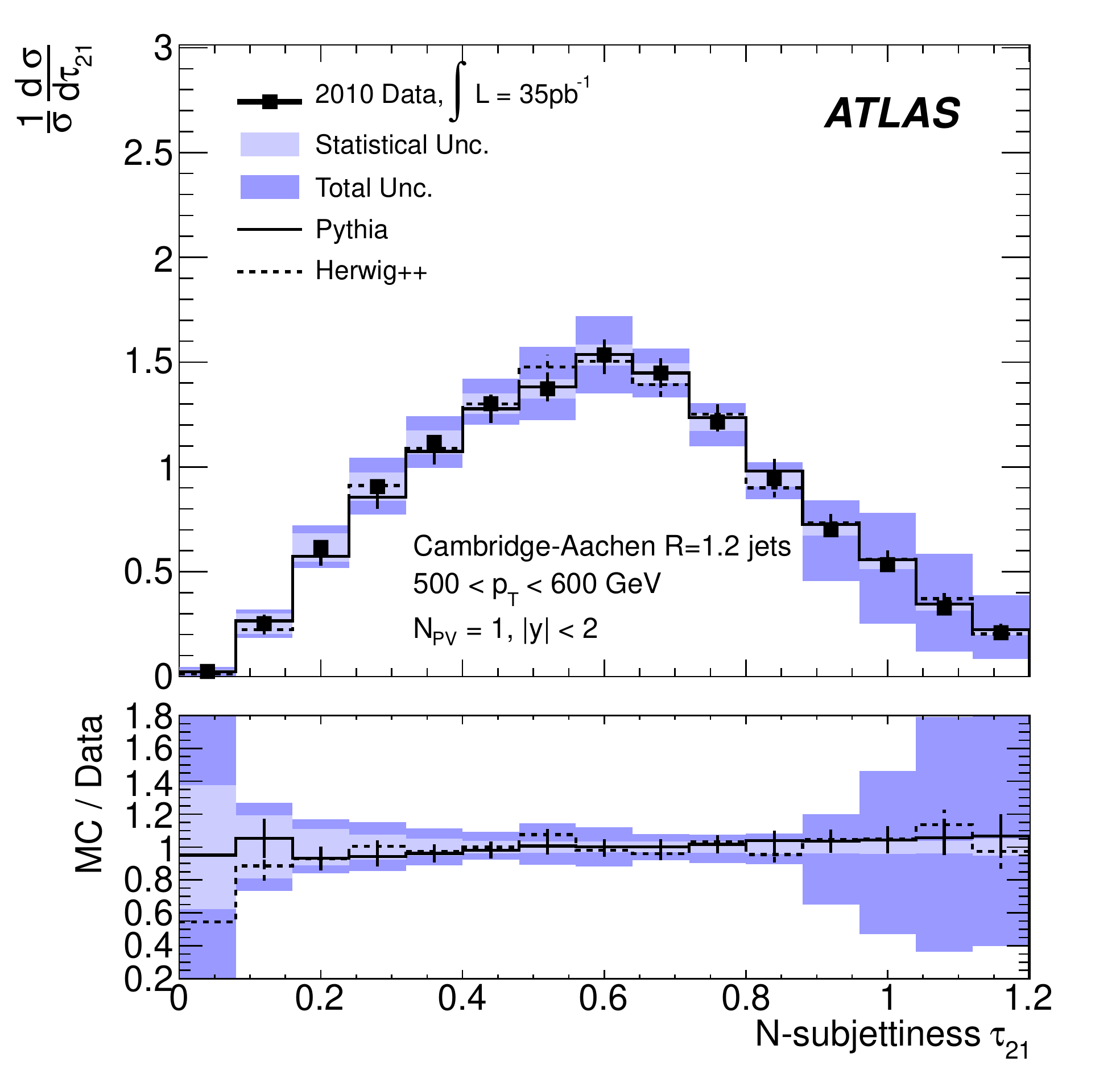}
    \caption{Normalised cross-sections as functions of \tauTwoOne of \ca\ jets with
      $R=1.2$ in four different \pt\ bins.\label{fig:final_ca12t21}}
  \end{center}
\end{figure}

\begin{figure}
  \begin{center}
    \includegraphics[width=0.49\linewidth]{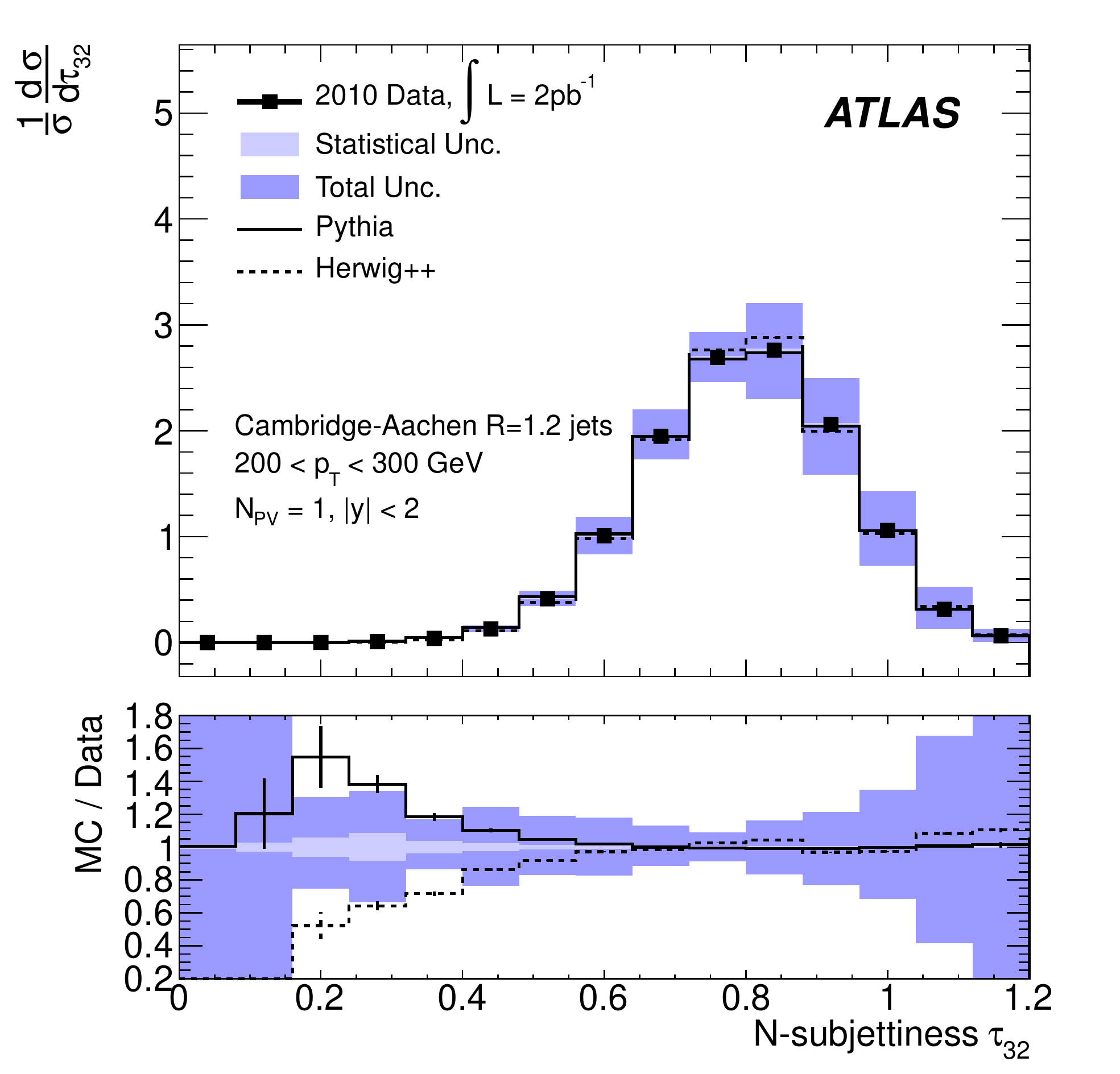}
    \includegraphics[width=0.49\linewidth]{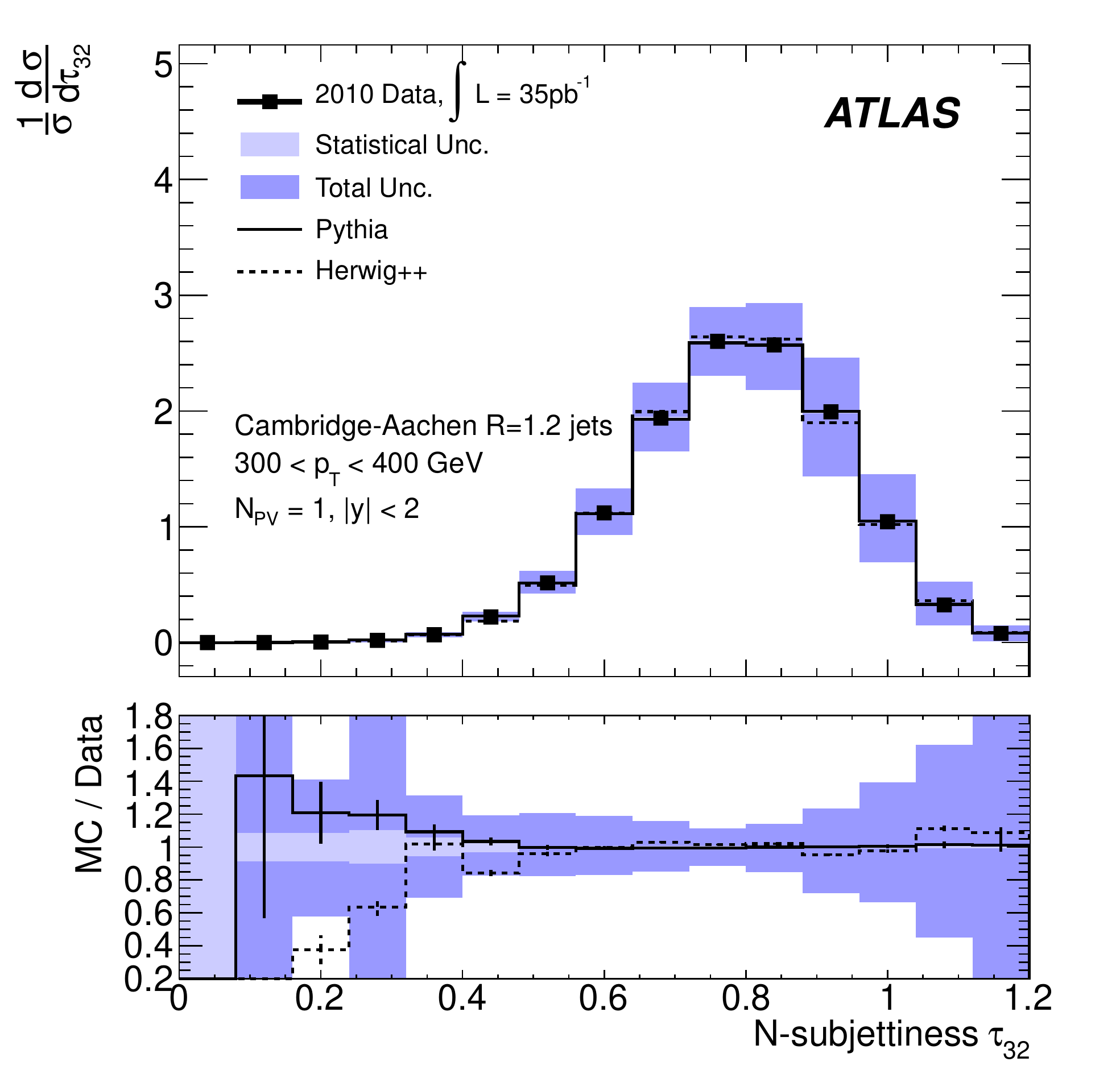}
    \includegraphics[width=0.49\linewidth]{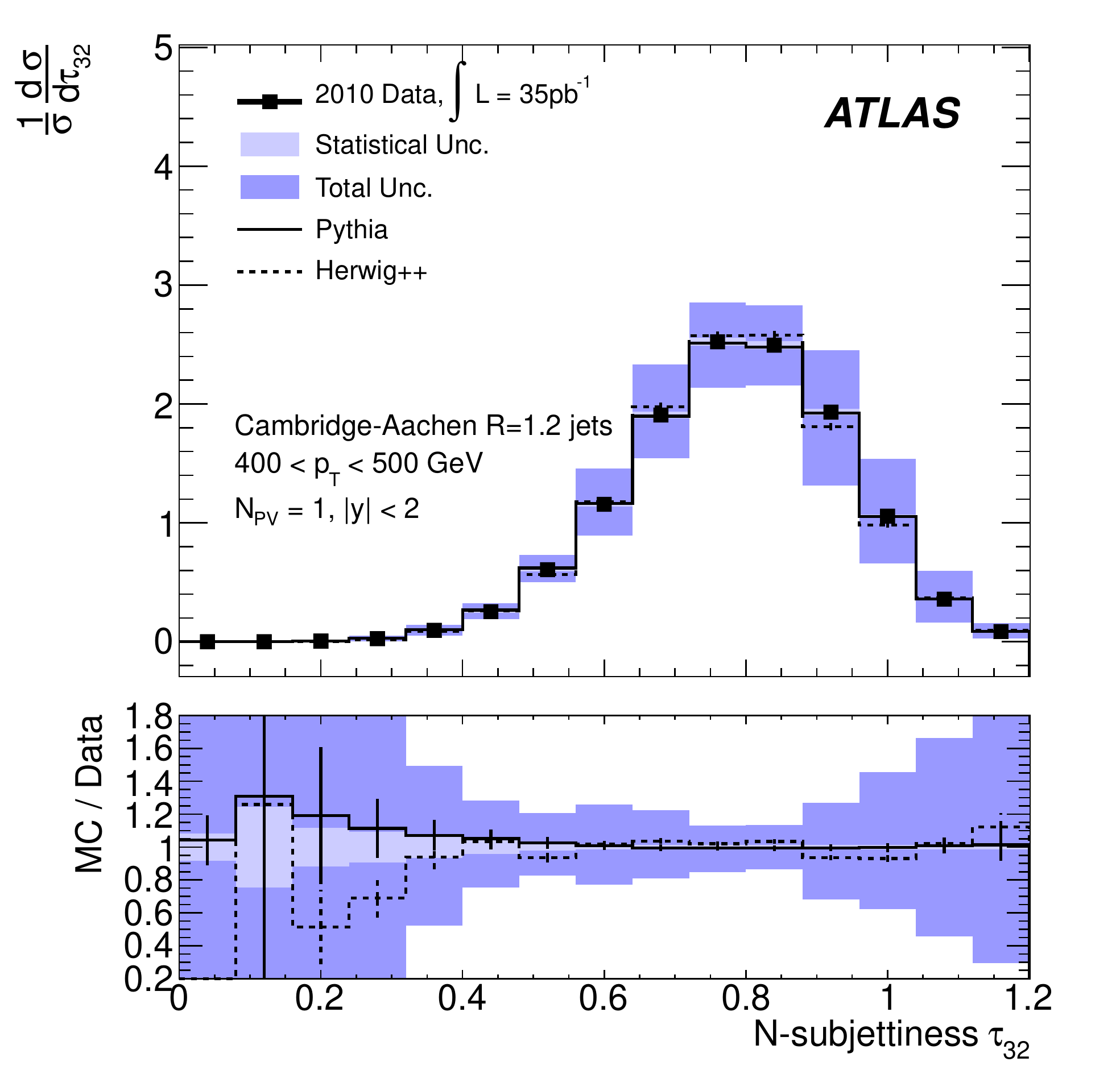}
    \includegraphics[width=0.49\linewidth]{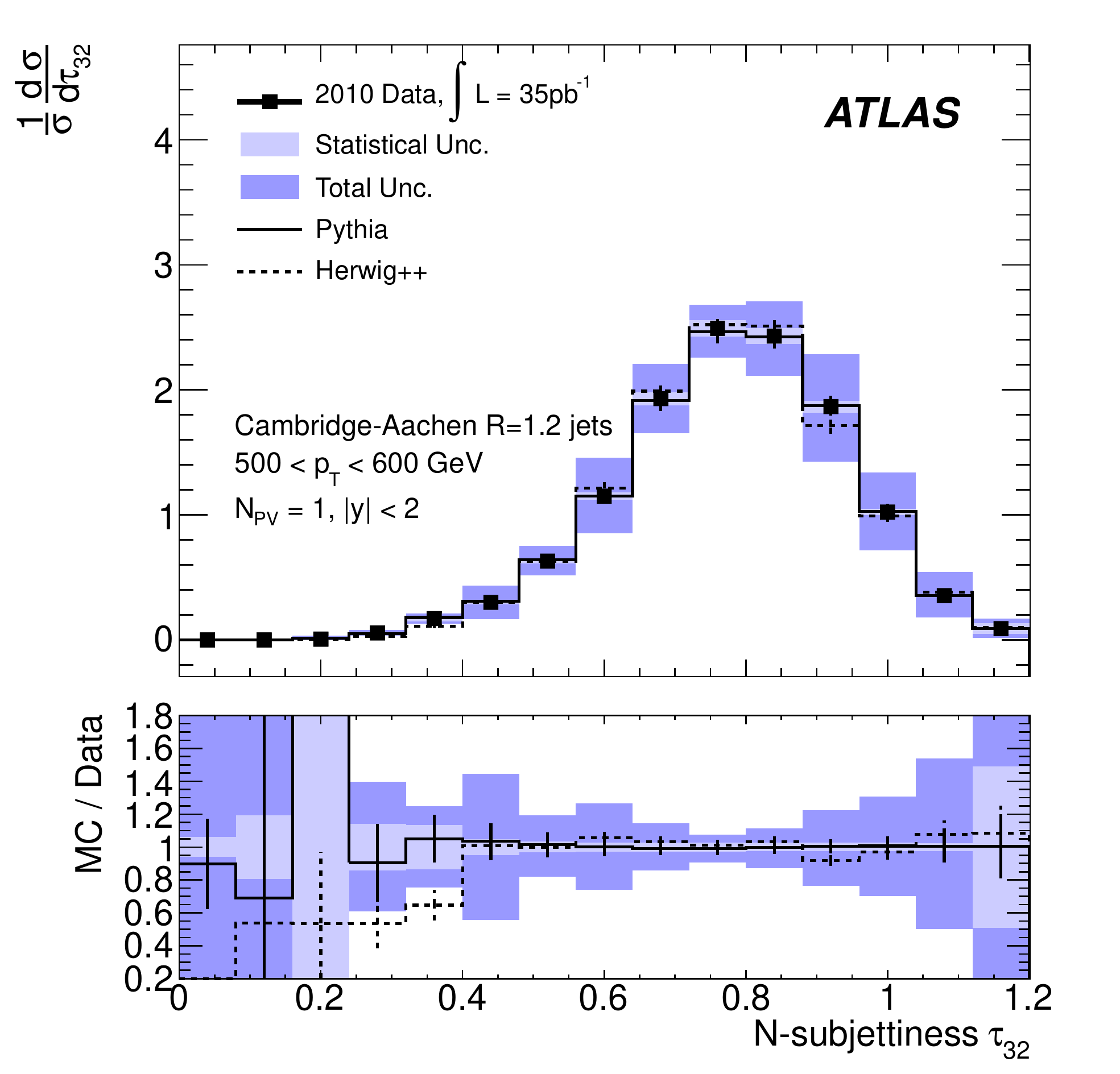}
    \caption{Normalised cross-sections as functions of \tauThrTwo of \ca\ jets with
      $R=1.2$ in four different \pt\ bins.\label{fig:final_ca12t32}}
  \end{center}
\end{figure}

\begin{figure}
  \begin{center}
    \includegraphics[width=0.49\linewidth]{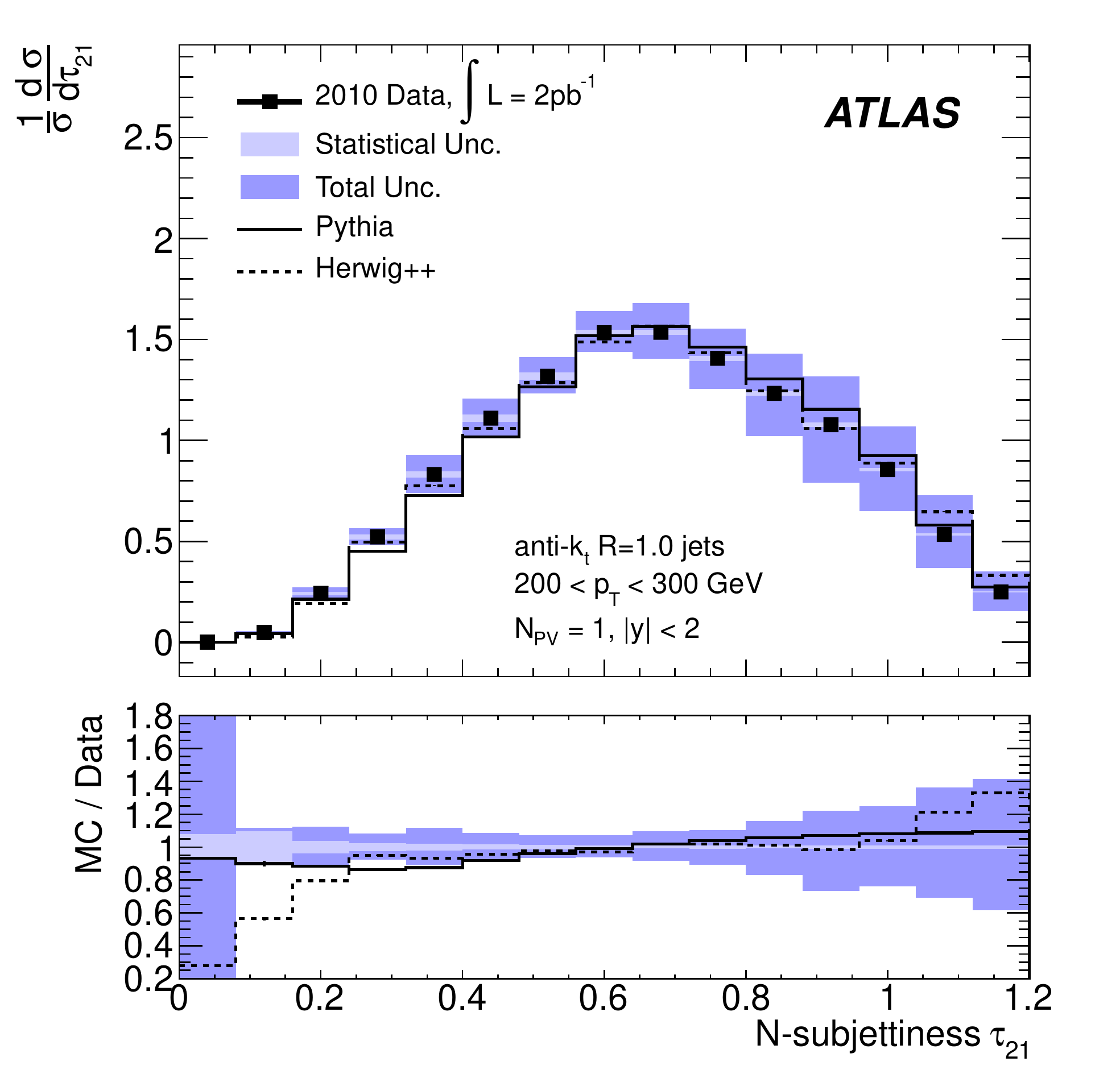}
    \includegraphics[width=0.49\linewidth]{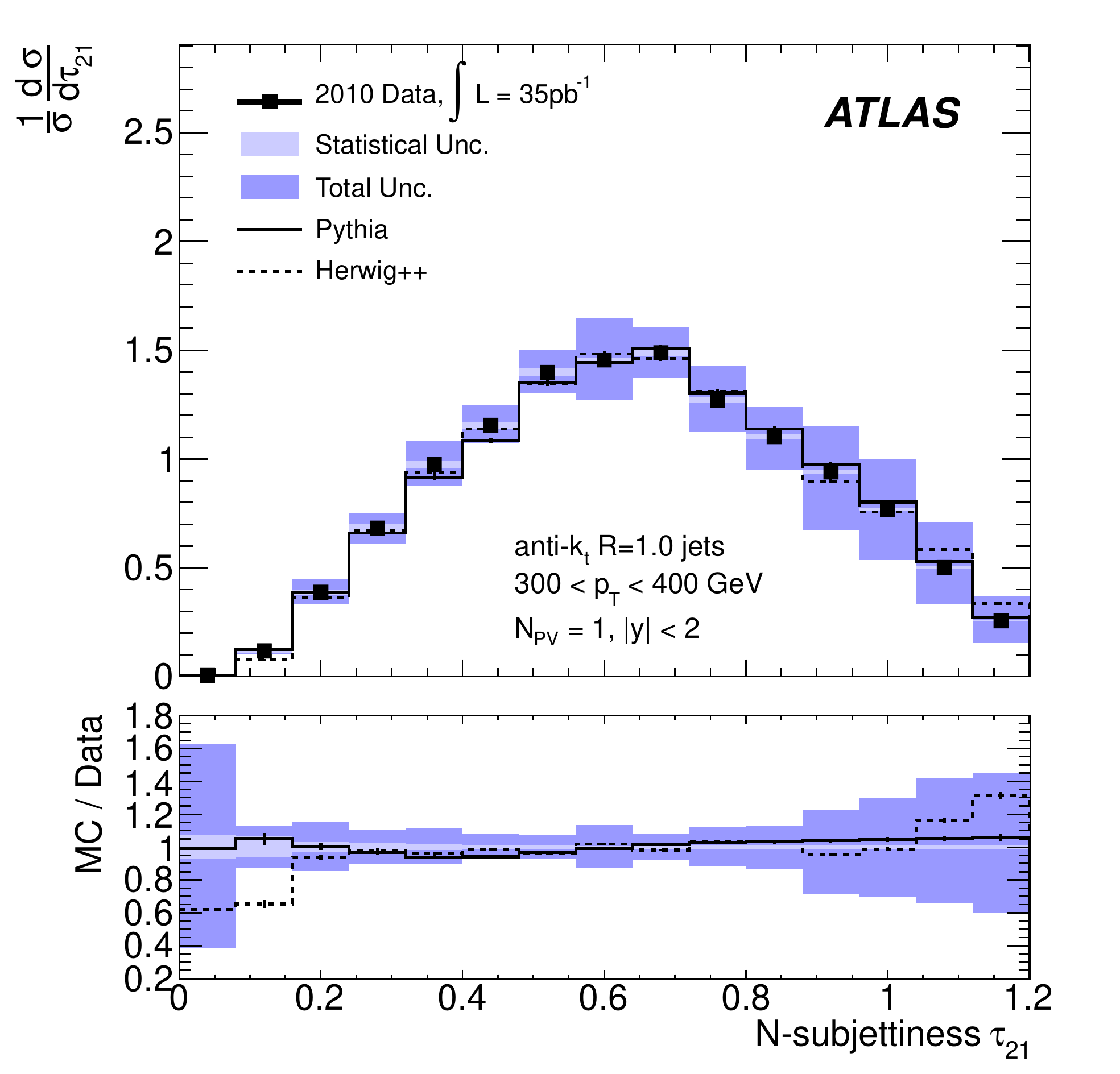}
    \includegraphics[width=0.49\linewidth]{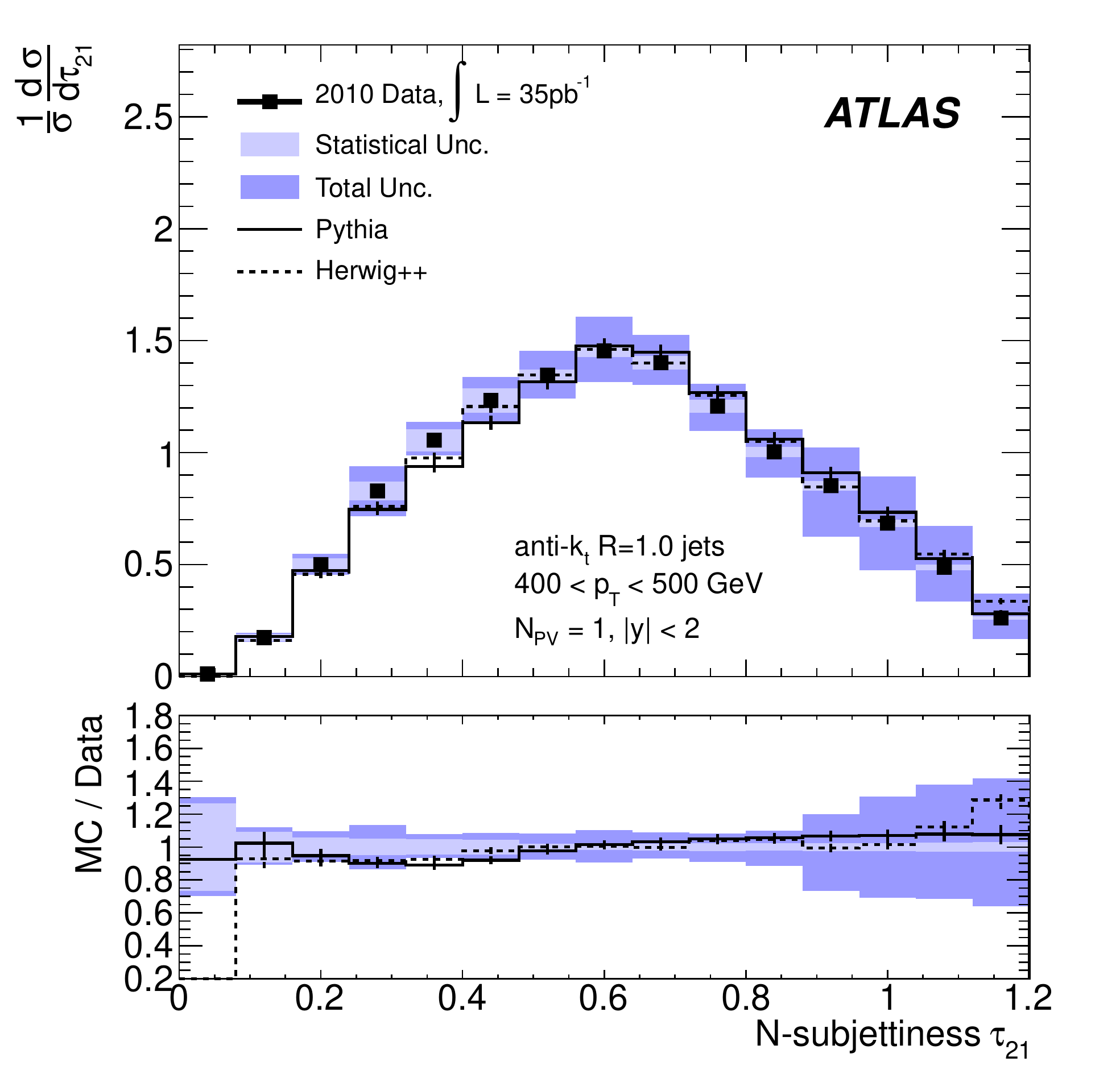}
    \includegraphics[width=0.49\linewidth]{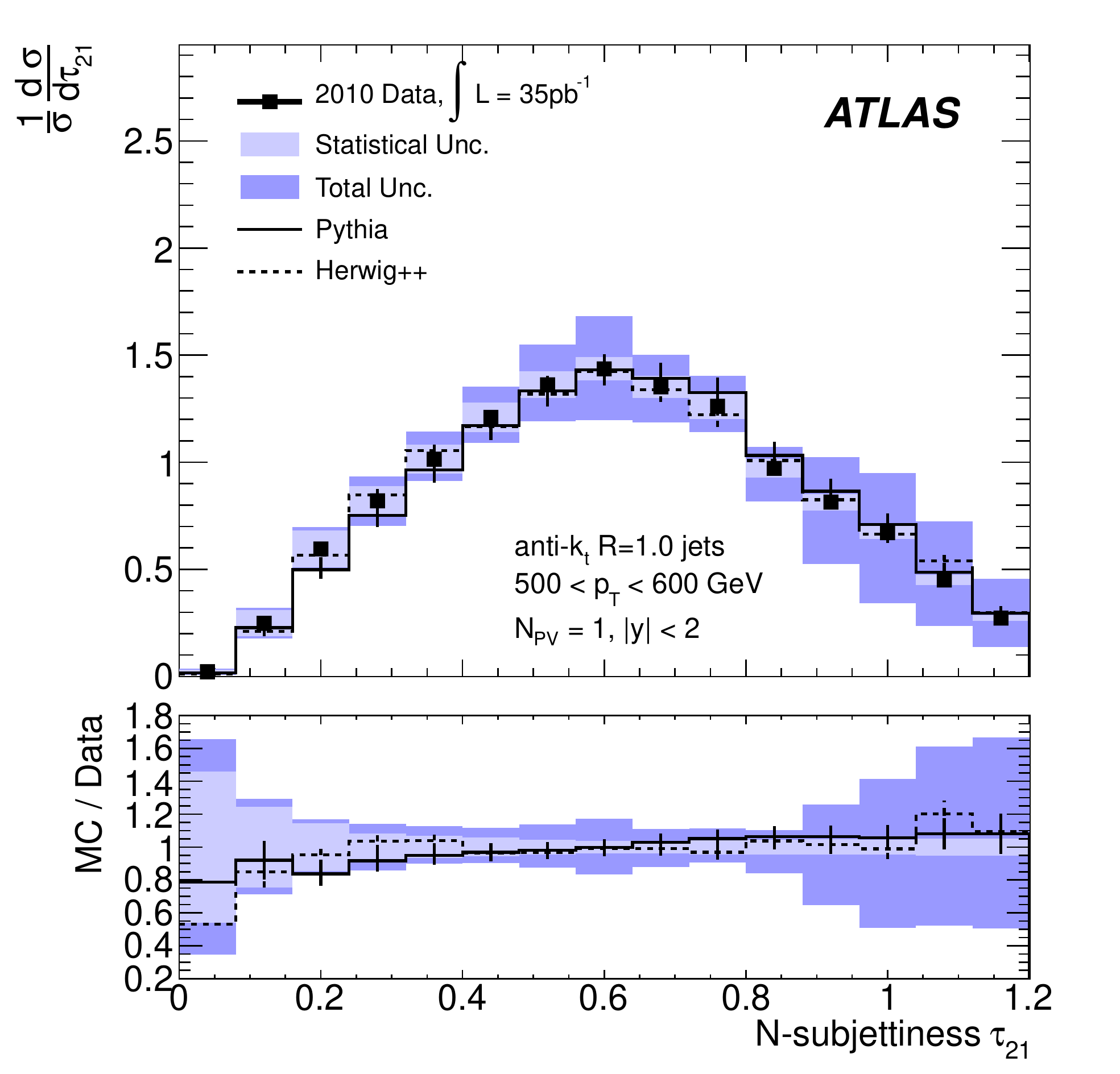}
    \caption{Normalised cross-sections as functions of \tauTwoOne of \antikt\ jets with
      $R=1.0$ in four different \pt\ bins.\label{fig:final_akt10t21}}
  \end{center}
\end{figure}

\begin{figure}
  \begin{center}
    \includegraphics[width=0.49\linewidth]{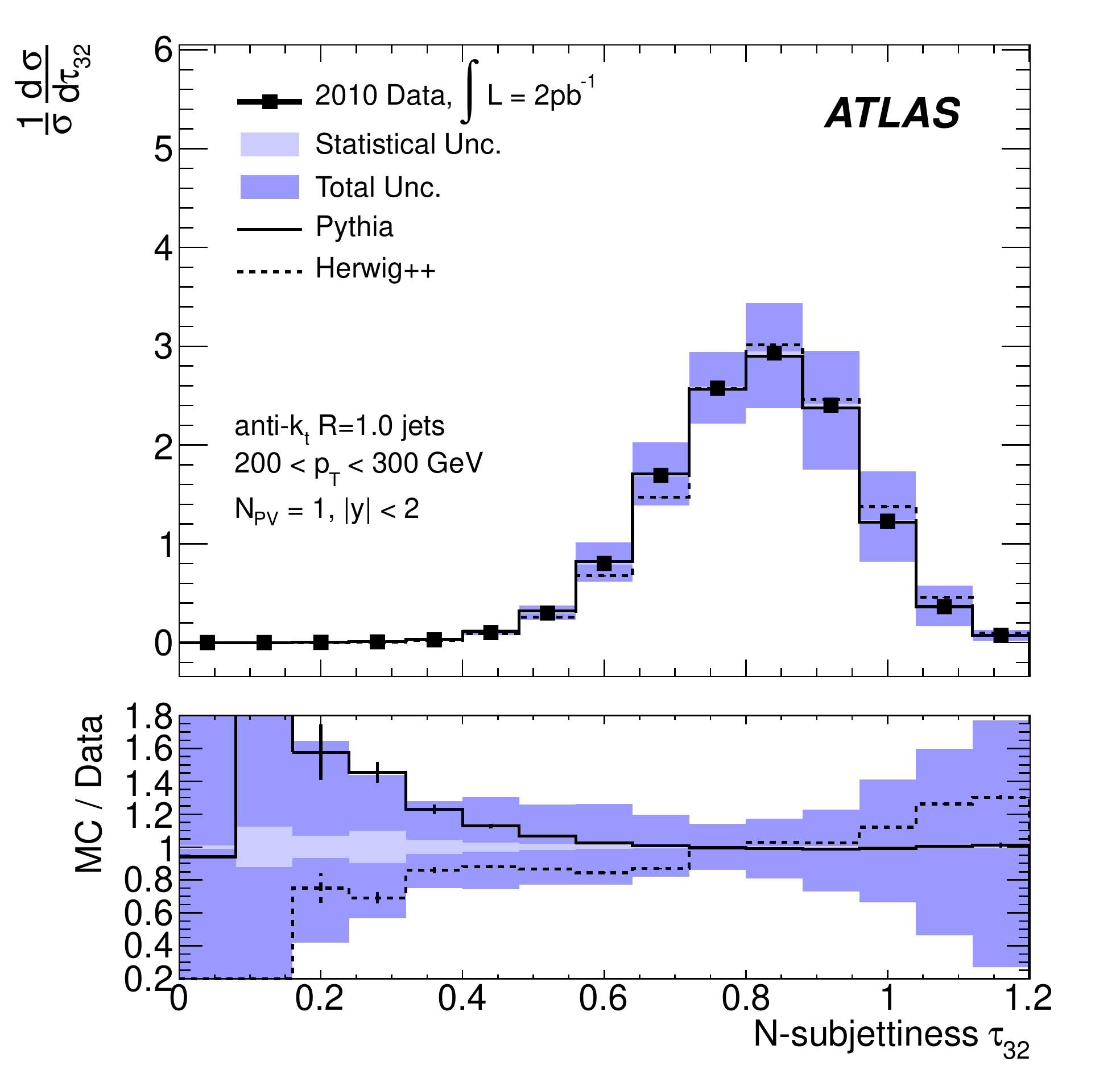}
    \includegraphics[width=0.49\linewidth]{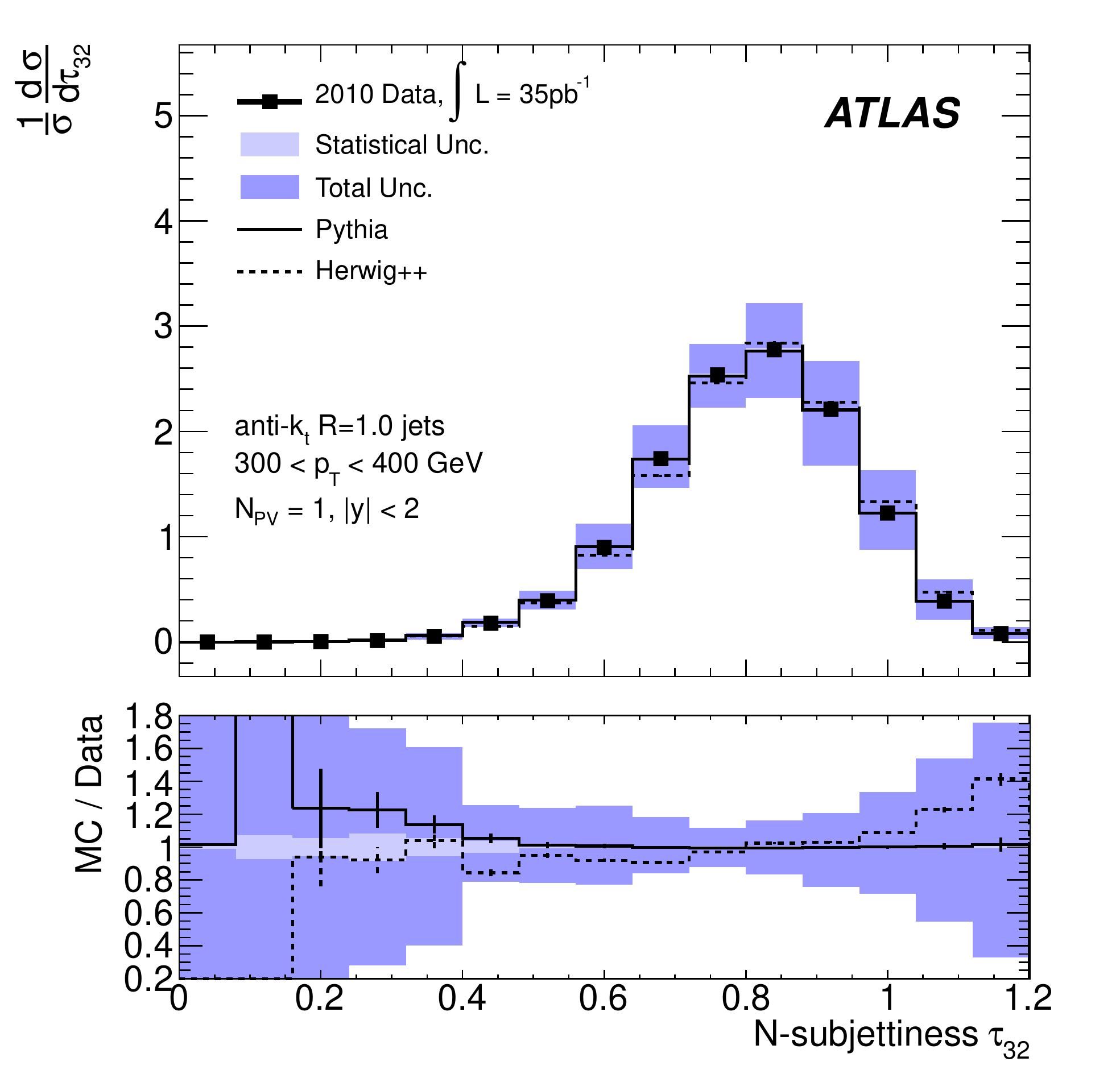}
    \includegraphics[width=0.49\linewidth]{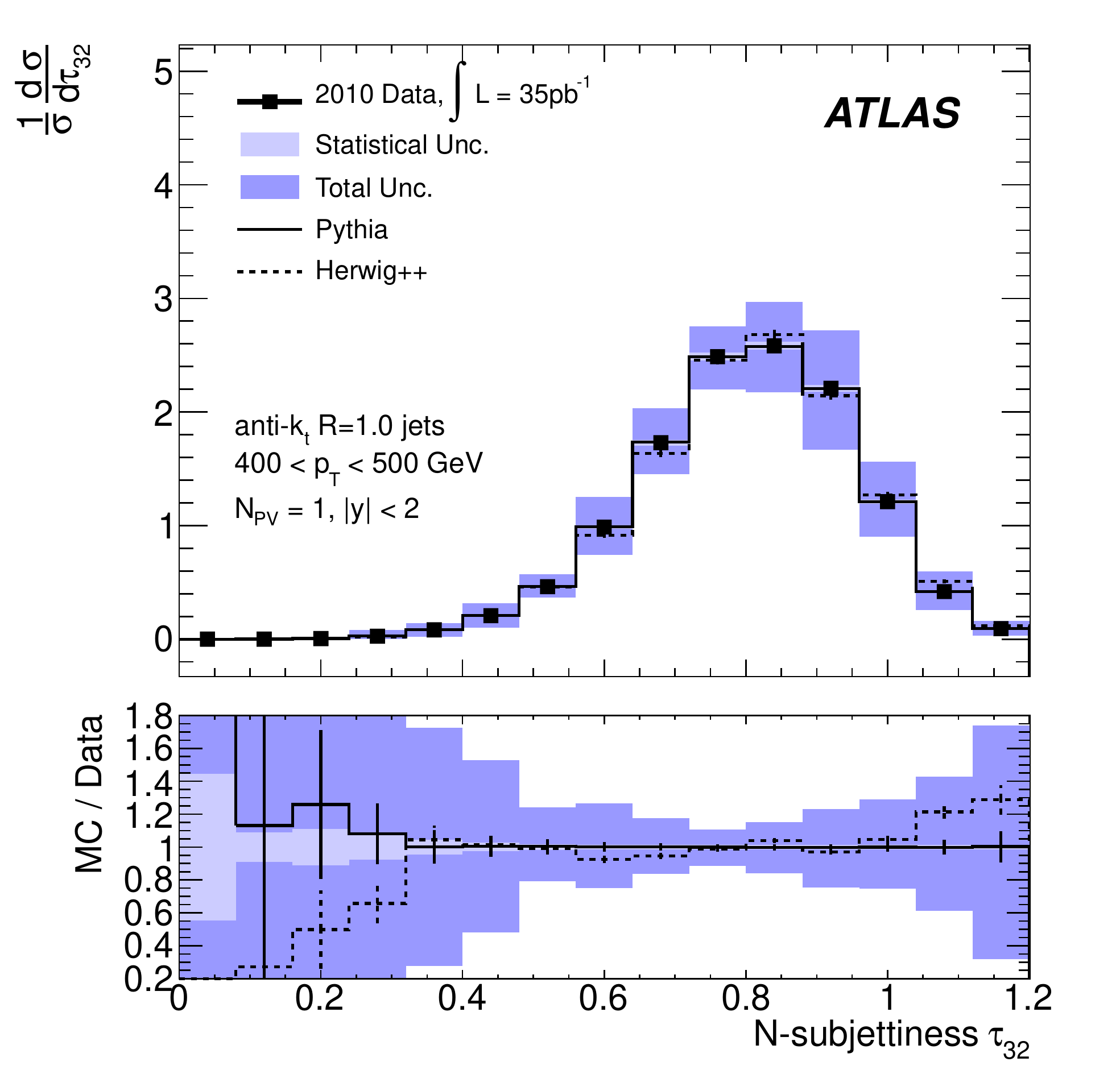}
    \includegraphics[width=0.49\linewidth]{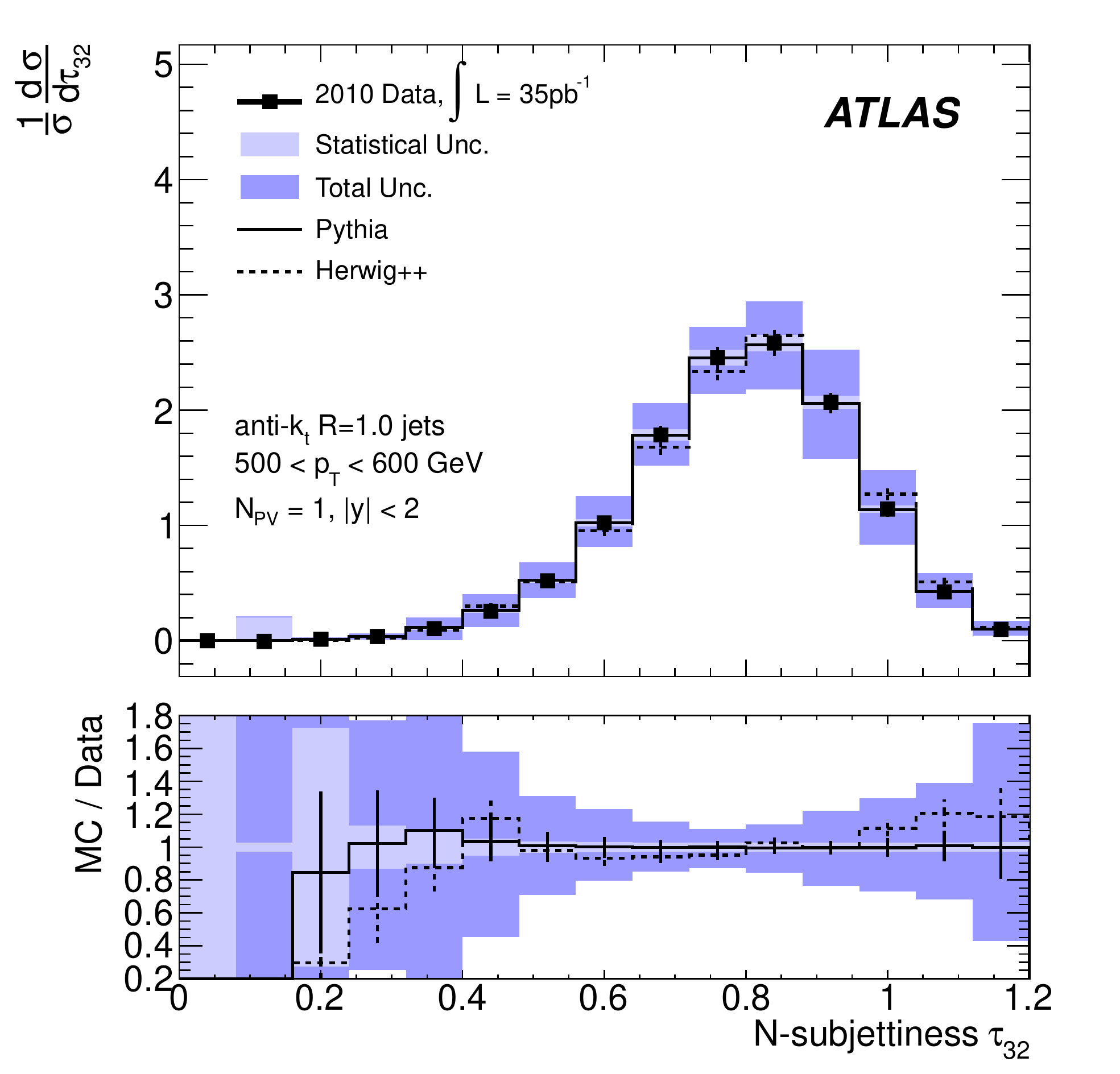}
    \caption{Normalised cross-sections as functions of \tauThrTwo of \antikt\ jets with
      $R=1.0$ in four different \pt\ bins.\label{fig:final_akt10t32}}
  \end{center}
\end{figure}

Using the analysis techniques outlined above, measured normalised
cross-sections are shown in
Figures~\ref{fig:final_ca12mass}-\ref{fig:final_akt10t32}. In ratio
plots, the statistical uncertainty on Monte Carlo predictions does not
include the data statistical uncertainty. Although in some cases the
Monte Carlo predictions are not in agreement with the data, the shapes
of the distributions are correctly reproduced. For jet mass the
distributions produced by \pythia\ tend to be too soft, while those
from \herwigpp\ are too hard. Notably, the \ca\ jet mass after
splitting and filtering, as shown in
Figure~\ref{fig:final_ca12filtmass}, is the only variable for which
the Monte Carlo predictions are in agreement to within statistical
uncertainties, both with each other and the data. The substructure
variables exhibit generally better agreement with Monte Carlo
predictions than mass, with all but a few bins correctly described by
both \pythia\ and \herwigpp. In the higher \pt\ bins statistical
fluctuations begin to limit the precision of the measurements, but the
level of agreement in all variables appears to remain approximately
constant between \pt\ bins.

The unfolding technique used introduces correlations between the
bins. The statistical uncertainty in these results represents the
diagonal element of the covariance matrix only; therefore, comparison
to alternative predictions requires use of the full covariance
matrices. These matrices are available, along with the full results
presented here, in {\sc HepData}~\cite{HEPDATA}. In particular the
structure at a jet mass of 150-180~\GeV\ in the 500-600~\GeV\ bin of
Figures~\ref{fig:final_ca12mass} and~\ref{fig:final_akt10mass} is
consistent with statistical fluctuations.

\section{Mean Mass With Multiple Proton-Proton Interactions}
\label{sec:pu}
The results presented so far have been for events containing only one
\pp interaction; however even in this early period of running, the
data contain events with multiple simultaneous \pp interactions
(pile-up)~\cite{OffsetNote}. These additional collisions are
uncorrelated with the hard-scattering process that typically triggers
the event. They therefore present a background of soft, diffuse
radiation that offsets the energy measurement of jets and will impact
jet-shape and substructure measurements. It is essential that future
studies involving jet-substructure variables, such as those
investigated here, be able to understand and correct for the effects
of pile-up. Methods to mitigate these effects will be essential for
jet multiplicity and energy scale measurements.

Substructure observables are expected to be especially sensitive to
pile-up~\cite{boost2010}. This is true in particular for the invariant
mass of large-size jets. Techniques such as the splitting and
filtering procedure used in this study reduce the effective area of
large jets and are therefore expected to reduce sensitivity to pile-up.

\begin{figure}
        \begin{center}
          \includegraphics[width=0.45\linewidth]{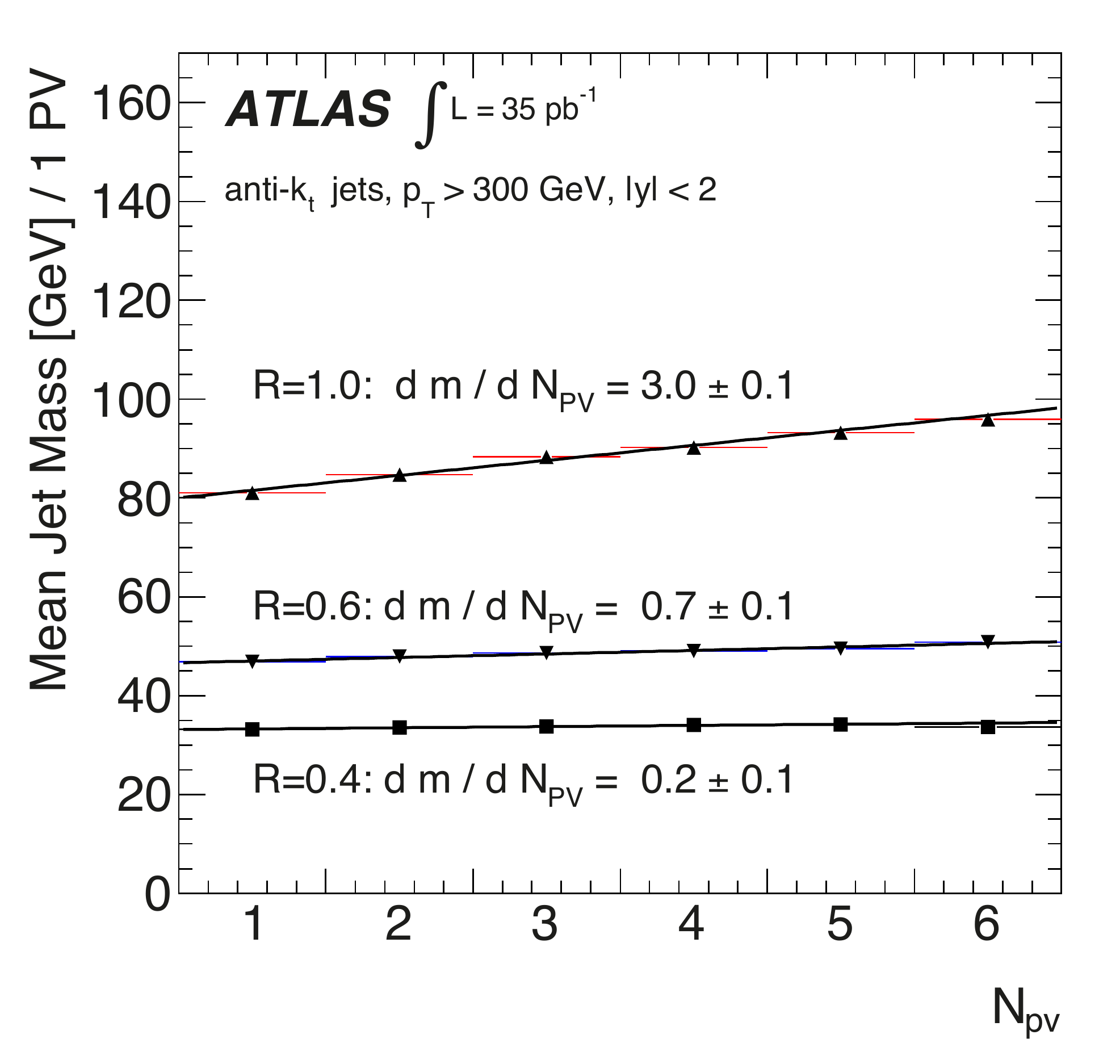}
          \includegraphics[width=0.45\linewidth]{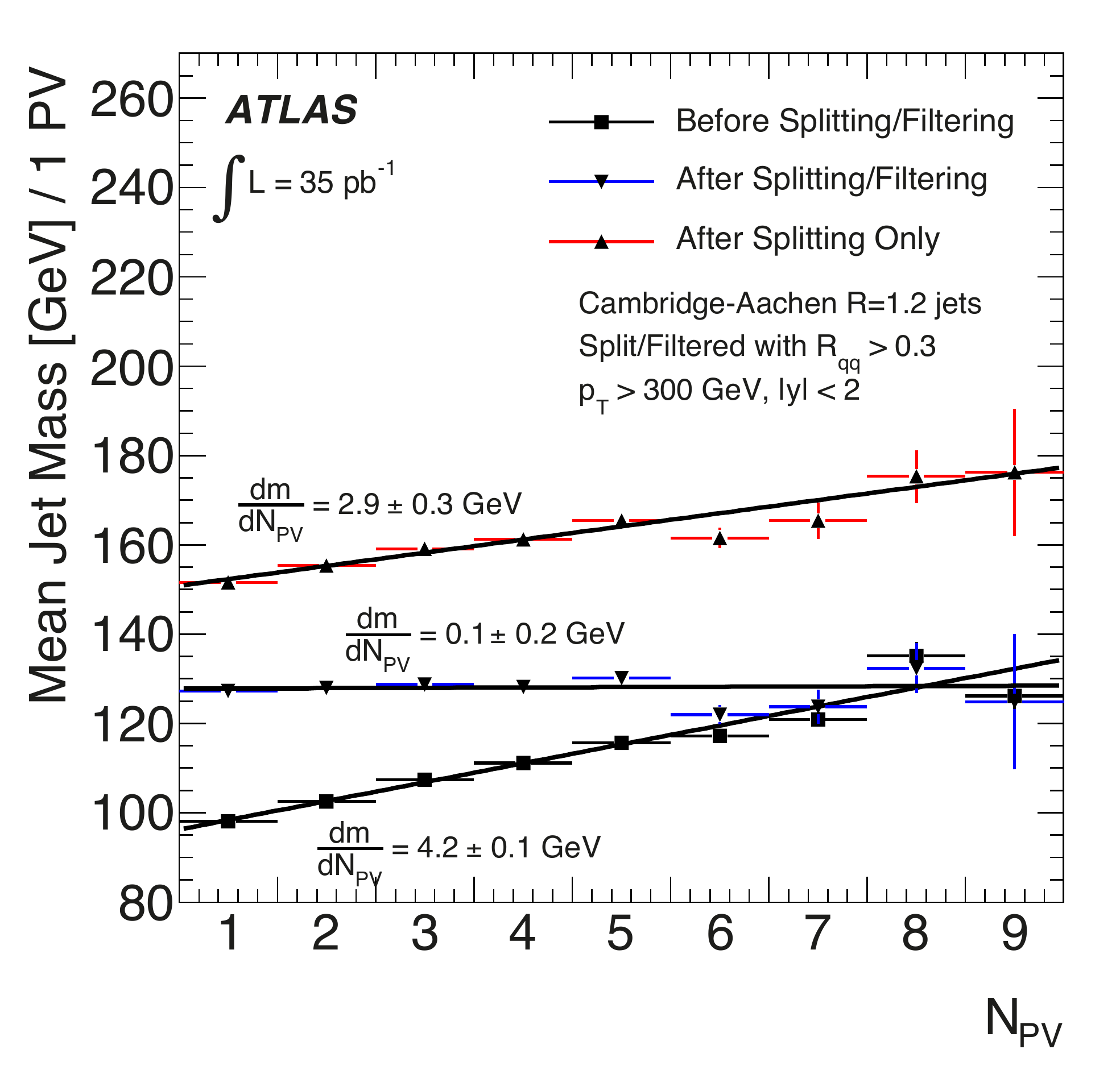}
          \caption{The mean mass for jets with $\pt > 300\GeV$ as a
            function of the number of primary vertices identified in
            the event. Comparisons show the effect for \antikt jets
            with different $R$-parameters (left) and \ca\ $R=1.2$ jets
            with and without splitting and filtering procedure
            (right). Each set of points is fitted with a straight
            line. }
    \label{fig:pileupeffect}
  \end{center}
\end{figure}

The sensitivity of mean jet mass to pile-up is tested in this
dataset. The correlation of the mean jet mass of \antikt\ jets with
the number of reconstructed primary vertices is presented in
Figure~\ref{fig:pileupeffect} (left). All jets with a $p_T$ of at
least 300 \gev\ in the rapidity range $|y|<$ 2 are considered. The
mean mass of jets in the absence of pile-up and the variation with
pile-up activity show the expected dependence on the jet size. The
mean mass in the $\Npv=$ 1 bin increases linearly with $R$. The ratios
of the fitted slopes $s_R$ are found to be:
\begin{align}
                   s_{1.0}/s_{0.6}  &=  4.3 \pm 0.5 & ((1.0/0.6)^3 &= 4.6), \\
                   s_{1.0}/s_{0.4}  &=  13 \pm 3 & ((1.0/0.4)^3 &= 15.6), \\ 
                   s_{0.6}/s_{0.4}  &=  3.0 \pm 0.8 & ((0.6/0.4)^3 &= 3.4), 
\end{align}
in good agreement with the ratio of the third power of the jet
$R$-parameter. This is in agreement with predictions of scaling of the
mean mass~\cite{Dasgupta:2007wa,BOOST2011}. This behaviour can also be
qualitatively explained by two factors. Firstly the jet area in the
$y-\phi$ plane grows roughly as $R^2$. Moreover, the
contribution of these particles to the jet mass scales with the
distance between them approximately as $R/2$, giving another power of $R$.

Figure~\ref{fig:pileupeffect} (right) shows the dependence on \Npv of
the mean jet mass before and after the splitting and filtering
procedure for \ca\ jets. Since the angular requirement $R_{jj} >
0.3$ is imposed, the splitting steps of this procedure naturally select more
massive jets. Since the splitting procedure selects a kinematically
biased subset of jets, a third line shows the mean mass prior to
filtering of jets that pass the splitting. The filtering step
significantly reduces the impact of pile-up on mean jet mass. In fact,
the slope of the straight line fitted to the filtered jet data points
is statistically consistent with zero.

Altogether, this demonstrates that the pile-up dependence of mean jet
mass in real LHC conditions matches expectations. Additionally, jet
substructure techniques that reduce the area of jets are promising
for suppressing the effects of pile-up.

\section{Conclusions}
\label{sec:conc}
Jet mass and several jet substructure variables have been
measured. This is the first particle-level measurement of these
variables at the LHC and in many cases the first at any
experiment. There is broad agreement between data and leading-order
parton-shower Monte Carlo predictions from \pythia\ and \herwigpp,
although there is some scope to improve this. Jet mass has generally
been found to exhibit the largest disagreements with Monte Carlo
simulations. However, in contrast to this, the masses of jets after
the \ca\ splitting and filtering procedure display good agreement both
with and between Monte Carlo simulations. The substructure variables
\DOneTwo, \DTwoThr, \tauTwoOne and \tauThrTwo are all reasonably well
reproduced by Monte Carlo predictions. Additionally, the effects of
pile-up on mean jet mass have been found to match phenomenological
expectations for $R$-parameter dependence. Splitting and filtering has
also been found to reduce the impact of pile-up significantly.

Generally these results show that jet mass and substructure quantities
can be successfully reproduced by leading-order parton-shower Monte
Carlo. This result bodes well for future analyses aiming to make use
of jet substructure techniques.


\section{Acknowledgements}

We thank CERN for the very successful operation of the LHC, as well as the
support staff from our institutions without whom ATLAS could not be
operated efficiently.

We acknowledge the support of ANPCyT, Argentina; YerPhI, Armenia; ARC,
Australia; BMWF, Austria; ANAS, Azerbaijan; SSTC, Belarus; CNPq and FAPESP,
Brazil; NSERC, NRC and CFI, Canada; CERN; CONICYT, Chile; CAS, MOST and NSFC,
China; COLCIENCIAS, Colombia; MSMT CR, MPO CR and VSC CR, Czech Republic;
DNRF, DNSRC and Lundbeck Foundation, Denmark; ARTEMIS and ERC, European Union;
IN2P3-CNRS, CEA-DSM/IRFU, France; GNAS, Georgia; BMBF, DFG, HGF, MPG and AvH
Foundation, Germany; GSRT, Greece; ISF, MINERVA, GIF, DIP and Benoziyo Center,
Israel; INFN, Italy; MEXT and JSPS, Japan; CNRST, Morocco; FOM and NWO,
Netherlands; RCN, Norway; MNiSW, Poland; GRICES and FCT, Portugal; MERYS
(MECTS), Romania; MES of Russia and ROSATOM, Russian Federation; JINR; MSTD,
Serbia; MSSR, Slovakia; ARRS and MVZT, Slovenia; DST/NRF, South Africa;
MICINN, Spain; SRC and Wallenberg Foundation, Sweden; SER, SNSF and Cantons of
Bern and Geneva, Switzerland; NSC, Taiwan; TAEK, Turkey; STFC, the Royal
Society and Leverhulme Trust, United Kingdom; DOE and NSF, United States of
America.

The crucial computing support from all WLCG partners is acknowledged
gratefully, in particular from CERN and the ATLAS Tier-1 facilities at
TRIUMF (Canada), NDGF (Denmark, Norway, Sweden), CC-IN2P3 (France),
KIT/GridKA (Germany), INFN-CNAF (Italy), NL-T1 (Netherlands), PIC (Spain),
ASGC (Taiwan), RAL (UK) and BNL (USA) and in the Tier-2 facilities
worldwide.

\bibliographystyle{JHEP}
\bibliography{substructure}

\clearpage
\begin{flushleft}
{\Large The ATLAS Collaboration}

\bigskip

G.~Aad$^{\rm 48}$,
B.~Abbott$^{\rm 110}$,
J.~Abdallah$^{\rm 11}$,
S.~Abdel~Khalek$^{\rm 114}$,
A.A.~Abdelalim$^{\rm 49}$,
A.~Abdesselam$^{\rm 117}$,
O.~Abdinov$^{\rm 10}$,
B.~Abi$^{\rm 111}$,
M.~Abolins$^{\rm 87}$,
O.S.~AbouZeid$^{\rm 157}$
% ...
\bigskip

% ...
$^{10}$ Institute of Physics, Azerbaijan Academy of Sciences, Baku, Azerbaijan\\
$^{11}$ Institut de F\'isica d'Altes Energies and Departament de F\'isica de la Universitat Aut\`onoma  de Barcelona and ICREA, Barcelona, Spain\\
% ...
$^{48}$ Fakult\"{a}t f\"{u}r Mathematik und Physik, Albert-Ludwigs-Universit\"{a}t, Freiburg i.Br., Germany\\
$^{49}$ Section de Physique, Universit\'e de Gen\`eve, Geneva, Switzerland\\
% ...
$^{87}$ Department of Physics and Astronomy, Michigan State University, East Lansing MI, United States of America\\
% ...
$^{110}$ Homer L. Dodge Department of Physics and Astronomy, University of Oklahoma, Norman OK, United States of America\\
$^{111}$ Department of Physics, Oklahoma State University, Stillwater OK, United States of America\\
% ...
$^{114}$ LAL, Univ. Paris-Sud and CNRS/IN2P3, Orsay, France\\
% ...
$^{117}$ Department of Physics, Oxford University, Oxford, United Kingdom\\
% ...
$^{157}$ Department of Physics, University of Toronto, Toronto ON, Canada\\
% ...
\end{flushleft}


\end{document}